\documentclass[fleqn,10pt]{wlscirep}
\usepackage[utf8]{inputenc}
\usepackage[T1]{fontenc}
\usepackage{authblk}
\usepackage{amsmath,amssymb}
\usepackage[convert-footnotes=true]{notes2bib}

\usepackage{lastpage,fancyhdr,graphicx}
\usepackage{epstopdf}
\usepackage{verbatim}
\usepackage{algorithm}
\usepackage{algpseudocode}

\usepackage{braket}
\usepackage{graphicx}
\usepackage{slashed}
\usepackage{sidecap}
\usepackage{amsthm}
\usepackage{float}
\usepackage{hyperref}
\usepackage{cite}
\usepackage{comment}
\usepackage{booktabs}
\usepackage{tabularx}
\usepackage{soul}
\usepackage[normalem]{ulem}

\newcommand\logten{\ensuremath{\log_{10}}}
\usepackage{geometry}
\usepackage{hyperref}
\hypersetup{colorlinks=true, citecolor=orange, urlcolor=blue, linkcolor=magenta}

\newcommand\fab[1]{{\color{black}#1}}
\newcommand\man[1]{{\color{black}#1}}

\title{Bow-Tie Structures of Twitter Discursive Communities}
\author[1,2]{Mattia Mattei}
\author[1,3]{Manuel Pratelli}
\author[4,5,1]{Guido Caldarelli}
\author[3,1]{Marinella Petrocchi}
\author[6,1,7,*]{Fabio Saracco}

\affil[1]{IMT School For Advanced Studies Lucca, p.zza San Francesco 19, 55100 Lucca, Italy}
\affil[2]{Alephsys Lab, Universitat Rovira i Virgili, Av. Paisos Catalans 26 Tarragona,
Tarragona E-43007, Catalonia, Spain}
\affil[3]{Institute of Informatics and Telematics, National Research Council,
  via Moruzzi 1,
  56124
  Pisa,
  Italy
}
\affil[4]{Ca' Foscari University of Venice, Department of Molecular Sciences and Nanosystems, Ed. Alfa,
  Via Torino 155,
  30170, Venezia Mestre, Italy}
\affil[5]{
  European Centre for Living Technology (ECLT),
  Ca' Bottacin, 3911 Dorsoduro Calle Crosera,
  30123
  Venice,
  Italy}
\affil[6]{Institute for Applied Mathematics ``Mauro Picone'', National Research Council, via dei Taurini 19, 00185 Roma, Italy}
\affil[7]{Centro Ricerche ``Enrico Fermi", via Panisperna 89 A, 00184 Roma, Italy}
\affil[*]{Corresponding author: fabio.saracco@cref.it}
\begin{abstract}
\fab{Bow-tie structures were introduced to describe the World Wide Web (WWW): in the direct network in which the nodes are the websites and the edges are the hyperlinks connecting them, the greatest number of nodes take part to a \emph{bow-tie}, i.e. a Weakly Connected Component (WCC) %\fab{divided in 7 sectors, according to the ability of the nodes in each sector to access the other ones. In the WWW, the bow-tie permits to characterize websites as search engines or authorities (as Wikipedia) due to the sector they belonged to.}\\
composed of 3 main sectors: IN, OUT and SCC. SCC is the main Strongly Connected Component of WCC, i.e. the greatest subgraph in which each node is reachable by any other one. The IN and OUT sectors are the set of nodes not included in SCC that, respectively, can access and are accessible to nodes in SCC. In the WWW, the greatest part of the websites can be found in the SCC, while the search engines belongs to IN and the authorities, as Wikipedia, are in OUT.\\
In the analysis of Twitter debate, the recent literature focused on discursive communities, i.e. clusters of accounts interacting among themselves via retweets. In the present work, we studied discursive communities in 8 different thematic Twitter datasets in various languages. Surprisingly, we observed that almost all discursive communities therein display a bow-tie structure during political or societal debates. Instead, they are absent when the argument of the discussion is different as sport events, as in the case of Euro2020 Turkish and Italian datasets.\\
We furthermore analysed the quality of the content created in the various sectors of the different discursive communities, using the domain annotation from the fact-checking website Newsguard: we observe that, when the discursive community is affected by m/disinformation, the content with the lowest quality is the ones produced and shared in SCC and, in particular, a strong incidence of low- or non-reputable messages is present in the flow of retweets between the SCC and the OUT sectors. In this sense, in discursive communities affected by m/disinformation, the greatest part of the accounts has access to a great variety of contents, but whose quality is, in general, quite low; such a situation perfectly describes the phenomenon of infodemic, i.e. the access to ``\emph{an excessive amount of information about a problem, which makes it difficult to identify a solution}", according to WHO).}
\end{abstract}

\begin{document}

\maketitle

\section{Introduction}
Since their first introduction, Online Social Networks (OSN) have been deeply  investigated for %the study of 
possible implications of the online public debate on political processes~\cite{Adamic2005}. In the last decade, the centrality of OSN for political communications and debates  has steady increased: OSN represent one of the most used tool for citizens to get an opinion~\cite{eurobarometer2019}. It is not surprising, then, that political parties use them extensively to carry out a sort of never-ending propaganda. 

% While there are different opinions on the role (if any) of different partitioning of users in partisan groups in shaping offline behaviours~\cite{Dubois2018,Valensise2021a,Gallotti2021}, 

Although in the literature there are different opinions on the impact that a particular grouping of users in OSN can have on their offline behavior~\cite{Dubois2018,Valensise2021a,Gallotti2021}, 
it is undeniable that the online social environment is strongly polarized. The origin of such a polarization has been deeply discussed in the sociological literature ~\cite{Urman2019,Yarchi2020} and seems to be extremely dependent on country's party systems~\cite{Barbera2015}. In particular, the concepts of \emph{selective exposure},  \emph{confirmation bias},  \emph{echo chambers} and \emph{filter bubbles} have had a great relevance in the literature.

% commenti da Walter
%In Online Social Networks produced a drastic shift to a disintermediated information system, in which every users can potentially be at the same time the producer and the target of novel pieces of news. Nevertheless
%In real data it was observed that

Selective exposure leads people to prefer information that confirms their preexisting beliefs~\cite{Lazer1094,gangware2019weapons}, while confirmation bias makes information consistent with one's preexisting beliefs more persuasive~\cite{DelVicario2016}.

% Online users tend to focus on messages that adhere to their pre-existing belief, by ignoring dissenting information, a process that has been called \emph{confirmation bias}~\cite{DelVicario2016}. 

Such phenomena imply the formation of groups of users, characterised by following the same information in terms,  e.g., of news outlets and personal opinions. These groups are thus closed in so called \emph{echo chambers}: ``a bounded, enclosed media space that has the potential to both magnify the messages delivered within it and insulate them from rebuttal"~\cite{jamieson08echo,Garrett2009,DelVicario2016}. Echo chambers, by being  impervious to information coming from outside that may contradict the pre-existing views of the chamber members, are believed to strongly contribute to the polarization of the online debate~\cite{Zollo2017}.
%In fact, message amplification occurs because people in an echo chamber seek information that reinforces their pre-existing beliefs. 
%The non-controversial nature of this information is the result of the absence of data in the chamber that contradicts the pre-existing views of the chamber members. 

Polarization \fab{may be} also fomented by \emph{filter bubbles}. This paradigm was first introduced by the activist Eli Pariser in 2011~\cite{Pariser11hide}: personalised results provided by search engines and shown in social media feeds can make users be trapped in a bubble of information they like and away from data and viewpoints considered less valuable, but that could challenge their beliefs. \fab{Although the user may not be affected in real life by the virtual bubble, due to the various communication channels he or she can take advantage of (see Ref.~\cite{Bruns2019}), the customization of algorithms may contribute to the formation of a virtual world apart.}

\paragraph{Discourse and discursive communities} Whether circulating within an echo chamber or suggested by recommendation algorithms, the type of information users come across online is fundamental to reinforcing or not the division into `closed' groups. Nevertheless, also the study of the interactions between users is of absolute interest to detect polarization phenomena.
The term \emph{discourse community} was coined in 1982 and it indicates `groups that have goals or purposes, and use communication to achieve these goals'~\cite{borg03discourse}.
%The analysis conducted in~\cite{porter92audience} is rather interesting: 
A discourse community is itself immaterial, and this tends to project it onto the forum on which it operates~\cite{porter92audience}. 
Thus, with the advent of OSN, discourse communities were projected onto the platforms themselves~\cite{kehus10definition}: `A discourse community can be viewed as a social network, built from participants who share some set of communicative purposes'. %The concept of the evolution of communities in the social sphere is also very interesting. 
According to Berkenkotter~\cite{berkenkotter93}, `just as the digital world is constantly evolving, discourse communities continually define and redefine themselves through communications among members'. 
% In ambito linguistico, nel 1982 e' stato coniato per la prima il termine `discourse community', definito come `groups that have goals or purposes, and use communication to achieve these goals'~\cite{borg03discourse}. 
% Interessante e' l'analisi condotta in~\cite{} (Porter, J. (1992). Audience and Rhetoric: An Archaeological Composition of the Discourse Community. New Jersey: Prentice Hall.). Una communita' discorsiva e' per se' immaterial, e questo tende a proiettarla sul forum su cui essa opera. Con l'avvento delle social networks, le comunita' discorsive si sono spostate, e sono proiettate su, le piattaforme stesse~\cite{} (Kehus, Marcella1, Kelley2 Walters, and Melanie3 Shaw. "Definition And Genesis Of An Online Discourse Community." International Journal of Learning 17.4 (2010): 67–85. Education Source. Web. 30 Oct. 2015.) `A discourse community can be viewed as a social network, built from participants who share some set of communicative purposes'. Interessante anche il concetto di evoluzione, legato all'ambiene social: Just as the digital world is constantly evolving, "discourse communities continually define and redefine themselves through communications among members", according to Berkenkotter~\cite{} 

% e allora...o vediamola un po' l'evoluzione delle discourse community: 
%{\bf QUA un cappellino su quali comunita' scientifiche hanno definito queste communities?}
In the discourse community definition, we implicitly know the identities of the individuals forming the community. Actually, in the case of Twitter, it is just partially true, since we have trustworthy information only about a small minority of accounts. For this reason, we prefer to use the term \emph{discursive communities}, as it was introduced in Ref.~\cite{Radicioni2021a} to identify group of users that are connected by non-trivial pattern of discourse, but for which we have limited information about the identity of the group itself. Nevertheless, since we can \emph{infer} the discourse community of the discursive community by looking at a set of non-trivial data characterising the group, as the most frequent keywords used therein, the difference is more formal than substantial. Therefore, in the following we will use the two terms interchangeably.

%\textcolor{red}{In the field of Computational Social Science, discourse communities can be often found under the name of discursive communities, nevertheless they represent the same concept: in the following we will use the two terms interchangeably}. 
To detect discursive communities in OSN, the first contributions applied mixed approaches to the political debate on Twitter~\cite{Conover2011, Conover2011a, Conover2012}. The work considered political debate on Twitter about the US presidential election campaign, i.e. a `perfectly polarized' one in which two opposite fronts face each other. %The authors manually annotated the political orientation of the more active users on Twitter via the keywords they used. 
The authors manually annotated the most frequent keywords characterizing Republicans and Democratics' narratives and use them to infer the political orientation of accounts using them. The orientation of accounts not using hashtags was later inferred using a label propagation algorithm~\cite{Raghavan2007b}. 
Remarkable, a clear partition in two distinct groups of users, supporters of the two political parties, was observed in the \textit{retweet network} only (the network of users sharing content created by others). 
%
%
% it was observed  that, among all the possible interactions amongst users, only in the \textit{retweet network}, i.e., the network of users sharing content created by others, there was a clear partition in two different groups, related to political orientation.
%
Finally, using a label propagation algorithm on the retweet network, the authors were able to successfully assign  to all accounts the proper political orientation, that can be translated in the present context to the correct discursive community. %\fab{@FABIO: altri da citare, per non citare solo gli sloveni che lo hanno chiesto espressamente?}\\  

Every country has a different way in which opinions are polarised. This is due to the various party systems and electoral laws and, in principle, there could be more than just two fronts~\cite{Barbera2015}. 
% Thus, the proper partition in more discursive communities should rise directly from data, instead of being based on \emph{a priori} manual annotation.
A methodology for detecting discourse communities less susceptible to human error should therefore rise from the data directly, rather than being based on \emph{a priori} manual annotation.
The approach firstly proposed in Ref.~\cite{Becatti2019} meets the desired property:
% overcomes this issue related to the method above: 
the idea is to infer the various discursive communities starting from accounts whose identity is certified by the social network itself. In Twitter, these are the so called \textit{verified} accounts. 
%
% verified accounts, i.e. those users whose identity is certified by Twitter platform itself. 
%
This class of accounts tend to produce new content rather than retweet the one created by others~\cite{Becatti2019}. Since we trust information regarding their identities, the issue is the identification of the discursive communities anchored to them, something that can be done using their interactions with `standard' users (based on the results by Conover et al.~\cite{Conover2011, Conover2011a, Conover2012}, in terms of retweets).

Let the reader consider a pair of verified Twitter users. If they share a large number of retweeters, it is reasonable to think that they attract `similar' users.
%
% In fact, any pair of verified accounts is claimed to involve similar users if they share a significant number of retweeters: 
%
In this sense, that pair of accounts are perceived to belong to the same discursive community, sharing similar views and ideas. Nevertheless, it is hard to state \emph{a priori} how many common retweeters two verified users should have in order to be considered as `similar'; in this sense, a maximum entropy null-model is used as benchmark~\cite{Cimini2018}. (More technical details on this construction can be found in the Section~\ref{sec:disc}.)
The labels for verified users are then propagated, following the same approach as in Ref.s~\cite{Conover2011, Conover2011a, Conover2012}. % (more details on this method can be found in Section~\ref{sec:disc}). 
In the present paper, we are going to follow this approach, that has great performances on manually annotated datasets~\cite{Saracco2022}.\\

%In the present manuscript, we address discourse communities in social networks. Taking Twitter as a benchmark, we consider  communities of accounts interacting among themselves via expression of interest, like retweets. 

The recent literature has extensively analysed online debate and discourse communities, focusing, from time to time, on coordinated activities in discursive communities~\cite{Becatti2019, Caldarelli2020,Bruno2021}, on the semantic network associated to the various discursive communities~\cite{Radicioni2021a,Patuelli2021, Radicioni2021b}, on their exposure to disinformation campaigns~\cite{Caldarelli2021, Mattei2021}, \fab{and on their dynamical evolution~\cite{Sluban2015, Cherepnalkoski2016, Uyheng2021, Evkoski2021plosone,Evkoski2022plosone}.}

% In the analysis of online debates, the recent literature analysed  discursive communities, i.e., communities of accounts interacting among themselves via expression of interest, like retweets in the case of Twitter.
%i.e. sharing the messages produced by other accounts
In the present paper, we tackle the analysis of the network structure of discursive communities: we collect and study 8 thematic Twitter datasets, on topics ranging from sports, to Covid-19, to political elections and immigration policies. Our main result is that  \textit{almost  all the  discourse  communities  therein features a bow-tie structure.}\\ 
 
\paragraph{Bow-ties} Bow-tie structures were initially introduced by Broder et al. in order to study the structure of World Wide Web (WWW)~\cite{Broder2000}. 
%Such structure is a partition of  nodes in a directed network into different sectors. 
The authors represented WWW as a directed network in which webpages are the nodes and the hyperlinks connecting them are the edges. Broder et al. noticed that the network displays a huge Weakly Connected Component (WCC), i.e., the maximal subgraph in which all nodes can be reached by any other one in the same subgraph, disregarding the direction of the link. This WCC includes more than 75\% of all nodes.

WCC breaks into three main pieces: a Strongly Connected Component (SCC), in which each node can be reached by any other one in the same block, following the direction of the links; a group of nodes that can reach SCC, without being reached by it (called IN); a group of nodes that can be reached by SCC, but that cannot reach it (the OUT block). The observation is that SCC is the most populated sector, followed by the IN and the OUT sectors.  Most of the websites can be found in the SCC, linking between each other; the IN sector was instead mostly composed by search engines, while the OUT one includes authorities, as Wikipedia. 

\begin{figure}[ht!]
     \centering
     \includegraphics[scale=0.8]{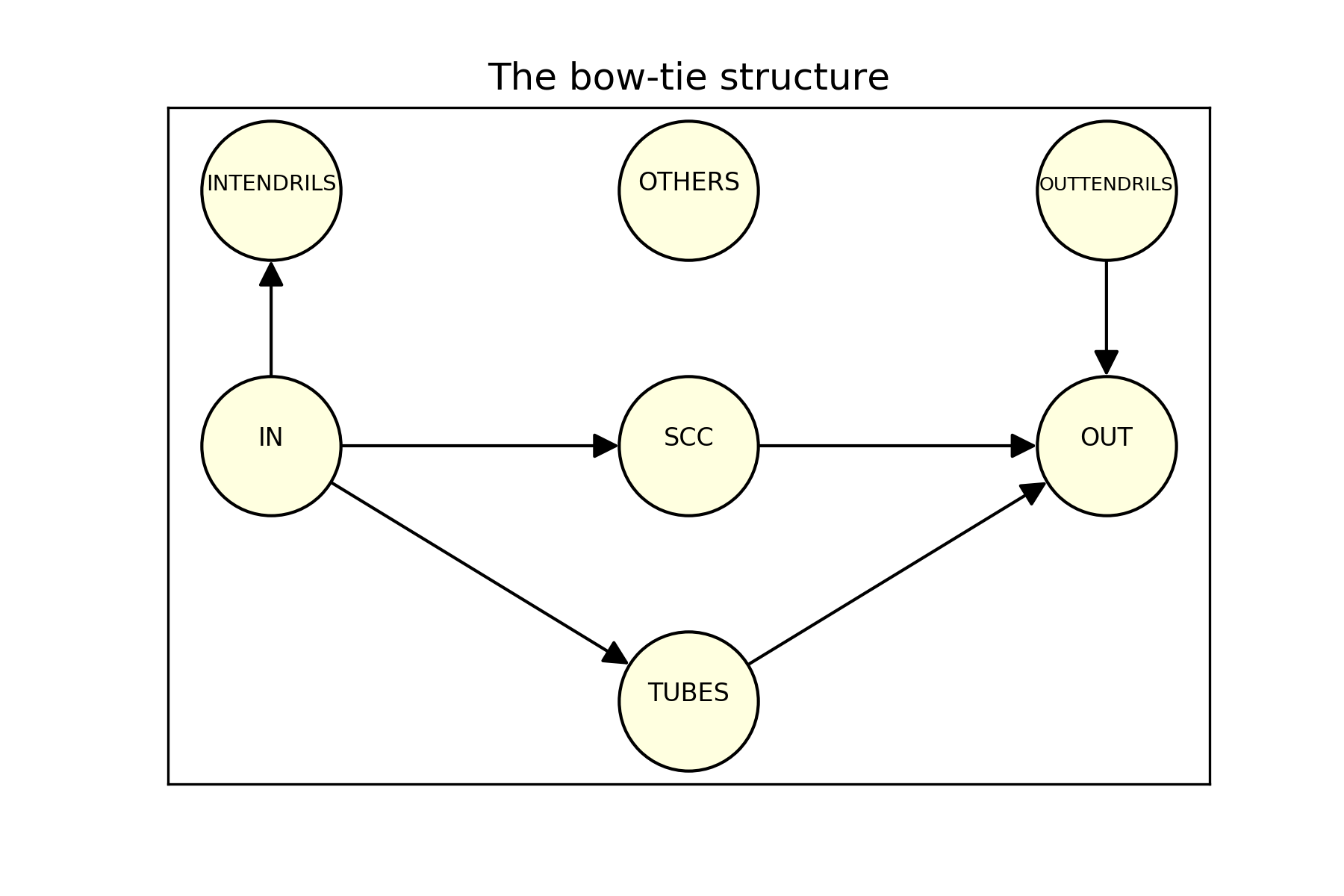}
     \caption{\textbf{The seven sectors of Yang's bow-tie structure}}
     \label{fig:structure}
 \end{figure}

Yang et al.~\cite{Yang2011} refined the partition of the  structure in~\cite{Broder2000}, introducing INTENDRILS, OUTTENDRILS, TUBES and OTHERS. %TUBES are nodes that can be accessed by the IN sector and that can access OUT, but they do not form a SCC; INTENDRILS are nodes that can be accessed by nodes in IN and  have no access to nodes in the other blocks;  OUTTENDRILS are nodes that can access nodes in OUT, but no other blocks~\cite{Yang2011}. 
The entire situation is pictorially represented in Fig.~\ref{fig:structure}.\\ \fab{Remarkably, the bow-tie structure was detected also in control network of transnational corporations, having deep implications on financial stability~\cite{Vitali2011}.}

\paragraph{Results in a nutshell} In the case of our 8 thematic datasets, we find that  a bow-tie structure is present in those discourse communities debating 1) about politics, like in the case, e.g., of election campaigns, and 2) about society, e.g., on 
%‘how to handle a pandemic?’  or ’how to manage migration fluxes?’,
the proper response to the pandemic or the appropriate management of migration fluxes. Instead, bow-ties 
are hardly present when the debate is about less socially relevant topics as sports (this  confirms what observed in Ref.~\cite{Barbera2015}). 

There are two relevant points in observing the presence of bow-tie structures in discursive communities: how big the bow-tie is respect to the entire discursive community (the greatest the accounts
in the non-OTHERS blocks, the more informative the bow-tie structure is)
% (a feature that is called \textit{uninformative}, \textit{weak} or \textit{strong} bow-tie in the following, and ) 
and how random the presence of this structure is (i.e., its statistical significance). Regarding the first point, when the bow-tie is informative, even in the worst case, it represents more than 80\% of all nodes in the discursive community. Regarding the second point, in order to be sure that the observed bow-ties are not due to a random organization of links only, we compare the observed quantities with a maximum entropy null-model for directed network, conserving the in- and out-degree sequences~\cite{Mastrandrea2014}. The results show that the dimension of most of the bow-tie sectors are statistically significant, i.e., they carry a signal that cannot be due to the degree sequence only. In this sense, the presence of a bow-tie structure is an extremely non-trivial feature of the system.\\

We can add more detail to the analysis of this structure. When the bow-tie is informative, we observe two cases: the OUT-dominant and the INTEND-dominant ones, depending on which sector is the largest (respectively, OUT and INTENDRILS).  The OUT sector has access to all information produced in the discursive community and, in particular, to the one produced by the most active block, SCC. 
% Thus, in principle, the OUT-dominant bow-tie should be more informed regarding the content shared in the discursive community. 
Instead, in the INTEND-dominant bow-ties, the most crowded sector is the one of INTENDRILS, i.e., the retweeters of  IN  that are not retweeted by anyone else and that cannot access to all content created by SCC.\\

% In principle, it should be desirable to have an OUT-dominant bow-tie,  since the access to more information may permit the related accounts to have a knowledgeable opinion. 
In principle, it should be desirable to have an OUT-dominant bow-tie: when the OUT sector is the most populated, there are many accounts that are exposed to information from all other sectors. This should give the accounts a multi-faceted, pluralistic knowledge. However, for the investigated datasets,  we carry out an analysis on the production and  quality of content in the various sectors of the bow-ties, and the outcome returns a different picture. \fab{In fact, regarding content production, SCC is the source  of the greatest flux of content. When the discursive community is affected by m/disinformation, the incidence of links from non-reliable sources shared by SCC is much greater than by any other sector and is particularly considerable in the flux between SCC and OUT. In those cases when OUT-sectors dominate the bow-tie, we observe an \emph{infodemic}\footnote{According to WHO, ``\emph{infodemics are an excessive amount of information about a problem, which makes it difficult to identify a solution}". Coronavirus disease 2019 (COVID-19)
Situation Report – 45:~\url{https://www.who.int/docs/default-source/coronaviruse/situation-reports/20200305-sitrep-45-covid-19.pdf?sfvrsn=ed2ba78b_4}},  since OUT, i.e. the widest block, is directly hit by the huge amount of messages of questionable quality produced by SCC.}
%\fab{Second, the analysis on the quality of content highlights that, in communities where links of untrustworthy news sources circulate,   these links are more likely to be produced by SCC. Thus, OUT-dominant bow-ties are exposed to m/disinformation.}
%
% which, in discursive communities  affected by m/disinformation, 
% is responsible of the greatest flux of contents from non reliable sources. In this sense, since the greatest block OUT is directly hit by questionable content produced by SCC, the OUT-dominant bow-ties are exposed to m/disinformation campaigns. This creates local infodemics. According to WHO, ``\emph{infodemics are an excessive amount of information about a problem, which makes it difficult to identify a solution}"\footnote{Coronavirus disease 2019 (COVID-19)
% Situation Report – 45:~\url{https://www.who.int/docs/default-source/coronaviruse/situation-reports/20200305-sitrep-45-covid-19.pdf?sfvrsn=ed2ba78b_4}}.
\\

Summarising, our contribution is twofold: 
\begin{itemize}
    \item almost all the discursive communities in the  8 investigated datasets of Twitter debates, on different topics in different countries, display a bow-tie structure which is statistically significant; 
    %\item \fab{in discursive communities where political and societal issues are debated, where an OUT-dominant bow-tie is present, and where the dissemination of posts with unreliable content exists, the majority of users is exposed to m/disinformation.}
    \item \fab{when the bow-tie is affected by m/disinformation and it is OUT-dominant, the majority of users (i.e. those in the OUT block) is exposed to the flux of m/disinformation. In this sense, the bow-tie structure fuels the phenomenon of infodemic.}
    % \item relating the presence of the bow-tie with the concept of infodemics and the production of controversial content, in most cases the majority of users is exposed to untrustworthy information. 
\end{itemize}

We would like to remark that the results in this manuscript do not represent  the only contribution that connects the diffusion of m/disinformation to the network structure (see, for instance, work in~\cite{Artime2020,Guarino2021,  Valensise2021, Castioni2021}, just to consider some of the most recent contributions). 
% However, this is the  first time that the diffusion of m/disinformation is related to the presence of the bow-tie structure in discursive communities.
\fab{However, to the best of our knowledge, for the first time the bow-tie structure emerges in online discursive communities. Moreover, its presence and its peculiarity permit do have a proper description of the phenomenon of infodemic.}
\section{Results}
\subsection{Datasets}
In order to make our analysis as general as possible, we consider several Twitter datasets across different countries and about different topics. \fab{The data collected using the Twitter Streaming API are publicly available for further research and reproducibility and can be found at the following URL: \url{https://toffee.imtlucca.it/datasets}}. In detail:
\begin{itemize}
    \item \textbf{Covid-19 datasets}: we explore Twitter posts containing keywords related to the Covid-19 pandemic\footnote{In particular, the keywords for tweets collection were  ``coronavirus", ``ncov", ``covid", ``SARS-CoV2", ``\#coronavirus", ``\#coronaviruses", ``\#WuhanCoronavirus",
``\#CoronavirusOutbreak", ``\#coronaviruschina", 
``\#coronaviruswuhan", ``\#ChinaCoronaVirus", ``\#nCoV", ``\#ChinaWuHan", ``\#nCoV2020", ``\#nCov2019", ``\#covid2019", ``\#covid-19", ``\#SARS\_CoV\_2", ``\#SARSCoV2", ``\#COVID19". The subset of Italian messages has been matter of investigation in  Ref.~\cite{Caldarelli2021} too.}, in different languages and therefore diffused in different countries. In particular, we consider the \textbf{Italian}, \textbf{German} and \textbf{French} %and \textbf{Spanish}
    debates about the pandemic, in the period between February and April 2020. The Italian dataset consists of \man{4,470,648} tweets published between February 17 and April 23. The German dataset contains \man{1,552,106} tweets posted between February 10 and April 23, the French one has \man{3,052,708} posts published between March 23 and April 7. %and the Spanish one contains 1,330,813 posts shared between February 24 and March 24 For the Spanish dataset, we selected only those tweets geolocated in Spain (i.e., tweets whose authors have made their geographical position known on Twitter at the time of publication) or whose author indicates `España', `Spain', `Madrid' or `Barcelona' as user location. This selection has been chosen since the Twitter API permits to select posts only according to the language and Spanish is spoken in many different countries.
    The different time frames for data collection have been chosen according to the intensity of the Twitter traffic.
    
    \item \textbf{Dutch elections dataset}: we collect Twitter posts about the national elections in the Netherlands in 2021. The keywords used for downloading data were ``tweedekamer", ``verkiezingen", ``kabinet", ``coalitie", ``stem", ``stembus", ``verkiezingen2021"\footnote{Respectively, ``House of representatives", ``reconnaissance", ``cabinet", ``coalition", ``vote", ``ballot box", ``explorations".} and only messages in Dutch were selected. The dataset contains \man{1,002,499} tweets posted between February 2 and March 31, 2021.

    \item \textbf{Italian debate on migrants}: we select Twitter posts shared in Italy with keywords regarding the discussion about the migration flows from Northern Africa to the Italian coasts. The dataset consists in \man{1,081,780} posts, published between January 23, 2019 and February 22, 2019.  The dataset is described in more details in Ref.~\cite{Caldarelli2020}.
    
    \item\textbf{Italian debate on the Astrazeneca vaccine}: we examine \man{583,236} Twitter posts published in Italian,  regarding the discussion about the safety of the Astrazeneca vaccine against Covid-19: the keywords used for the download were ``astrazeneca", ``aifa", ``ema", ``trombosi"\footnote{Respectively, ``astrazeneca", "Italian Medicines Agency", ``European Medicines Agency" and ``thrombosis".}. The dataset contains posts shared between March 15, 2021 and May 15, 2021.
    \item\textbf{Italian and Turkish EURO2020}: we analyze \man{144,725} Italian tweets and \man{430,374} Turkish ones about the European Football Championship EURO2020; the keyword used for the download was simply ``\#euro2020". The tweets were published between, respectively,  June 11-13 and June 11-23, 2021.
\end{itemize}
So as not to burden the presentation, in the following we will present the results about the Italian Covid-19, Italian EURO2020 and Turkish EURO2020 datasets. We will show the results related to the other datasets wherever there will be something substantially different, compared with the Italian dataset. However, all graphics and results about the other datasets can be found in the Supplementary Material.

\subsection{Discursive communities}\label{sec:disc}
Our analysis focuses on the structure of  networks of retweets, for each dataset. Retweeting a post is one of the possible ways in which people can interact on Twitter and it consists in sharing the content of a tweet written by another user. It usually means endorsing the post content \fab{as }it has also the effect of raising \fab{the} visibility \fab{of the original post. It was also shown that, among all possible interactions, retweets are the best performing to infer the political orientation of the various accounts}~\cite{Conover2011, Conover2011a, Conover2012}. 

We start by distinguishing between \emph{verified} and \emph{non-verified} accounts. The former ones denote Twitter users whose identity has been verified by the social platform. This procedure is usually adopted to certify the accounts of renowned people and organizations and figures of public interest in general, as politicians, journalists, political parties, newspapers and TV-channels. We place the verified accounts on one layer of a bipartite network\footnote{In a bipartite network, nodes belong to two different sets, called layers. An edge can exist only between vertices placed on different layers.} and the non-verified ones on the other one, again considering links as retweets between them\footnote{In the present construction, we disregard the information about the direction of the retweet, since we are interested in the interaction between the two class of users. Nevertheless, as mentioned above, verified users tend more to create new contents (i.e tweet) than to share it with her followers (retweet).}. 
The main idea is to anchor the definition of discursive communities on verified users since they usually introduce new content and posts: as observed in many other studies~\cite{Becatti2019,Caldarelli2020, Caldarelli2021,Radicioni2021a, Radicioni2021b, Mattei2021, Gonzalez2021}, verified users are, on average, much more retweeted than common users. Such a procedure obtains great performances, since it can be observed that the various discursive communities are coherent in terms of verified users belonging to the same political front; in a further analysis we are comparing this procedure with annotated datasets, better quantifying our performances~\cite{Saracco2022}.
%This time, we ignore the direction of the edges, because verified users are those who usually introduce new content and posts and therefore they are intensely retweeted by common users. 

Following the methodology introduced in Becatti et al.~\cite{Becatti2019}, we count the common neighbors of each pair of verified users or, in simpler words, the number of non-verified users that have interacted (by retweeting or being retweeted) with the same pair of verified ones. The aim is projecting the bipartite network into the layer of the verified accounts, establishing an edge between two of them if the number of their common neighbors is significantly higher than what expected by a proper null-model. When this happens, we can assert that the two verified users refer to the same audience and, therefore, they probably share similar content and opinions. The statistical significance of the number of common neighbors can be established only comparing it with the predictions of an accurate benchmark, which, in this case, is represented by the Bipartite Configuration Model (\emph{BiCM},~\cite{Saracco2015a}), an entropy-based model suited for bipartite networks. A complete description of the model and the projecting procedure can be found in Section~\ref{sec:netmethods}. 

The result of the above procedure is a monopartite network of verified users. We further obtain a partition in communities implementing the Louvain algorithm~\cite{Blondel2008} for the optimization of the modularity, with a slight modification. \fab{In fact, the standard definition of the modularity~\cite{Newman2004} implements the Chung-Lu null-model~\cite{Chung2002}, which can be considered as a sparse matrix approximation of the entropy-based null-model defined in~\cite{squartini2011analytical} and it is known to return wrong results in the presence of strong hubs~\cite{Cimini2018}.} We thus replaced the Chung-Lu null-model in the modularity with the unipartite configuration model (\emph{UCM}) defined in Ref.~\cite{squartini2011analytical}. \fab{Furthermore, we  correct for the node ordering bias that affects Louvain algorithm, independently on the objective function chosen. In fact, we perform multiple runs, each time reshuffling the order of the nodes: we finally select the partition displaying the greatest value of the (UCM-modified) modularity.} More details can be found in Section~\ref{sec:netmethods}.

For all the datasets, looking at the members of each discursive community, we can \emph{a posteriori} associate the latter to a political wing, using the available information for verified users. We thus obtain clusters of users (even if we cannot characterise them on the basis of other topological quantities\cite{zlatic2009rich}) which represent the main wings of the political scenario of each of the examined countries. In addition, in almost all the datasets,  we identify also a Media cluster, with official accounts of newspapers, TV-channels, radio and other media.\\ 

%Once the labels for verified users are obtained using the approach above, we propagate them on the retweet network, using the method by Raghavan et al.~\cite{Raghavan2007b} as in Ref.s~\cite{Conover2011, Conover2011a, Conover2012}. This second network is a monopartite and directed one in which nodes represent users and a link between them indicates that one user has retweeted the other one at least once: the single edge starts from the retweeted user and is directed towards the one who retweets.\\

In the Section~\ref{ssec:disccomm}, the interested reader can find  a complete description of all the discursive communities for the Italian Covid-19 dataset. For the other datasets, a brief description of their discursive communities is in the Supplementary Material.

\subsubsection{Political orientation of non-verified users}
 The next step in our procedure consists in extending the discursive communities to non-verified accounts. %The idea is to consider the tag obtained for verified account and propagate them on the retweet network.
 %The idea is to  to the interactions in the retweet network between the accounts in the former discursive communities with the other users. 
 More in details, following the approach in Ref.~\cite{Caldarelli2020}, we use the membership of verified users as (fixed) seeds for the label propagation algorithm proposed by Raghavan et al.~\cite{Raghavan2007b} on the retweet network. This network is a monopartite and directed one in which nodes represent users and \fab{ links start from the retweeted users and are directed towards the one who retweets}. Let us remind that, in case the algorithm cannot find a dominant label for a specific vertex (i.e., in case of a tie), it randomly removes some of the edges attached to that vertex and repeats the procedure: for this reason, we run the label propagation 500 times and assign to each node the most frequent label (actually, the noise in the assignment of the labels is extremely limited). 
 
\begin{figure}
    \centering
    \includegraphics[scale=0.35]{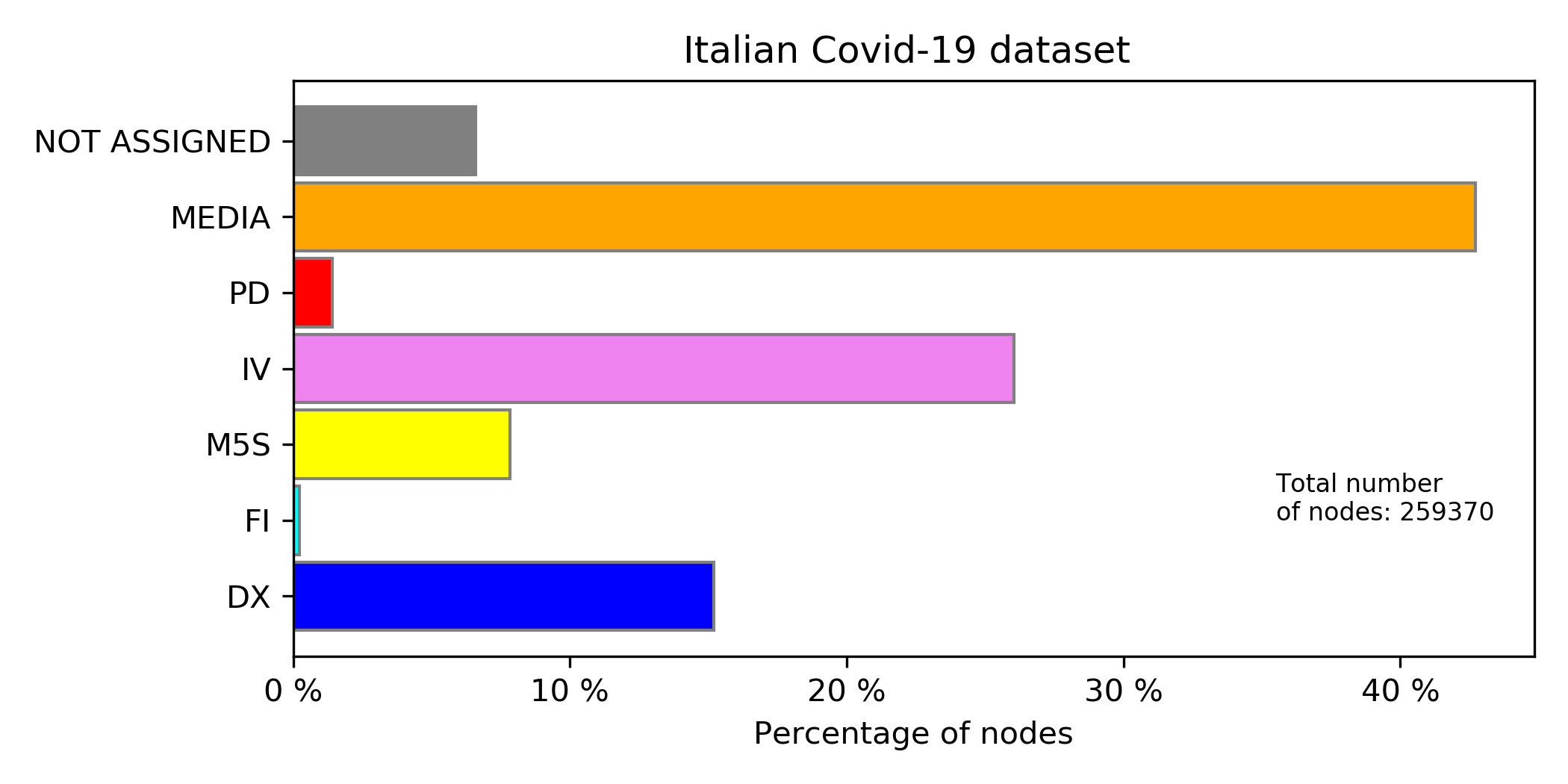}
    \caption{\textbf{Percentages of nodes in each discursive community, Italian Covid-19 dataset.} Due to the presence of politicians and political parties from a specific political area, the various discursive communities are called following their political alignment. ``PD" stays for the Italian Democratic Party (\emph{Partito Democratico}); \emph{Italia Viva} (``IV") is the political party of the former prime Minister and former PD secretary Matteo Renzi, while M5S is the ``Movimento 5 Stelle", a political movement born on the web and being the most represented party in the Italian parliament at the time of the data collection. ``FI" stays for \emph{Forza Italia}, the political party of the former Prime Minister Silvio Berlusconi, while the ``DX" (\emph{Destra}) community includes right wing parties as Lega and Fratelli d'Italia. The most crowded discursive community is the one of Media in which there are most of the online news outcasts and newspapers. The accounts for which it was not possible to assign a discursive community are in grey.}
    \label{fig:disc}
\end{figure}

Fig.~\ref{fig:disc} shows the percentages of nodes placed in the various discursive communities for the Italian Covid-19 dataset (a detailed description of the various communities can be found in the caption of the figure). Considering also the other datasets, in almost all the cases, the label propagation procedure could assign a label to approximately 90\% of the nodes. As we could expect, in the Covid-19 datasets, the Media community is always the most numerous one: updates on the spread of the pandemic, written by the official accounts of various media, received a great amount of retweets. %Nevertheless, in all the examined datasets,  even if the dimension of the different political factions are variable,  there is always a couple of communities significantly bigger than the others. 
\\
As highlighted in other works~\cite{Becatti2019,Caldarelli2020,Caldarelli2021, Radicioni2021a,Radicioni2021b,Bruno2021,Mattei2021}, the presence of well-defined discursive communities is the signal that users on Online Social Networks (OSNs) are strongly polarized, i.e., they tend to tend to split into groups, which one with same opinions and political orientation. 

\subsection{The bow-tie structure}\label{sec:bowtie_strut}
% The `bow-tie structure' was initially introduced by Broder et al. in order to study the structure of World Wide Web (WWW)~\cite{Broder2000}. 
% %Such structure is a partition of  nodes in a directed network into different sectors. 
% Broder et al. represented the World Wide Web (WWW) as a directed network in which the nodes are webpages and the edges are the hyperlinks connecting them. They noticed that the network displays a huge Weakly Connected Component (WCC), i.e., the maximal subgraph in which all nodes can be reached by any other one in the same subgraph, disregarding the direction of the link. This WCC includes more than 75\% of all nodes. 

% WCC breaks into three main pieces: a Strongly Connected Component (SCC), in which each node can be reached by any other one in the same block, following the direction of the links; a group of nodes that can reach SCC, without being reached by it (called IN); a group of nodes that can be reached by SCC, but that cannot reach it (the OUT block). The observation is  that the SCC is the most populated sector, followed by the IN and the OUT sectors.  Most of the websites can be found in the SCC, linking between each other; the IN sector was instead mostly composed by search engines, while the OUT one includes authorities, as Wikipedia.

The original concept of bow-tie by Broder et al.~\cite{Broder2000} sees WWW divided into 3 main sectors: a Strongly Connected Component (SCC), in which each node can be reached by any other one in the same block, following the direction of the links; a group of nodes that can reach SCC, without being reached by it (called IN); a group of nodes that can be reached by SCC, but that cannot reach it (the OUT block).

The description by Broder et al. was subsequently refined by Yang et al. \cite{Yang2011}, who split the network in seven distinct parts\footnote{In the following, we will call the various part \emph{blocks} or \emph{sectors} interchangeably.}:
\begin{itemize}
    \item the greatest Strongly  Connected Component (\textbf{SCC});
    \item the \textbf{IN} block;%, collecting all those nodes which point to the SCC, without being reached by nodes in SCC;
   \item the \textbf{OUT} block;%, containing all those nodes reachable from the SCC, but that cannot reach nodes in SCC;
   \item the \textbf{TUBES} sector, including nodes reachable from IN and accessing OUT, but not being part of SCC;
   \item the \textbf{INTENDRILS} group, collecting all those nodes pointed by IN that cannot reach the OUT block;
   \item the \textbf{OUTTENDRILS} sector, containing all those nodes pointing to OUT that cannot reach nodes in IN;
  \item the \textbf{OTHERS} group, including all those nodes that cannot be placed in one of the previous six sectors.
 \end{itemize}
 In Fig.~\ref{fig:structure} there is a schematic representation of the bow-tie structure defined in Ref.~\cite{Yang2011}. The seven groups of nodes are mutually disjointed. 
 
 We remark that every directed network can be divided in blocks using the bow-tie decomposition. Nevertheless, as a rule of thumb, the bow-tie representation is {\it informative} about the network structure if the number of nodes in blocks other than OTHERS is greater or of the same order of those in OTHERS: the greatest the impact of the non-OTHERS blocks, the more informative the bow-tie structure is. %, otherwise no strong structure is present.
% The aim of this project is to see if an analogous structure can be found also in a network of retweets. Our idea is to observe the collocation of the users within these seven sectors for all the directed retweets’ sub-networks described as \emph{discursive communities} in the last section. This could also tell us some information about the structure of the “conversation” on Twitter for each discursive community.
 \subsubsection{The bow-tie structure of the discursive communities}
 In the present manuscript, we investigate the presence of a bow-tie structure in the discursive communities of the retweet network, i.e., in the network composed by Twitter accounts (the nodes) and retweets (the links connecting the original author to the retweeter). %Our results show that a strong bow-tie structure is present in almost all the discursive communities of our datasets, extremely depending on the argument of the debate. 
 %To evaluate our results, we say that the bow-tie is informative if the OTHERS block contains less than 50\% of nodes of the entire network.
 
 Results show that, when considering political online debates, a bow-tie structure is informative in almost every discursive community of our datasets, while for non-political debates (as the case of Euro2020), the bow-tie structure is less informative. Euro2020  itself records the extreme case 
  in which more than one half of the nodes are in the OTHERS sector. We state  that this bow-tie structure is {\it uninformative} -- see,  for example,  the case of the Turkish debate during Euro2020 in Fig.~\ref{fig:tur_euro2020}.
 %In fact, generally, only few nodes are located in the OTHERS sector, while the great majority are placed in the remaining sectors of Yang's partition. 
 We remark that the presence of informative bow-ties in many of the discorsive communities  here investigated is not a trivial result. Indeed, there are no evident reasons for expecting such distribution of the nodes \emph{a priori}.
 
 When a bow-tie structure is informative, we observe two recurrent situations in the investigated datasets and, according to them, we classify the bow-tie into two different categories: 
%  according to the sector which contains the most part of the vertices. 
%  As stated above, the identification of the bow-tie structure is informative once the WCC (SCC+IN+OUT) of the bow-tie includes an amount of nodes greater or of the same order of OTHERS. 
%   Then, we can identify two different configurations:
 \begin{itemize}
     \item When the OTHERS block is smaller than SCC,  %In these cases we observed that most of the nodes are placed in the OUT sector (around 40-50\% of nodes on the total). %When this is the case, the second biggest sector is usually the INTENDRILS group. % and the OTHERS one is generally small. 
     we will refer to  \textbf{strong} bow-tie structures;
     \item  When the OTHERS block is greater than SCC,   %In this case, we often have that the most part of the nodes are located in the INTENDRILS sector, i.e., when most part of the users retweets accounts from the IN zone and has little to no interaction with the users in the other sectors.  
     we will refer to
     %to these networks as those with a 
     \textbf{weak} bow-tie structures.
 \end{itemize}

 Furthermore, when the bow-tie is informative, may it be weak or strong, we can categorize it in two different ways, that we called respectively \textbf{OUT-dominant} and \textbf{INTEND-dominant}. In OUT-dominant bow-ties, most of the nodes  of the bow-tie are placed in the OUT sector. As a rule of thumb, OUT-dominant bow-ties are more frequent when the bow-tie is strong,  but we can find some counter-examples. The INTEND-dominant bow-tie is a bow-tie structure in which instead the most part of the nodes is located in the INTENDRILS sector, i.e., when most part of the users retweets accounts from the IN zone and has little to no interaction with the users in the other sectors. INTEND-dominant bow-ties are in general more frequent in weak bow-ties.
 
%  Before looking at the composition of the various blocks, let us remark 
 
 We highlight that it is not so strange that the most crowded blocks in the bow-ties are OUT and INTENDRILS: it was already observed in Ref.~\cite{Gonzalez-Bailon2013} that the greatest number of users tend to mostly retweet content created by others and limit their production of new messages. 
 The difference between OUT-dominant and INTEND-dominant bow-ties is the {\it access to information}: OUT-dominant bow-ties are those in which the majority of users can access almost all messages exchanged over the discursive community, while in the INTEND-dominant ones the majority of users limits their retweets to the content produced by accounts in the IN block. %We will see that this difference is going to be crucial in the following.
 Otherwise stated, the main difference between INTEND- and OUT-dominant bow-tie structures is that the former displays a more `hierarchical' structure, i.e., %the conversation consists simply in 
few accounts (those in the IN sector) introduce new content and many others just share it (the INTENDRILS sector). 
Instead, in OUT-dominant bow-ties, the greatest part of the users (i.e., the OUT block) not only shares posts by accounts in the IN block, but also it retweets  content by users in SCC, OUTTENDRILS and TUBES. We argue that this behaviour, while more `democratic', is, at the same time, more risky. 

In fact, we will see in Subsection~\ref{ssec:user_loc} that users with high visibility and which introduce new content on Twitter can be found mostly in the IN sector: typically, they are verified accounts. As observed in other studies, see, e.g. Ref.~\cite{Caldarelli2021}, verified users tend to limit the spreading of low-quality content. We may argue, then, that users interacting mostly with verified users are safer from m/disinformation campaigns. In the following, we will see that the reputability of information shared confirms our hypothesis and we will come back on the matter.\\

\begin{figure}
    \centering
    \includegraphics[scale=0.3]{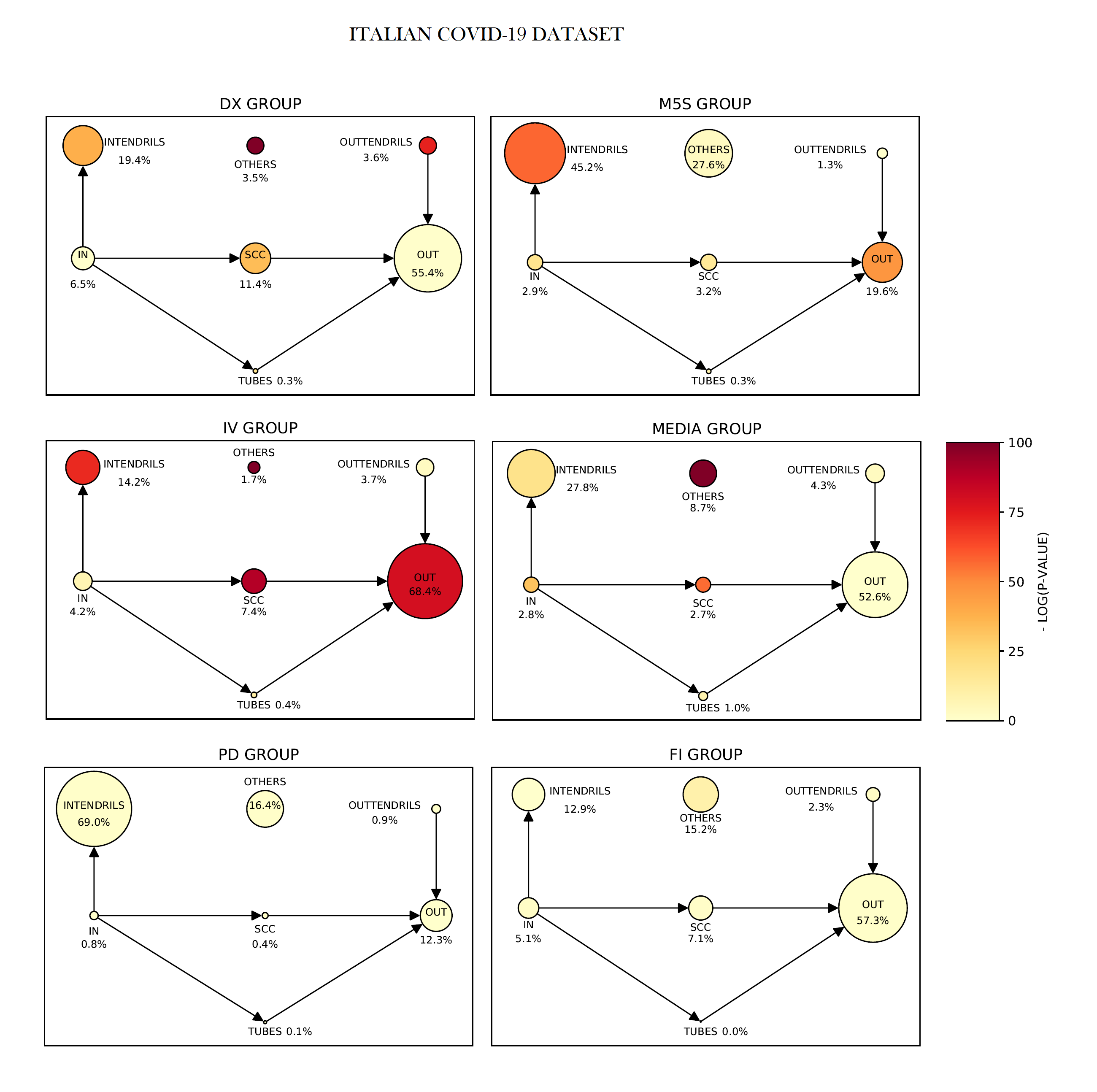}
    \caption{\textbf{The bow-tie structure of the discursive communities in the Italian Covid-19 dataset.} The dimension of the sectors is proportional to the number of nodes: DX and IV discursive communities have strong bow-ties (the OTHERS block is smaller than SCC), while the others are weak (the OTHERS block is greater than SCC, still being smaller then bow-tie WCC). The DX, IV, FI and MEDIA groups display a OUT-dominant bow-tie structure, with the most part of the nodes located in the OUT sector.  M5S and PD communities have a INTEND-dominant bow-tie structure, the INTENDRILS sector being the dominant one.\\ The colour of the blocks quantifies the distance between the observed dimensions and those predicted by the Direct Configuration Model (DCM). 
    % Considering the comparison with the predictions of the DCM, 
    The observed dimension for the OTHERS sector is significantly less numerous (considering a significance level at $\alpha=0.01$) for all the communities, besides PD. Remarkably, for INTEND-dominant bow-ties, also other sectors, as SCC and  INTENDRILS, are usually bigger than what we expect from the model.}
    \label{covid_ita}
\end{figure}

Fig.~\ref{covid_ita} displays the bow-tie structure of each discursive community for the Italian Covid-19 dataset (analogous plots for the other datasets can be found in the Supplementary Material). A single node represents one bow-tie sector and its dimension is proportional to the number of accounts in it. 
First, according to the definitions given above, the bow-tie structure is informative in all the discursive communities. In the cases of DX and IV, the bow-tie is particularly informative: its blocks include respectively 96.5\% and 98.3\% of the entire discursive community. 
Second, different discursive communities display bow-ties with different strengths. For instance, DX and IV discursive communities display strong bow-ties, 
%the ratios between the dimensions of the SCC and the OTHERS blocks are respectively 3.26 and 4.35. As a comparison, consider than in WWW the same ratio was 3.38~\cite{Broder2000}.
%(bow-tie blocks include respectively 96.5\% and 98.3\% of the entire discursive community. As a comparison, consider than in WWW the bow-tie blocks include nearly 75\% of nodes~\cite{Broder2000}). 
while, M5S, Media, PD and FI have weak ones, since their SCCs are relatively small (and smaller than OTHERS).

Third, the graph shows that the DX, IV, MEDIA and FI communities display OUT-dominant bow-ties, in which the OUT sector is the biggest one; considering all the investigated datasets, OUT-dominant bow-ties represent the most frequent configuration, being 21 out of 31 communities. Instead, 6 out of 31 discursive communities are INTEND-dominant bow-ties (as PD and M5S in Fig.~\ref{covid_ita}). %, followed by OTHERS sector, which is the largest in 5 out of 37. %The former is the case, for example,  of the M5S, and, especially, the PD community in Fig.~\ref{covid_ita}. 

We remark that, in all our datasets, all the right wing discursive communities display bow-ties with an OUT-dominant structure; in most of the cases, these bow-ties are also strong. The colours of the nodes in Fig.~\ref{covid_ita} are going to be explained in the following section. 

\subsubsection{Statistical significance of the bow-tie structure}\label{ssec:statistical_significance}

It may be argued that the bow-tie structures featured by the discursive communities in our datasets are just an accident, due to the different role of the various users in the debate. In fact, those accounts that have high out-degrees and low in-degrees are naturally in the IN sector; those that, viceversa, have high in-degrees and low out-degrees are in the OUT sector, and so on. To test whether the presence of bow-ties is merely attributable to the behavioral characteristics of the accounts, we compare the dimensions of the different sectors, as observed in the real network, with those in a randomised system in which the in- and out-degree sequences are fixed. If the partition in the various bow-tie sectors were just a matter of the degree sequence, none of the dimensions of the various blocks should be statistically significant. Otherwise, we should observe a significant mismatch with respect to the expectation of the null-model.

%We thus compare the observations with an entropy-based null-model preserving the in- and out-degree sequences. 
In order to have an unbiased benchmark, we build an entropy-based null-model that preserves the in- and out-degree sequences, being  maximally random for all the rest (see  Ref.~\cite{Cimini2018} for a review on the subject). Summarising, starting from a real network, we consider the set of all possible graph realizations (the graph \emph{ensemble}) having the same number of nodes as in the real system. Then, we assign to each representative of the ensemble a different probability of realization by maximising the entropy of the ensemble, but constraining the average value of some topological property of the real network (in our case, the in- and out-degree sequences). In this way, even if the single realization of the ensemble does not display the network properties that we would like to preserve, the entire ensemble, on average, does.

%The choice of an entropy-based framework is justified by its unbiased-by-construction nature~\cite{Cimini2018,Squartinia}. 
In the last years, such procedure has been adopted to analyse financial and economic systems~\cite{squartini2011analytical,Squartini2013a,Squartini2013b, Picciolo2013, Mastrandrea2014, Gualdi2016a, Saracco2015a, Saracco2016, Saracco2016a, DiGangi2018, Squartini2018, Bardoscia2021, Straka2017, Gabrielli2019, Adam2019, Bruno2018, Cimini2021, DiVece2021,Lin2020,Vallarano2021}, biological networks~\cite{Straka2018, PayratoBorras2019, Bruno2020, Caruso2021} and online social networks~\cite{Becatti2018, Becatti2019, Caldarelli2020, Caldarelli2021, Radicioni2021a, Radicioni2021b, Mattei2021, Patuelli2021, Bruno2021} and it was shown to be effective to extract the relevant structure from a real network~\cite{Parisi2020,Neal2021}.

Here, we implement the Direct Configuration Model (\emph{DCM}), firstly introduced in Ref.~\cite{Mastrandrea2014} and implemented in the Python module \href{https://nemtropy.readthedocs.io/en/master/}{\texttt{NEMtropy}}~\cite{Vallarano2021}. More details on the exact derivation of DCM can be found in Subsection~\ref{ssec:DCM}.\\

Going back to Fig.~\ref{covid_ita}, the colour of the circles indicates the agreement between the actual size of the bow-tie sectors and the size predicted by the DCM: we are interested in detecting both too ``big" and too ``small" blocks.
In particular, the darker the colour of the sectors in Fig.~\ref{covid_ita}, the larger the $-\logten(\text{p-value})$ (so the lower the p-value) and the greater the disagreement of the real system from the randomization. For each sector, the two-tailed p-value has been calculated looking to a sample of 1000 graphs generated by the DCM. 

% for the dimension of each block the two-tailed p-value is calculated looking to a sample of 1000 graphs generated by the DCM: in the figure, the darker is the color,  % and, therefore, it quantifies the distance from the predicted dimension.

The p-value tells us about the existence of a  disagreement, but not about the direction of the disagreement. For instance, looking at the DX bow-tie in Fig.~\ref{covid_ita}, both the dimensions of OTHERS and SCC have a really small p-value, thus they do not agree with the randomization, but the OTHERS block is smaller than predicted by the DCM, while SCC is larger. % For more details, consult Subsection~\ref{ssec:DCM}.

%The color of the nodes indicates that the DCM predicts discursive community subgraphs with a weaker or even totally absent bow-tie structure: for 19 on 27 discursive communities the OTHERS group result significantly less numerous than in the graphs of the theoretical ensemble (considering a significance level of 1\%). 
%The strongest argument in support of this hypothesis is that for 19 on 27 discursive communities the OTHERS group result significantly less numerous than in the graphs of the theoretical ensemble (considering a significance of 1\% and therefore a p-value $<$ 0.01). 

\begin{table}
\begin{tabular}{c||c|c|c|c|c|c|c}
     & \textbf{SCC} & \textbf{IN} & \textbf{OUT} & \textbf{TUBES} & \textbf{INTE.} & \textbf{OUTTE.} & \textbf{OTHERS}  \\
    \hline
    \hline
    \textcolor{orange}{\textbf{DX}}\textcolor{red}{$\bullet$} & $10^{-35}$* & 0.7 & 0.4 & $10^{-18}$* & $10^{-39}$* & $10^{-74}$* & 0*\\
    \hline
    \textcolor{teal}{\textbf{M5S}}\textcolor{blue}{$\bullet$} & $10^{-15}$* & $10^{-18}$* & $10^{-48}$* & $10^{-9}$* & $10^{-58}$* & 0.5 & 0.0006*\\
    \hline
    \textcolor{orange}{\textbf{IV}}\textcolor{red}{$\bullet$} & $10^{-90}$* & $10^{-8}$* & $10^{-81}$* & $10^{-12}$* & $10^{-72}$* & 0.0004* & 0*\\
    \hline
    \textcolor{teal}{\textbf{PD}}\textcolor{blue}{$\bullet$} & 0.04 & 0.7 & 0.9 & 0.08 & 0.1 & 0.8 & 0.1\\
    \hline
    \textcolor{teal}{\textbf{FI}}\textcolor{red}{$\bullet$} & 0.03 & 0.01 & 0.1 & 0.4 & 0.5 & 0.005 & $10^{-10}$*\\
    \hline
    \textcolor{teal}{\textbf{MEDIA}}\textcolor{red}{$\bullet$} & $10^{-57}$* & $10^{-33}$* & 0.9 & $10^{-9}$* & $10^{-19}$* & 0.0002* & 0*\\
    \hline
\end{tabular}
\caption{\textbf{P-values related to the various bow-tie sectors in the Covid-19 Italian dataset.} In orange strong bow-ties, while in teal weak ones;  red dots indicate OUT-dominant bow-ties,  blue dots indicate INTEND-dominant ones. If we set the statistical significance to $\alpha=0.01$, then we then have to correct for multiple hypothesis testing per block. In the present case, we used the False Discovery Rate method (FDR,~\cite{Benjamini1995}). In the table,  validated p-values are marked by an asterisk `$*$'. The OTHERS block is statistically significant (in particular it is smaller than in the randomization) for all discursive communities but the PD one. It is remarkable that the dimension of SCC is significant in all strong bow-ties, while the one of OUT is significant only for IV bow-tie.}\label{tab:pvalues}
\end{table}

Table~\ref{tab:pvalues} reports the exact p-values of the different blocks for the various bow-ties of Fig.~\ref{covid_ita}. The significance of the blocks for each bow-tie can be assessed by using the False Discovery Rate (FDR) correction~\cite{Benjamini1995}, setting the statistical significance level to $\alpha=0.01$. In the present case the correction is limited, due to the small number of blocks in the bow-tie. 

It is interesting to observe that, in both strong and weak bow-ties, the OTHERS block is statistically significant in all the discursive communities but PD. In particular, the dimension of the OTHERS block is much smaller then predicted by the null-model and the presence of the bow-tie is not due to the degree sequence only. %The only cases in which the OTHERS sector is not significant are those with weak or absent bow-tie structure: it is the case, for instance, of the PD group in the Italian Covid-19 dataset.% (note that, viceversa, for the DX, IV and MEDIA communities the dimension of OTHERS sector is instead significant, having a really small p-values and being colored by an intense red). 
%Instead, the SCC, TUBES, INTENDRILS and OUTTENDRILS sectors are usually bigger than what we expect from the DCM model, or, at least, in agreement with it. Moreover, the dimension of the IN group is usually in agreement with the predictions (for 19 communities on 27 the p-values are greater than 0.01) while in the case of the OUT block it fluctuates (for 10 communities it is significantly bigger, for 9 ones significantly smaller, and for 8 communities in agreement). 

SCC is statistically significant (and bigger than expected) for all bow-ties but FI and PD. The IN block is often statistically significant and smaller than expected.
We may notice that in the strong bow-tie of IV discursive community the dimensions of all sectors are statistically significant, while none are in the PD bow-tie, which is the smallest discursive community.
It is worth noting that also the dimension of the discursive community has a role: due to the limited possible variability, smaller bow-ties feature more agreement with the model. %The exact p-values for each sector of all the communities of the Italian Covid-19 dataset can be found in Table~\ref{tab:pvalues}.

\subsection{Verified users' distribution}\label{ssec:user_loc}
\begin{figure}
    \centering
    \includegraphics[scale=0.3]{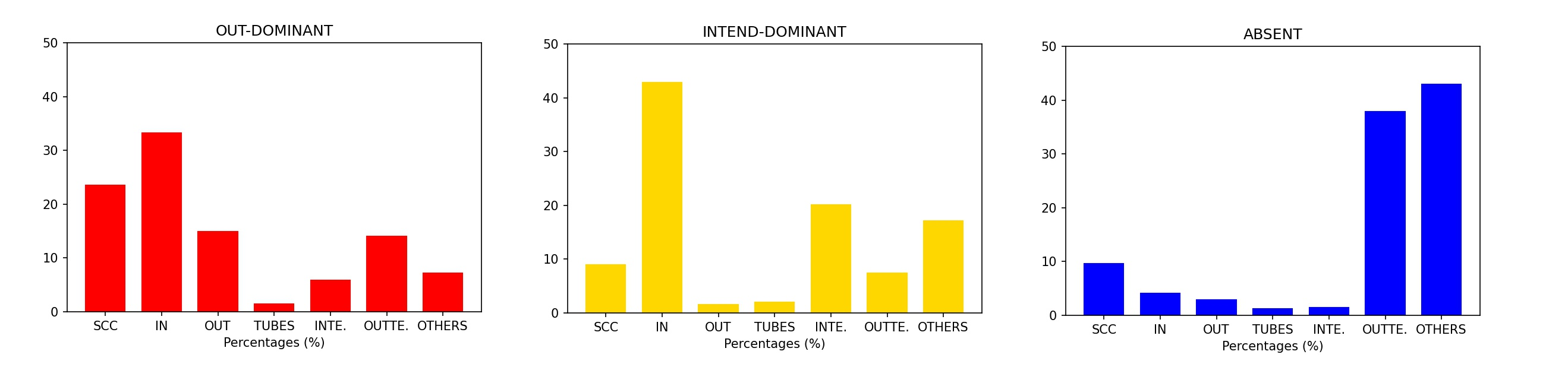}
    \caption{\textbf{Distribution of the percentage of verified users in each sector of the discursive communities with, respectively, OUT-dominant, INTEND-dominant and not informative bow-ties.} Each bar-chart displays the average percentage of verified users in a specific sector, calculated respectively for all the  OUT-dominant, INTEND-dominant and not informative bow-ties. In the cases of  OUT-dominant and INTEND-dominant  bow-ties, the highest percentages of verified accounts can be found in the IN group. Moreover, in OUT-dominant bow-ties, we can found a relevant percentage of verified accounts also in the SCC. Naturally, for those communities with no bow-tie structure the verified accounts are mostly placed in the OTHERS sector and, to less extent, in the OUTTENDRILS one.}
    \label{fig:histo_verified1}
\end{figure}

%\paragraph{Verified users}
%Histograms in Fig.s~\ref{fig:histo_verified1},  ~\ref{fig:histo_verified2} and ~\ref{fig:histo_verified3} show the distributions of the percentages of verified accounts in each bow-tie sector, for all the discursive communities, and for the cases of OUT-dominant, pupps and absent bow-tie structures, respectively. 
%As already described above in Subsection~\ref{sec:disc}, verified users are those for which Twitter has certified the authenticity of the account owner. 
Usually, verified accounts on Twitter belong to public characters and organizations, such as journalists, politicians, actors, political parties, media, and VIPS in general.
Previous studies testify that verified users tend to introduce new content and have high visibility on the platform~\cite{Gonzalez-Bailon2013, Becatti2019, Caldarelli2020, Caldarelli2021, Radicioni2021a}. Thus, we expect to find them in the IN block. 
% \textcolor{red}{IL COMMENTO SEGUENTE E' AD UNA FIGURA CHE NON C'E'.}
The results in Fig.~\ref{fig:histo_verified1} confirm this intuition: in the case of OUT-dominant bow-ties (leftmost panel), the 33.2\% of verified users, on average, are in the IN sector. High percentages of verified users are also in the SCC block (23.5\%). In the case of INTEND-dominant bow-ties (central panel), the percentage of verified users in the IN group increases to 42.8\%; the second block per percentage of verified users is INTENDRILS (20.1\%).
%It is possible to note that the distribution for the IN sector is more shifted to the right and the peak is between 20\% - 40\%. %Verified users usually receive a great amount of retweets (especially by the SCC and the INTENDRILS) and, at the same time, they tend  not to retweet other users. %However, when they also retweet some other accounts, then they are probably placed in the SCC sector. 
%Nevertheless, for some discursive communities the major part of verified users are located in the SCC: they debate,  by retweeting messages from other accounts too. 
In those communities where the bow-tie structure is not informative (right panel, Fig.~\ref{fig:histo_verified1}), a high percentage (42.9\%) of verified users, on average, is in the OTHERS sector.  In a few cases of not informative bow-ties, it happens that verified users are mostly in the OUTTENDRILS sector. In this last case, their messages hardly reach a big audience and are simply retweeted by a group of strong retweeters (OUT sector), not catching the interest of the accounts in the SCC. Let us remark that in the case of non-informative bow-ties the dimension of OUT  and SCC blocks is nevertheless limited.
%When the OUT sector is numerous despite the SCC being quite small, it could happen that the verified accounts are placed in the OUTTENDRILS sector; this is the case for just few communities.

\begin{figure}[ht!]
    \centering
    \includegraphics[scale=0.33]{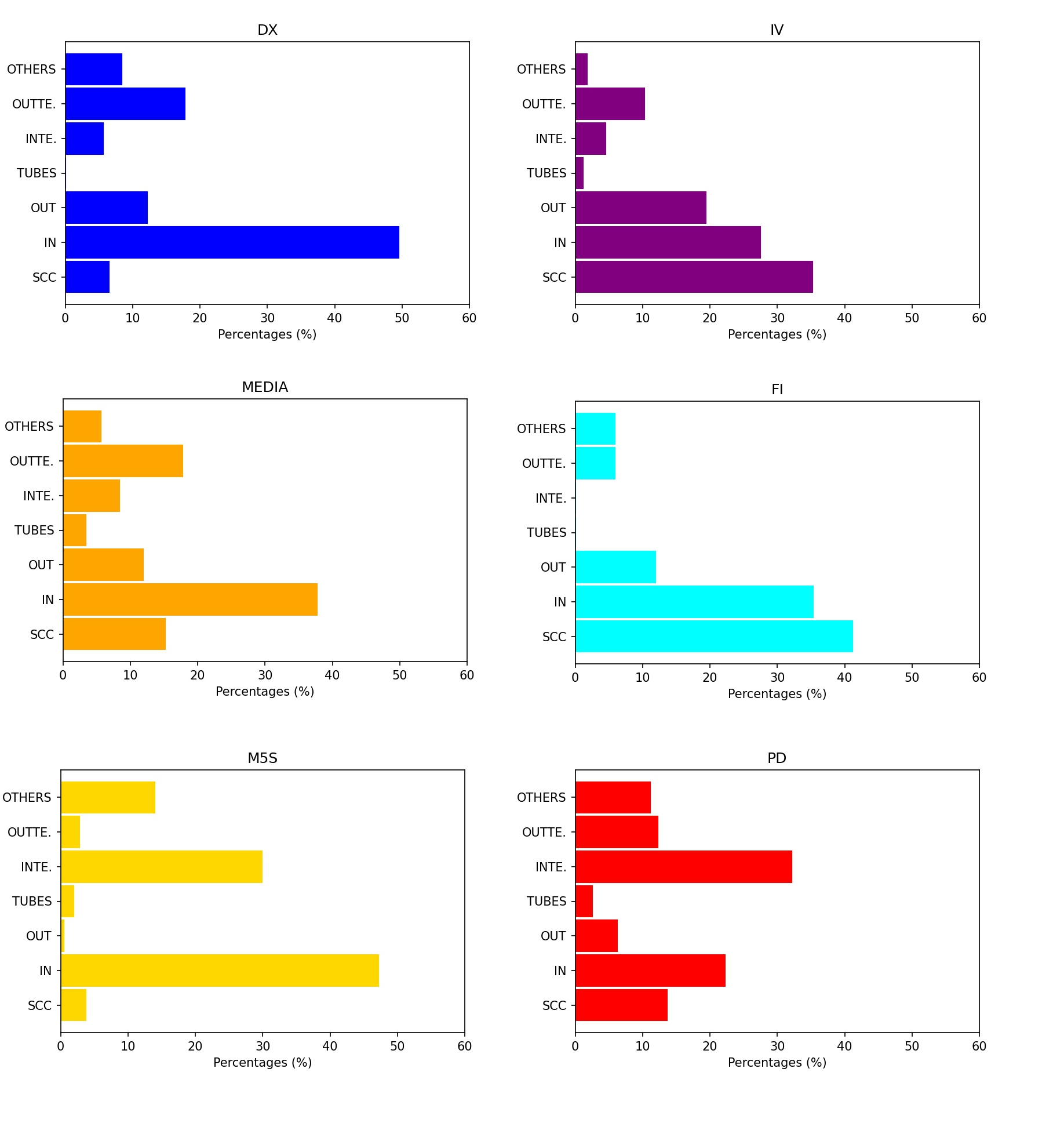}
    \caption{\textbf{Percentage of verified accounts in the bow-tie sectors for each discursive community of the Covid-19 dataset.} The bar-charts confirm that verified accounts are mainly located in the IN sector and, to a  less extent, in the SCC one. Only for the PD group, which has a INTEND-dominant bow-tie structure, verified accounts are mostly placed in the INTENDRILS block. %This is a recurrent situation also for the others datasets.
    }
    \label{fig:ver_covid_ita}
\end{figure}

Fig.~\ref{fig:ver_covid_ita} reports the same bar-chart, about the presence of verified users, for the bow-ties of the Covid-19 Italian dataset. It is possible to observe that in OUT-dominant bow-ties - i.e., DX, IV, FI and MEDIA - verified users are mainly in IN and SCC sectors. Also,  in INTEND-dominant bow-ties, the INTENDRILS sector contains quite a number of verified users. Other user characterizations of the bow-tie blocks can be found in the Section~\ref{ssec:bot}.
%\textcolor{red}{PER IL DATASET COVID-19 ITALIANO FAREI ANCHE UNA FIGURA DIVERSA, OVVERO LA FREQUENZA DI UTENTI VERIFICATI NEI VARI SETTORI COME $\dfrac{N(\text{verified users in IN})}{N(\text{IN})}$. IL PROBLEMA E' COME RENDERE QUESTA FIGURA: FORSE LA COSA MIGLIORE E' UNA BARCHART, ANCHE SE LE VARIE BARRE NON SONO CONFRONTABILI TRA DI LORO (LE FREQUENZE SONO CALCOLATE AVENDO DIVERSI DENOMINATORI).}

\subsubsection{Conservatives groups}

\begin{figure}
    \centering
    \includegraphics[scale=0.7]{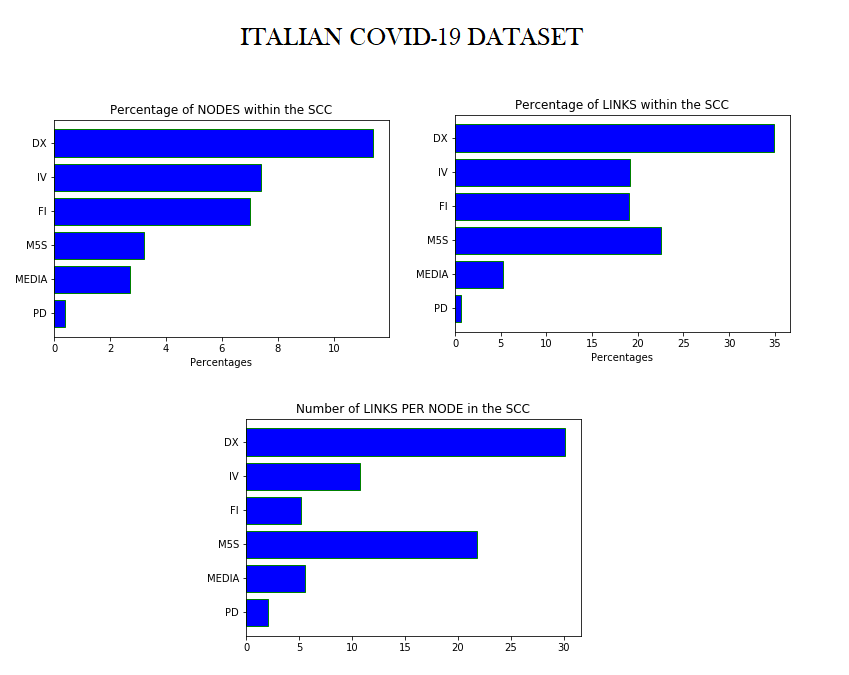}
    \caption{\textbf{Percentage of nodes and edges in SCC for the communities in the Italian Covid-19 dataset.}\\In the Italian Covid-19 dataset,  the conservative and right-oriented discursive community (DX) has more numerous and denser SCCs, as it is displayed in the highest two graphics. In the lowest graphic, it can be seen that, also considering the number of links per node in SCC, DX results again the first discursive community. These results hold for all the conservative groups in all the datasets under investigation.}
    \label{fig:conservatives}
\end{figure}

The bar-charts in Fig.~\ref{fig:conservatives} show the percentage of nodes, the percentage of edges and the number of edges per node in the Strong Connected Component, for each discursive community of the Italian Covid-19 dataset. Not only DX is the one with the greatest number of nodes and the greatest number of links in SCC, but also the link density of SCC in DX  is much greater than that of any other discursive community. Thus, the number of links in SCC of DX  is not proportionate to the number of nodes, and it results in a greater average degree per node. We found very similar behaviours also for the right-oriented communities of the other datasets.\\

In fact, in all our datasets, the discursive communities of conservative groups (i.e., DX in the Italian dataset, AfD in the German one, Conservatories in the Dutch one) are those with the highest percentage of nodes and, especially, of edges within SCC. This peculiar feature signals the presence of a common (self-)organization of accounts in line with conservative ideas on Twitter.

\begin{figure}
    \centering
    \includegraphics[scale=0.7]{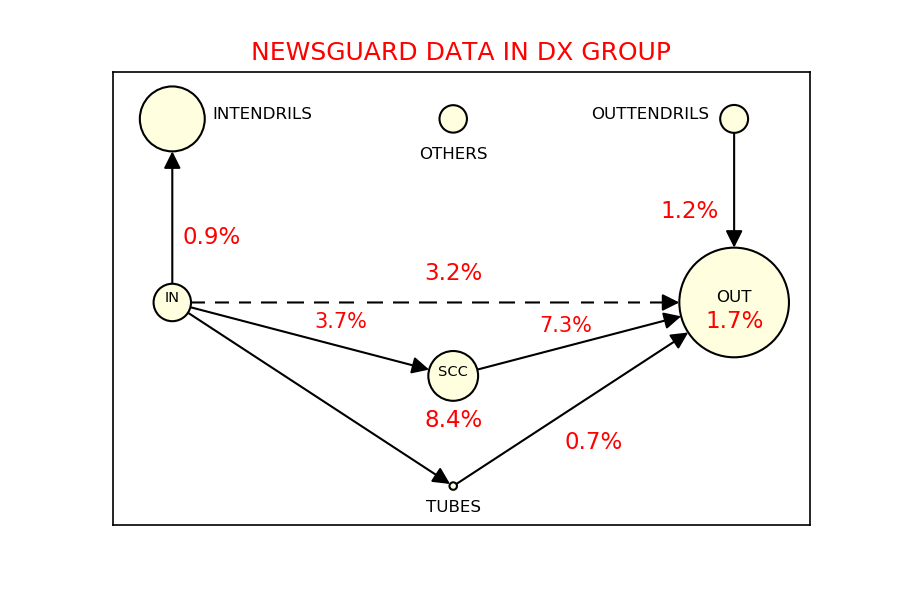}
    \caption{\textbf{Bow-tie structure of the DX group and percentages of retweets containing URLs of untrustworthy webpages.} The DX community in the Italian Covid-19 dataset presents the highest number of retweets containing a link to untrustworthy webpages. Most of them origin from SCC:  8.4\% of the retweets in SCC and 7.3\% of the retweets between SCC and OUT contain not reliable URLs. In the diagram, we also insert  the link between IN and OUT (the dashed line), which, considering the definition of each sector, is not forbidden a priori.}
    \label{fig:newsguard}
\end{figure}

%A%cross our datasets, the conservative discursive communities show another common feature.\\
NewsGuard\footnote{\url{https://www.newsguardtech.com/it/}} is an independent software toolkit that monitors the quality and transparency  of several news websites worldwide.  
% its analysts classify them according to the trustworthiness of the contents they share. 
% Through the tags that Newsguard has assigned to news sites whose links appear in the retweets othe amount f the our communities, we are able to quantify
Through the tags that NewsGuard has assigned to news sites whose links appear in the retweets of our communities, we are able to quantify the amount of retweets containing untrustworthy URLs.
% We observed if the retweets in our networks contained URLs of the untrustworthy pages indicated by Newsguard. 

The recurrent situation is that  almost only the conservative discursive communities display retweets with such URLs. For the Italian Covid-19 dataset, the DX group has 26,318 retweets with links to untrustworthy webpages of news sites, many more than in other communities: 1,356 retweets for M5S, 78 retweets for IV, 20 retweets for MEDIA, 9 retweets for FI and 0 for the PD group. A very similar situation has been found for the other datasets, see Supplementary Materials.

Another interesting aspect is that the most part of retweets containing not reliable URLs has origin in the strongly connected component. Fig.~\ref{fig:newsguard} shows  in red the percentage of retweets containing URLs of untrustworthy news pages within and between the sectors of the bow-tie structure for the DX group. The highest percentage can be found in SCC and between SCC and OUT.  Again, this is a recurrent situation also for the conservative communities of the other datasets under investigation. 

\subsubsection{The case of EURO2020}
Here, we devote a specific section to comment about the case of the European football championship (EURO2020) dataset\footnote{We do this for academic reasons, and not because Italy  won the  Euro2020 championship.}. This dataset features a less divisive, less debated, and less discussed tweets topics.
The topics of all the other datasets either have a strong political nature or are debating with sharp different positions. We then analyze whether the fact that topics are less discussed/devated has anything to do with the presence -or absence- of a bow-tie structure in the EURO2020 dataset.

% Our aim was to observe if the discursive communities in these networks present strong, weak or absent bow-tie and, therefore, to understand if the nature of the topic has influenced this structure.\\

\begin{figure}
    \centering
    \includegraphics[scale=0.3]{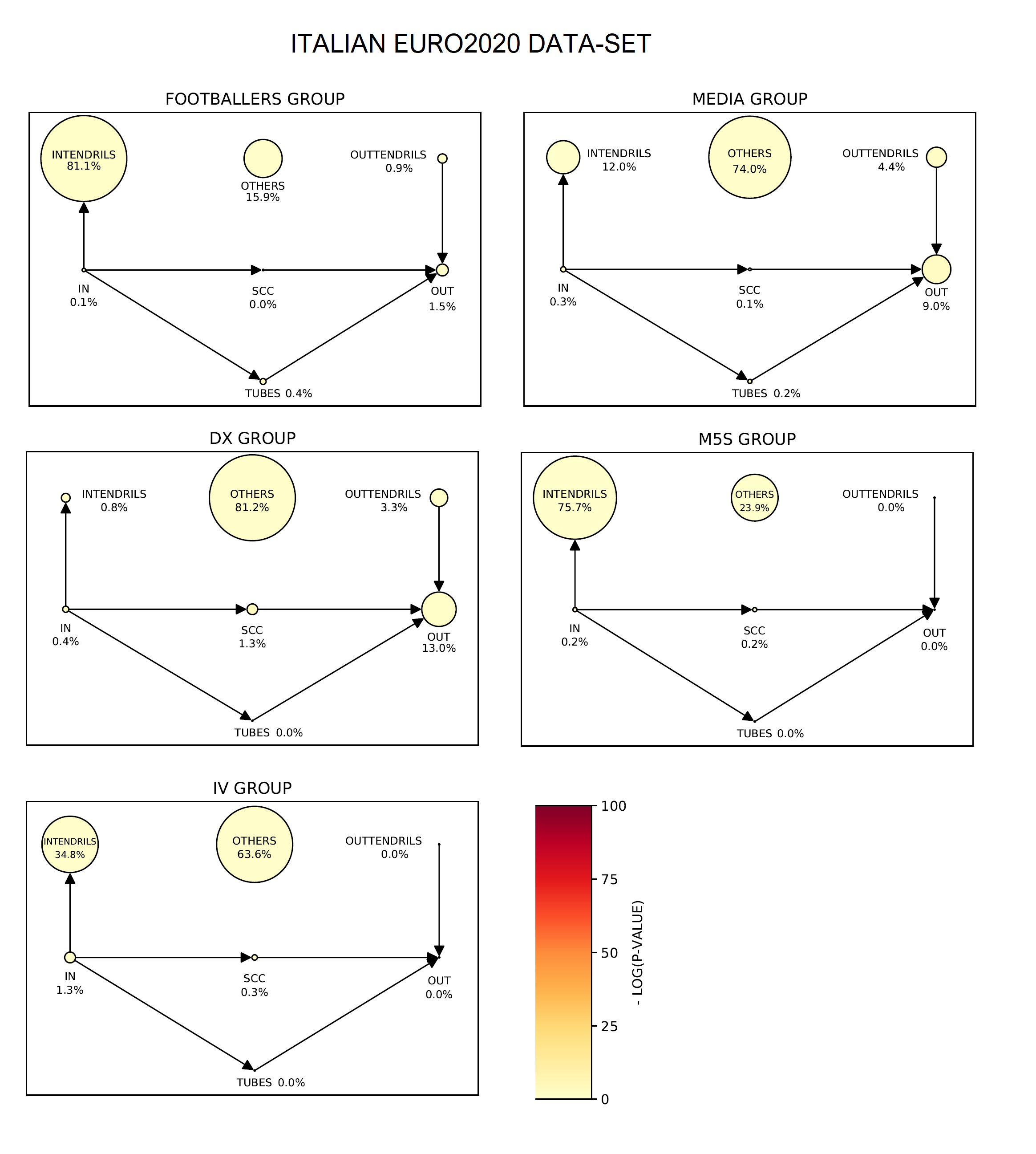}
    \caption{\textbf{The bow-tie structure of the discursive communities for the Italian EURO2020 dataset.}\\ The dimension of the sectors is proportional to the number of nodes therein and the color quantifies the distance between the observed and the predicted dimension. The main discursive community is formed by football players, sports newspapers and journalists. Then, we identified a MEDIA community, containing  accounts of Italian media, and  three small political communities (DX, IV, M5S). MEDIA, DX and IV do not display an informative bow-tie structure (respectively 74\%, 81.2\% and 63.6\% of the nodes in OTHERS), while FOOTBALLERS and M5S show a weak bow-tie (respectively 81.1\% and 75.7\% of nodes in INTENDRILS).}
    \label{fig:ita_euro2020}
\end{figure}

We identified 5 discursive communities for the Italian dataset and 2 discursive communities for the Turkish one. Of these 7, 4 do not have an informative bow-tie structure (in fact, most part of the nodes are in OTHERS), and the other three have a weak one (OTHERS is smaller than the weakly connected component of the bow-tie, but still greater than the strongly connected one). 
\begin{figure}
    \centering
    \includegraphics[scale=0.3]{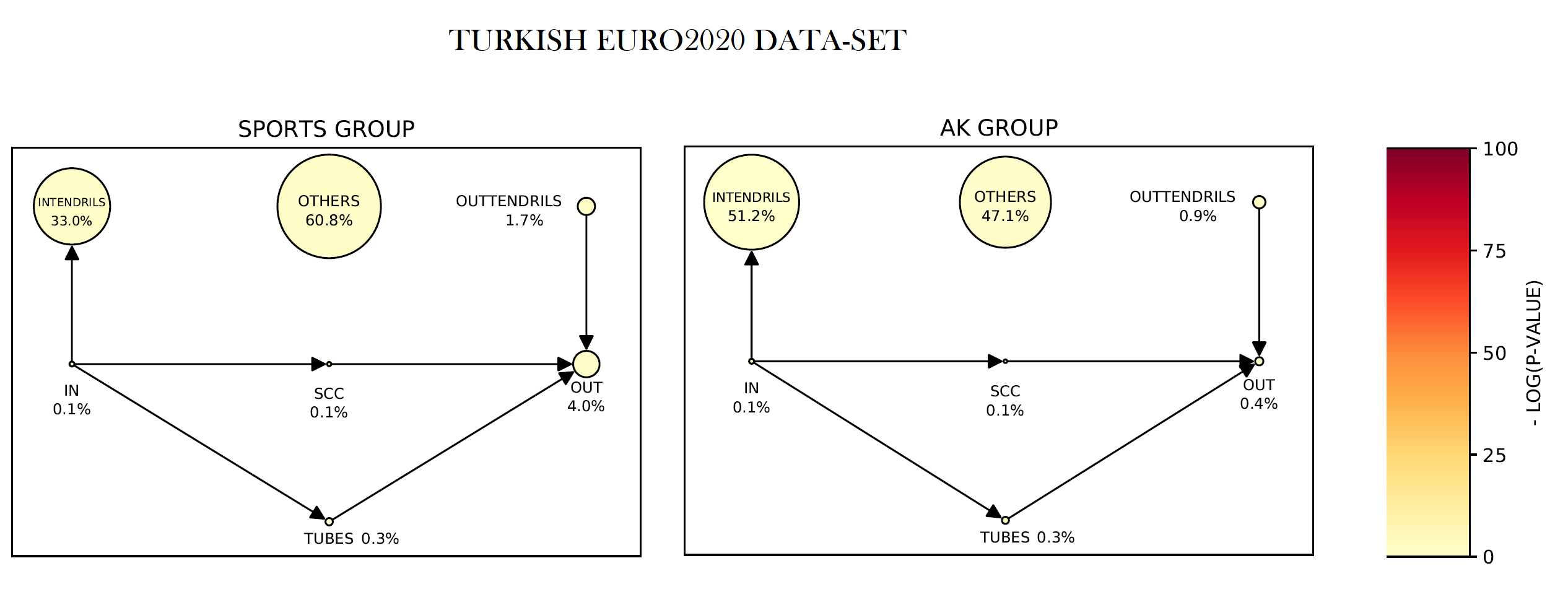}
    \caption{\textbf{The bow-tie structure of the discursive communities for the Turkish EURO2020 dataset.}\\
    The SPORTS group contains the official accounts of football players and clubs,  and sports newspapers, while AK refers to the Justice and Development Party (Turkish: Adalet ve Kalkınma Partisi, AKP), which is a conservative political party in Turkey, including President Erdogan and his ministries. The SPORTS discursive community does not display an informative bow-tie structure, while the AK one has an extremely weak (INTEND-dominant) bow-tie. The dimension of the sectors is proportional to the number of nodes therein and the color quantifies the distance between the observed and the predicted dimension. Looking to the color of the vertices, it is possible to see that the observed dimensions are not statistically significant.}
    \label{fig:tur_euro2020}
\end{figure}\\
Fig.~\ref{fig:tur_euro2020} reports the bow-tie structures of the two discursive communities in the Turkish dataset. The SPORTS group contains the official accounts of football players and clubs, and of sports newspapers. AK refers to the Justice and Development Party (Turkish: Adalet ve Kalkınma Partisi, AKP), which is a conservative political party in Turkey including President Erdogan and his ministries. %Both these communities do not present a strong bow-tie structure and their vertices are mainly located in the INTENDRILS and in the OTHERS sectors. 
While SPORTS does not display any informative  bow-tie, AK has a weak one. Following our interpretation, the latter displays a more hierarchical conversation on Twitter, in which the SCC is not numerous. Moreover,  the dimensions of the sectors are mostly not statistically significant. \\
For the Italian case (Fig.~\ref{fig:ita_euro2020}) the main discursive community is formed by football players, sports newspapers and journalists. There is also a MEDIA community, containing accounts of Italian media, and other three small political communities (DX, IV, M5S). MEDIA, DX and IV does not display an informative bow-tie structure (respectively, 74\%, 81.2\% and 63.6\% of the nodes are in OTHERS), while FOOTBALLERS and M5S show a weak bow-tie (respectively 15.9\% and 23.9\% of nodes in OTHERS).
\\
Euro2020 dataset is the only, among ours, in which no discursive communities have a strong bow-tie structure. 

\section{Discussion}

In the present manuscript, we analysed eight thematic Twitter datasets in different languages, related to various debates in Europe. \fab{We identified the discursive communities in the retweet networks and we investigated the presence of bow-tie structures in such communities. In previous works, discursive communities were shown to mirror the political orientation of users~\cite{Conover2011,Conover2011a,Conover2012,Becatti2019, Caldarelli2020,Caldarelli2021, Radicioni2021a, Radicioni2021b, Bruno2021,Mattei2021}, thus the analysis of their structure is of utmost importance to infer the way  opinions create and circulate}.

\paragraph{Discursive communities and bow-ties} %A first result of the analysis carried out in this work is that, in almost all the discursive communities extracted from the eight datasets,  \fab{the Weakly Connected Component WCC of the bow-tie includes the great majority of the accounts. WCC contains a Strongly Connected Component, whose accounts are all reachable to each other via retweets; an IN section, whose nodes are not part of the SCC but can be retweeted by the SCC; and an OUT section, which can retweet SCC nodes.} %Retweets may be undirected, that is, nodes in SCC are reached by those in IN either because the former directly retweet a node in IN or because they retweet a node in IN that has retweeted another node in IN. The same applies to retweets that modes in OUT make on tweets/retweets by nodes in SCC.}
%with some differences. 

We found that a bow-tie structure is present in those discursive communities debating about politics, like in the case, e.g., of election campaigns (it is the case of the Dutch elections dataset) or debating about Society, e.g., `how to handle a pandemic?' (it is the case of the Italian, German and French datasets about Covid-19) or 'how to manage migration fluxes?' (it is the case of the Italian online debate on migrants). Instead, a bow-tie structure is absent when the topics of the discussion are sportive ones, as in the case of Euro2020 Turkish and Italian datasets.
%da qui, evidenziare la peculiarita' della scoperta, ampliando le considerazioni qua sotto:
%
%se intendrils e' grande, gli arriva solo la roba dagli IN, che e' stato provato non sparano molte minchiate. 
%
%quando invece sono gli out a essere grandi, gli arriva tutta la schifezza. 

More in details, we state that the bow-tie is informative if the corresponding WCC includes more than one half of the nodes of the entire discursive community%, otherwise it is not informative
. In the present datasets, we found that bow-ties are informative in all the discursive communities debating about politics. In the case of the Euro2020 dataset, bow-ties are not informative, or, if present, they are extremely weak.
%More in details 
When the bow-tie is informative, we found essentially 2 cases: 1) the most crowded block is the OUT one; 2) the most crowded block is the INTENDRILS one. The former is typical of the discursive communities of right wing parties in all European political/societal debates of our datasets, while the latter is more common in less active political discursive communities in many political/societal datasets.\\ %We called the two case, respectively OUT-dominant and INTEND-dominant.

%In the first case, the dimension of SCC is relatively small, as compared to the portion of nodes not taking part to the bow-tie (in the bow-tie jargon, this latter block is call OTHERS). It is the case, for instance, of the discursive communities of media, i.e. the one including online newspapers, newscasts and journalists in political debates or any community during sport discussions.\\
%In the second case, the dimension of SCC is of the same order as in the OTHERS block. Nevertheless, in weak bow-ties the greatest number of nodes is included in INTENDRILS, i.e. the exclusive retweeters of accounts in the IN block. It is the typical case of less active political discursive communities in many political datasets, as for instance the Italian Democratic Party in the Italian Covid-19 dataset.\\ 
%In strong bow-ties, which is typical of the discursive communities of right wing parties in all European political debates of our dataset, the SCC sector is greater than the OTHERS one and OUT, i.e. the ``sink" of the retweet discursive community, is, by far, the most crowded block.\\

\paragraph{Which users in which bow-tie sectors and the exposure to m/disinformation} A closer inspection of the nodes in the various blocks and the quality of the shared content permit to better characterise the users in the bow-tie.
%a clearer idea on such a structure in the debate. 
The first observation is that the greatest part of the verified users, i.e., those accounts for which the identity of the owner has been certified by Twitter, in the IN sector, in each bow-tie. This finding is not surprising: as already observed in previous studies,  verified users create content and are less active in sharing messages written by others~\cite{Becatti2019, Caldarelli2020, Radicioni2020, Caldarelli2021, Radicioni2021a, Gonzalez2021}. Verified users are mostly politicians and official accounts of political parties, as well as journalists and official accounts of their newscasts and newspapers. In this sense, a discursive community displaying a INTEND-dominant bow-tie structure (where INTRENDILS is the most crowded block) may appear, at first sight, as a less democratic group: the content is created by a few accounts and shared by a group of followers that limit their interactions to sharing the messages coming from the IN block. Instead, in a OUT-dominant bow-tie, the greatest block is OUT and it can access the content created by all the other blocks in the bow-tie (with the only exception of INTENDRILS), so having the possibility to intercept every voice in the discursive community.  

Actually, the issue is on the quality of the content created in the various blocks, see Fig.~\ref{fig:newsguard}. Leveraging our ongoing collaboration with the NewsGuard organization\footnote{\url{https://www.newsguardtech.com/it/}}, we annotated the URLs that appear in tweets in our datasets, based on the reliability and transparency ratings of the news sites to which those URLs belong (such ratings have been assigned by NewsGuard). 
%
% using the domain annotation from the fact-checking website Newsguard\footnote{\url{https://www.newsguardtech.com/it/}.},
%
% we considered the reliability of the URLs shared inside the various blocks or between blocks. 
It turns out that the lowest reliable URLs, in a strong bow-tie, are the ones shared in SCC. The fact that verified accounts are not responsible for the vast majority of m/disinformation sharing was already observed in Ref.~\cite{Caldarelli2021} and, in the present context, it reflects the fact that accounts in IN are minimally responsible for the spreading of low quality/untrustworthy content. Otherwise stated, when the source of information is not identifiable, the average quality of the content is lowered down.%A largely populated OUT block implies that the greatest part of the accounts has access to a great variety of content. %\fab{and, in case of the discursive communities featuring a strong bow-tie, the greatest part of the accounts is exposed to low quality or no quality content. }
%its quality is lower than in the case of weak bow-ties.

\fab{An OUT-dominant bow-tie is, in this sense, more exposed to m/disinformation campaigns, as the majority of the accounts, i.e. those in the OUT block, is exposed to a great flow of content, in which the percentage of m/disinformation is quite high. On the other hand, the INTEND-dominant bow-tie is ``safer", since the greatest part of the accounts therein (i.e. the INTENDRILS nodes) accesses the messages from the IN sector that is less prone to m/disinformation campaigns.

It worth to be remarked that, due to the considerations above, the OUT-dominant bow-ties are at risk of infodemic. \emph{Infodemic} is a recently introduced neologism, that became particularly popular during the Covid-19 pandemic. According to the WHO, ``\emph{infodemics are an excessive amount of information about a problem, which makes it difficult to identify a solution.
Infodemics can spread misinformation, disinformation and rumors during a health emergency. Infodemics can
hamper an effective public health response and create confusion and distrust among people}\footnote{Coronavirus disease 2019 (COVID-19)
Situation Report – 45:~\url{https://www.who.int/docs/default-source/coronaviruse/situation-reports/20200305-sitrep-45-covid-19.pdf?sfvrsn=ed2ba78b_4}}". The effects of the present Covid-19 infodemic, even if debated~\cite{Valensise2021,Gallotti2021a}, may put at risk the countermeasures to the spread of an epidemic and it is worrisome for policy makers\footnote{See, for instance, the Joint Communication titled “Tackling COVID-19
disinformation - Getting the facts right” (June 10th, 2020), available at the following link: \url{https://ec.europa.eu/info/sites/default/files/communication-tackling-covid-19-disinformation-getting-facts-right_en.pdf}.}.\\

Finally, let us} consider also the peculiarity of right-wing discursive communities: for all those, the bow-tie is strong (i.e., the dimension of the OTHERS block is smaller than the SCC one) and it is neatly OUT-dominant. \fab{The structural exposure of the OUT-dominant bow-tie to infodemic} is even more emphasized by the extreme activity of the SCC: for instance, in the Covid-19 Italian dataset the link density in the right-wing bow-tie is at least 3 times greater than any other OUT-dominant strong bow-ties.\\

\begin{comment}
In the present work, \fab{we uncover the presence of bow-tie structures in Twitter discursive communities. Through the analysis of the activity of the various sectors of the bow-tie, we were able to identify some recurrent behaviours.}
%
% we relate the infodemic phenomenon to the specific structure of the discursive communities. 
%
As highlighted above, for the investigated datasets:
\begin{itemize}
    \item OUT-dominant bow-ties are the most affected by \fab{infodemics};% low quality contents;
    \item This effect is particularly amplified in right-wing discursive communities, due to the extraordinary activity in creating contents of their SCC sectors; 
    \item \fab{In right-wing discursive communities, SCC sectors strongly contribute to the creation and diffusion of m/disinformative contents and the most exposed to m/disinformation are the OUT sectors}.
\end{itemize} 
\end{comment}

\paragraph{Statistical significance of the analysis} Here, we remark an important aspect of our analysis, of uttermost importance. 
% This aspect have appeared as excessive zeal at a first sight, but it is of utmost importance. 
%
In the analysis of a complex network, it is necessary to consider what is being measured, and what is its baseline. 
% When analysing a complex network, we have to pay attention to what it is measuring and what is her baseline: 
A typical example is the modularity, i.e. one of the most used target function for community detection. The problem resides in stating what is the number of links inside a group of nodes that is  enough to form a community. In this case, we build a null-model, i.e., a model that shows part of the properties of the original system, being random for all the rest, to have a proper benchmark for our observations. We then compare the number of edges inside a group of nodes with the one expected by the null-model. Without the null-model, we could not know whether the number of links that bind a group of nodes are due to the degree sequence, or whether they are instead the genuine signal of the presence of a community. 
% Without the null-model, we would not know if the links inside a group of nodes may be an accident due to the degree sequence or either a genuine signal of the presence of a community.

In the present study, we used an entropy-based null-model as a benchmark for our analysis~\cite{Squartinia,Cimini2018}. An entropy-based null-model allows to have a benchmark that is tailored to the system under analysis. It fixes (on average) some topological quantities to the values observed in the real network and leaves all the rest completely random. Being based on the (Shannon) entropy maximisation, it guarantees that it uniformly considers all the possible configurations (it is `ergodic', using Statistical Physics jargon), thus it does not introduce any bias in the analysis. 

To strengthen the analysis, we study if the bow-tie structures are due to the degree sequence of the nodes in the various discursive communities. 
% Due to the peculiar structure (see again Fig.~\ref{fig:structure}), 
In fact, the size of IN and OUTTENDRILS could simply be due to the presence of many nodes with zero in-degree (an analogous consideration could be done for the OUT and the INTENDRILS blocks, considering, instead the out-degree).
%
% the dimension of the IN and OUTTENDRILS blocks, for instance, could have been related just to the presence of nodes with zero in-degree (analogous considerations can be done for the OUT and the INTENDRILS blocks, considering, instead the out-degree). 
Thus, strong, weak and not informative bow-ties could be due to degree sequence only, and do not carry any kind of 
 information on their own.
 
We thus used the Directed Configuration Model defined in Ref.~\cite{Squartini2013a} and implemented by the Python module \href{https://pypi.org/project/NEMtropy/}{\texttt{NEMtropy}}~\cite{Vallarano2021}. Our results show that the dimensions of the blocks in the bow-tie are very often statistically significant: the p-value of the observed dimensions of the various blocks against the null-model expected distribution are extremely small, such that they are not compatible with the degree sequence, or, otherwise stated, the dimension of the various blocks cannot be explained using the degree sequences only.

\paragraph{Limitations} Even if we have obtained strong results (see the null-model validation check on the dimension of the bow-tie sectors), we have nevertheless to remark few aspects of our analysis that can limit its generalization. 
% Limit: few datasets; limited in place (Europe); single social network platform.
First, the analysis is related to eight different thematic datasets in different languages, all referring to European debates, some of them of political nature.  Indeed, while the total amount of messages analysed is quite impressive, we are aware that, even if the spectrum of the arguments covered is various, our findings may be valid on our datasets only. In the near future, we are going to expand the countries covered by our analyses and expand the list of \fab{debated} arguments.

%Secondly, what we are observing a correlation: 
Following our jargon, OUT-dominant bow-ties \fab{expose the majority of their accounts to the risk of infodemic}. Nevertheless, it is not a causal relation: the presence of OUT-dominant bow-ties \fab{does not imply the presence of an infodemic or of a disinformation campaign. In fact, if the sources shared by SCC are reputable, we will not observe} any infodemic or m/disinformation signal. \fab{At the same time,} it is true that OUT-dominant bow-ties help the diffusion of m/disinformation, when present, since accounts in OUT are exposed to all contents \fab{-reliable and not reliable-} created by nearly every block in the discursive community.

%The point is that it depends on what are the sources of m/disinformation messages: if the sources were the IN block, then it was easier to find the effects in INTENDRILS, thus the structure that would have helped the diffusion of the m/disinformation messages would have been the weak bow-tie. As we saw, it is not the case: probably due to their recognizability, verified users share less non reputable URLs. In this sense, focusing on a few reliable sources seems to be a good strategy to limit the diffusion of m/disinformation. 
% A peculiar method for extracting discursive communities? In preparation a comparison between various approaches that shows that it works.
%  Finally, \fab{as far as we know, }the observed bow-ties appear just in discursive communities, a method that is recent and quite limited in the applications~\cite{Becatti2019,Caldarelli2020,Caldarelli2021,Radicioni2021a,Radicioni2021b,Mattei2021,Bruno2021}. \fab{Among the targets of future researches, we have the investigations about the presence of bow-tie in the discursive communities of other new online social debates, in Twitter and in other online social networks.}

%poi attenzione, quando la dimensione e' sia grande che statisticamente significativa...abbiamo un risultato ancora piu' forte (Diogo Pachego)

\fab{\paragraph{Final remarks} Let us conclude with some final remarks.
First, the bow-tie structure is present in the discursive communities of retweet networks. Let us recall that we build the retweet network by creating a direct link for every retweet, from the author of the original post to the retweeter. Then, from the retweet network we extract the subgraphs relative to the discursive communities obtained through the procedure described in Subsection~\ref{sec:disc}. With this procedure, we are recovering the flow of information \emph{inside} each discursive community; in this sense, we are disregarding the possible interactions among different discursive communities.\\ 
There is another limitation which is unavoidable, due to the nature of  Twitter's data: we have information regarding who retweeted who, but not on the ``chain" of retweets, i.e. we cannot distinguish if the retweeter retweeted directly the message of the original author of the post or through one of the retweets given by one of followers of the original author.\\ 

Given the structure of the retweet network, it is therefore natural to ask what is the meaning of the bow-tie structure. In particular, what is the sense of the reachability of nodes in the retweet network?\footnote{We are thankful to reviewer 2 for suggesting this reflection that permits to give a clearer frame to our results.}\\
%Actually, there two different points: first, we do not really know the exact order in which Twitter present messages in  a user timeline, but nevertheless the original messages, retweets and like by the followees are displayed. In this sense, even with the important caveat of the lack of information about the retweet chain, we can expect that the retweet network is good proxy of the follower network. 
%There are no information regarding the correlation between the follower network and the retweet network, even if we expect to find a correlation: even if the exact order in which Twitter present messages in a user timeline is not know, nevertheless the original messages, retweets and like by the followees are displayed. Thus we expected to find some signal, since it is more likely for users to be attracted by some message by their followees and retweet and/or like. 
In fact,  the retweet network describes an influence flow: users, by retweeting messages, testify that they are influenced by their opinions. In this sense, the opinions expressed in messages written by a user $A$ influence the ones of her/his retweeter $B$, that, on turn, influence the one of her/his retweeter $C$ and so on, even if $C$ has never directly retweeted any of the messages of $A$. 
Otherwise stated, what the bow-tie is capturing is a division of users in different sets: in IN we find the creators of contents, only partially influenced by the content created by others\footnote{Consider that, following Yang's definition, accounts in IN are allowed to retweet each other, even if they are not part of the greatest SCC. In this sense, they are not completely independent of the influence of accounts in the same block.}, in OUT the big audience of standard users, influenced by the contents created in IN, SCC, TUBES and OUTTENDRILS, in SCC the active users influencing each other, and so on.\\

Finally, even if we have not access to the chain of retweets, we still expect the retweet network to be a good proxy of the followership network. 
Indeed, users either retweet messages from the accounts they follow or they retweet some message when searching for a specific topic. However, we expect the first method to be way more frequent, since in the user's home the activity (i.e. new messages, likes and retweets) of the followed accounts are present. In this sense, we expect that the bow-tie will be present (and statistically significant) even in the followers network. Such an investigation is going to be part of future research.
%Remarkably, this division is not random, even given the activity (i.e. the in- and out-degrees) of each node. 
%Consider, for instance, the p-values relative to DX discursive community in Table~\ref{tab:pvalues}: the number of elements in SCC, TUBES, INTENDRILS and OUTTENDRILS (just to focus on the bow-tie WCC) are particularly far from what expected by the null-model (i.e. extremely low p-values). In this case, in fact, the bow-tie structure is particularly informative. 
}

\section{Methods}
\fab{\subsection{Bow-tie detection}\label{sec:bow_tie_yang}

In the following we briefly described the main steps of the detection of the bow-tie structure, following the procedure outlined in Ref.~\cite{Yang2011}. The first step is the identification of the greatest Strongly Connected Component (SCC) and then the identification of the nodes in the various sectors, using the bow-tie definition.\\ 
Let be $G_D(V, E)$ a directed graph where $V$ is the set of nodes and $E$ the set of links, and $G_D^T(V, E)$ its counterpart obtained reversing the direction of the edges. 
The functions for the identification of the greatest SCC and the depth-first searches (DFS in the following), used to identify nodes reachable from a given node, are implemented in many python modules, such as \href{https://igraph.org/python/}{\texttt{igraph}} or \href{https://networkx.org/}{\texttt{networkx}}; for the present analysis, we used the former python module. The algorithm is pretty straightforward and follows the definitions in Section~\ref{sec:bowtie_strut}: the pseudo code is presented in Algorithm~\ref{alg:btda}.\\

\noindent Philosophically, the algorithm works as follows. First, consider the greatest strongly connected component and call it $SCC$; then choose randomly a node $v\in SCC$. All nodes that can reach $v$ (identified via DFS), but that are not part of $SCC$ represent the $IN$ sector; an analogous line of reasoning takes to the identification of the $OUT$ sector. Regarding the remaining node, the crucial information to be calculated is if they can be reached by nodes in $IN$ and if they can reach nodes in $OUT$ (again, using DFS): based on this final pieces of information, we can identified all remaining sectors.

}

\begin{algorithm}
\fab{
\begin{algorithmic}
\State $SCC \gets \text{all the nodes in the greatest Strongly Connected Component}$
\State $v \gets \text{a randomly chosen node in } SCC$
\State $DFS_{G_D}(v) \gets $ all the nodes that can be reached by $v$
\State $DFS_{G_D^T}(v) \gets $ all the nodes that can reach $v$
\State $OUT\gets DFS_{G_D}(v)/S$
\State $IN\gets DFS_{G_D^T}(v)/S$
\State $remainingV \gets V/SCC/IN/OUT$\Comment{All nodes that are not in $SCC$, neither in $IN$, neither in $OUT$.}
\For{$w \in remainingV$}
    \State $IRW \gets (IN \cap DFS_{G_D^T} (w)\neq\emptyset)$\Comment{``$IN$ Reaches $w$?"}
    \State $WRO \gets (OUT \cap DFS_{G_D} (w)\neq\emptyset)$\Comment{``$w$ Reaches $OUT$?"}
\If{$IRW$ and $WRO$}
    \State $w\in TUBES$
\ElsIf{$IRW$}
    \State $w\in INTENDRILS$
\ElsIf{$WRO$}
    \State $w\in OUTTENDRILS$
\Else
    \State $w\in OTHERS$
\EndIf
\EndFor
\end{algorithmic}
\caption{Bow-tie detection algorithm}\label{alg:btda}
}
\end{algorithm}

\subsection{Entropy-based null-models for network analysis and their applications}\label{sec:netmethods}
\subsubsection{The Bipartite Configuration Model}\label{ssec:BiCM}
In order to create the various discursive communities we needed an appropriate null-model as benchmark for identifying those verified users that share the same audience. In this sense, it is necessary to compare the observed quantities with accurate predictions in order to state their significance: actually, the common audience may appear similar just due to the extreme activity of the considered verified users.\\
We represent the interaction between verified accounts -the ones whose identity is certified by Twitter platform- and unverified ones (i.e. all the others) via a bipartite undirected binary network in which a link connects a verified users to an unverified ones if there is at least a retweet between one and the other, or viceversa. Since the information about the number of different accounts interacting -via tweet or retweet- with a user is encoded, in this representation, in the degree sequence for nodes of both layers, we need a benchmark discounting it. The natural choice is to choose an entropy-based null-model, since it provides, by definition an unbiased framework~\cite{Cimini2018}: the null-model is maximally random, but for the constraints imposed on the system. The bipartite null-model discounting the degree sequence is the Bipartite Configuration Model (BiCM,~\cite{Saracco2015a}). In the present section we will briefly revise the steps of its definition.\\

Let us consider a bipartite network in which the two layers $\top$ and $\bot$ have dimension, respectively, $N_\top$ and $N_\bot$; in the following, Latin indices will be used to identify nodes on the $\top$ layer while Greek ones will be used for the $\bot$ layer. Then, the bipartite network can be represented by its biadjacency matrix, i.e. a $N_\top\times N_\bot$ matrix $\mathbf{M}$ whose generic entry $m_{i\alpha}$ is $1$ if the node $i\in\top$ is connected to the node $\alpha\in\bot$ and 0 otherwise.

Let us start from a real bipartite network $G_\text{Bi}^*$ (in the following, all quantities denoted by a $*$ will indicate those measured on the real network). First, let us define an ensemble of graphs, i.e. the set of all the possible bipartite graphs having the same number of nodes of $G_\text{Bi}^*$, but with all different topologies, from the fully connected to the empty ones. Then, we can define the Shannon entropy over the ensemble, by assigning a different probability to each of its elements:

\begin{equation*}
        S=-\sum_{G_\text{Bi}\in\mathcal {G}_\text{Bi}}P(G_\text{Bi})\ln{P(G_\text{Bi})};
\end{equation*}
where, $P(G_\text{Bi})$ is the probability of the generic element of the graph ensemble $G_\text{Bi}$. Let us now maximise the entropy, while constraining the network degrees: in particular, we want that the ensemble average of degrees to match the value observed on the real network, in order to have a null-model tailored to the real system. In term of the biadjacency matrix, the degree sequences of the $\top$ and $\bot$ layers respectively read $k_i=\sum_\alpha m_{i\alpha}$ and $h_\alpha=\sum_i m_{i\alpha}$. Using the method of the Lagrangian multipliers,  the constrained maximisation can be expressed as the maximisation of $S'$, defined as

\begin{eqnarray*}
    S'&=&S\nonumber\\
    &&+\sum_i\eta_i\left[k_i^*-\sum_{G_\text{Bi}\in\mathcal {G}_\text{Bi}}P(G_\text{Bi})k_i(G_\text{Bi})\right]+\sum_\alpha\theta_\alpha\left[h_\alpha^*-\sum_{G_\text{Bi}\in\mathcal {G}_\text{Bi}}P(G_\text{Bi})h_\alpha(G_\text{Bi})\right]\nonumber\\
    &&+\zeta\left[\sum_{G_\text{Bi}\in\mathcal {G}_\text{Bi}}P(G_\text{Bi})-1\right]
\end{eqnarray*}
where $S$ is the Shannon entropy defined above, $\eta_i$, $\theta_\alpha$ are the Lagrangian multipliers relative to the degree sequences, respectively, on $\top$ and $\bot$, and $\zeta$ is the one relative to the probability normalization.

Maximising $S'$ leads to a probability per graph $G_\text{Bi}\in\mathcal{G}_\text{Bi}$ that can be factorised in terms of the probabilities per link $p_{i\alpha}$~\cite{park2004statistical}, i.e.

\begin{equation}
\label{factorized}
    P(G_\text{Bi})=\prod_{i,\alpha}p_{i\alpha}^{m_{i\alpha}(G_\text{Bi})}\,(1-p_{i\alpha})^{1-m_{i\alpha}(G_\text{Bi})},
\end{equation}
where $p_{i\alpha}=\dfrac{e^{-\eta_i-\theta_\alpha}}{1+e^{-\eta_i-\theta_\alpha}}$. Nevertheless, at this level the above equation is just formal, since we do not know the numerical value of $\eta_i$ and $\theta_\alpha$. To this aim, we can then maximise the likelihood of the real network~\cite{Garlaschelli2008,squartini2011analytical}; it can be shown that the likelihood maximisation is equivalent to imposing

\begin{equation*}
    \langle k_i\rangle_\text{BiCM}=k_i^*,\,\forall\:i\in\top;\qquad\langle h_\alpha\rangle_\text{BiCM}=h_\alpha^*,\,\forall \alpha\:\in\bot.
\end{equation*}

\subsubsection{Validated projection of bipartite networks}
We want to infer similarities among nodes on the same layer. We can use as a measure of similarity the number of common neighbours - for each couple of verified users, the number of unverified users that have interacted, via tweet or retweet, with both. Let us assume, without loss of generality, that we want to project the information contained in the bipartite network onto the $\top$ layer and call $V_{ij}$ the number of common neighbors between nodes $i,j\in\top$\footnote{Following Ref.~\cite{Saracco2016}, we use the letter $V$ to indicate common neighbours, since this pattern appear in the bipartite network as a ``V" between the layer.}.

%
\begin{comment}
If we imagine of disposing the layer of hashtags above that of users, when two hashtags share the same user, a V-shaped form emerged, like in figure \ref{Figure14}.
\begin{figure}
    \centering
    \includegraphics[scale=0.4]{Figure_14.png}
    \caption{
     A simple representation of three different V-motifs between a pair of hashtags.}
    \label{Figure14}
\end{figure}

This structure is called a \emph{V-motif}. If we call $\mathbf{A}$ the adjacency matrix describing our bipartite networks, whose dimensions will be $N_H\times N_U$ (where $N_U$ and $N_H$ are respectively the total number of users and the total number of hashtags) and $a_{hu}$ its entries, then the total number of V-motifs between two nodes $h$ and $h'$ is
\end{comment}
%

In terms of the biadjacency matrix, $V_{ij}$ can be expressed as

\begin{equation*}
V_{ij}=\sum_\alpha V_{ij}^\alpha=\sum_\alpha m_{i\alpha}m_{j\alpha},
\end{equation*}
where we have defined $V_{ij}^\alpha= m_{i\alpha}m_{j\alpha}$;  $V_{ij}^\alpha=1$ if both $i$ and $j$ are connected to node $\alpha\in\bot$ and 0 otherwise. Let us now compare the observed $V_{ij}$ for each possible pair of nodes in $\top$ with the prediction of the BiCM. Since link probabilities are independent, the presence of each V-motif $V_{ij}^\alpha$ can be regarded as the outcome of a Bernoulli trial:

\begin{equation*}
\begin{split}
f_{\text{Ber}}(V_{ij}^\alpha=1)=&p_{i\alpha}p_{j\alpha},\\
f_{\text{Ber}}(V_{ij}^\alpha=0)=&1-p_{i\alpha}p_{j\alpha}.
\end{split}
\end{equation*}

In general, the probability of observing $V_{ij}=n$ can be expressed as a sum of contributions, running on the n-tuples of considered nodes (in this case, the ones belonging to the layer of users). Indicating with $A_n$ all possible nodes n-tuples among the layer of $\bot$, this probability amounts at

\begin{equation}\label{eq:PB}
f_{PB}(V_{ij}=n)=\sum_{A_n}\left[\prod_{\alpha\in A_n}p_{i\alpha}p_{j\alpha}\prod_{\alpha'\notin A_n}(1-p_{i\alpha'}p_{j\alpha'})\right],
\end{equation}
where the second product runs over the complement set of $A_n$. Eq.~(\ref{eq:PB}) represent the generalization of the usual Binomial distribution when the single Bernoulli trials have different probabilities, also known as Poisson Binomial distribution~\cite{Hong2013}.

We can, then, verify the statistical significance of the observed co-occurrences by calculating their p-value according to the distribution in Eq.~\ref{eq:PB}, i.e. the probability of observing a number of co-occurrences greater than, or equal to, the observed one:

\begin{equation}
\text{p-value}\big(V^*_{ij}\big)=\sum_{V_{ij}\ge V^*_{ij}}f_{PB}\big(V^*_{ij}\big).
\end{equation}

Repeating this calculation for every pair of nodes, we obtain $\binom{N_\top}{2}$ p-values. In order to state the statistical significance of the hypotheses belonging to this group, it is necessary to adopt a multiple hypothesis testing correction; in the present paper, we use the False Discovery Rate (FDR,~\cite{benjamini1995controlling}), since it controls the false positives rate.

\subsubsection{Direct Configuration Model}\label{ssec:DCM}
From the entire retweet network, in which the various accounts are represented as nodes in a direct network in which an arrow points the retweeter of a post, starting from its author, we extracted the various subgraphs of discursive community. Then, in order to compare the observed dimensions of the bow-tie sectors of these subgraphs and state their statistical significance, we adopted the \emph{Direct Configuration Model} (DCM), which is the entropy-based model suited for direct monopartite networks~\cite{Mastrandrea2014}. 
For directed networks, the adjacency matrix is (in general) not symmetric, and each node $i$ is characterized by two degrees: the out-degree $k^{\text{out}}_i=\sum_ja_{ij}$ and the in-degree $k^{\text{in}}=\sum_ja_{ji}$, where $a_{ij}$ is the generic entry of the (directed) adjacency matrix $\mathbf{A}$. The Directed Configuration Model (DCM) is therefore defined as the ensemble
of direct networks with given out-degree and in-degree sequences. Using the same machinery as in the previous subsection~\ref{ssec:BiCM}, it is possible to derive a probability per graph: if $G_D$ is the generic representative of the ensemble of directed graphs $\mathcal{G}_D$, then the probability per graph $P(G_D)$ reads:
\begin{equation*}
    P(G_D)=\prod_{i,j\neq i}q_{ij}^{a_{ij}(G_D)}(1-q_{ij})^{1-a_{ij}(G_D)}.
\end{equation*}
Thus, again the probability per graph factorises in terms of probabilities per link $q_{ij}$, which can be expressed in terms of Lagrangian multipliers 
\begin{equation*}
    q_{ij}=\dfrac{e^{-\gamma_i-\delta_j}}{1+e^{-\gamma_i-\delta_j}},
\end{equation*}
where $\gamma_i$ and $\delta_j$ are the Lagrangian multipliers associated, respectively to the out-degree of node $i$ and to the in-degree of node $j$. In order to get the numerical value of $\gamma_i$ and $\delta_j$, 
we can use the maximum likelihood as in the above subsection~\ref{ssec:BiCM}, which is equivalent to impose
\begin{equation*}
    \langle k_i^{\text{out}}\rangle_\text{DCM}=k_i^{*^{\text{out}}},\qquad\langle k_i^{\text{in}}\rangle_\text{DCM}=k_i^{*^{\text{in}}},\,\forall i.
\end{equation*}
Since the bow-tie decomposition is highly non linear, in order to calculate the statistical significance of the dimension of the various blocks, we generated a sample of 1000 different graphs for each discursive community, using the probabilities provided by the DCM.
Then, we obtained a distribution for the dimensions of the bow-tie sectors just looking to the decomposition of each graph in our ensemble. At this point, we could calculate a two-tailed p-value with a significance at $\alpha=0.01$ for estimating the distance between the dimensions observed with those reproduced by the ensemble. 

\subsubsection{Modularity and community detection}
In the present analysis, we inferred the discursive communities from the communities in the validated network of verified users. In particular, we used the modularity based Louvain algorithm~\cite{Blondel2008}.\\

The modularity~\cite{Newman2010} compares the number of edges within the actual communities with its expectation under a certain null-model. Modularity can be written as

\begin{equation}
    Q=\frac{1}{2m}\sum_{ij}\big(a_{ij}-p_{ij}\big)\,\delta(C_i,C_j)
\end{equation}
where $m$ is the total number of links of the network, $a_{ij}$ are the entries of the adjacency matrix, $p_{ij}$ is the probability to have a link between nodes $i$ and $j$ according to the chosen null-model, $C_i$ and $C_j$ are, respectively, the communities of nodes $i$ and $j$ and the Kronecker delta $\delta(C_i,C_j)$ selects all the pairs of nodes contained in the same community (equal to 1 if $C_i=C_j$ or 0 otherwise). In the original definition in Ref.~\cite{Girvan2002}, the null-model chosen is the Chung-Lu one~\cite{Chung2002}, which conserve the degree sequence, but it is known to be inconsistent for dense networks that present strong hubs~\cite{Cimini2018}. In the present paper we use instead the entropy-based Undirected Configuration Model (UCM) defined in~\cite{Garlaschelli2008,squartini2011analytical}: it can be shown that in the case of sparse network, the UCM can be approximated by the Chung-Lu null-model. 

\fab{Furthermore, Louvain algorithm is known to be order dependent, i.e. the resulting configuration depends on the order of the nodes given to the algorithm. In order to avoid this bias, we run 100 times the algorithm reshuffling each time the node order. At the end of 100 runs, we select the final partition displaying the maximum of the objective function (in our case, the modified modularity with the UCM).\\ The multiple runs approach is quite common~\cite{Saracco2016, Becatti2018, Becatti2019, Evkoski2021plosone, Evkoski2022plosone}, but different approaches are present in the literature for the final choice of the resulting partition in communities: for instance, in Ref.s~\cite{Evkoski2021plosone, Evkoski2022plosone} the authors, instead of choosing the partition with the greatest value of the modularity, prefer to choose the most common clusters, the choice being motivated by the specific profile of modularity~\cite{Good2010}. While the procedure proposed in Ref.s~\cite{Evkoski2021plosone, Evkoski2022plosone} is perfectly acceptable, we prefer ours, since it targets directly the objective function we are considering.}

\subsubsection{\texttt{NEMtropy}}
In the present paper, we implemented the BiCM, the DCM and the Louvain algorithm using UCM null-models via the Python module \href{https://pypi.org/project/NEMtropy/}{\texttt{NEMtropy}}, described in Ref.~\cite{Vallarano2021}.

\subsection{Discursive communities for the Italian Covid-19 dataset}\label{ssec:disccomm}
Here, we give a brief description of the discursive communities identified in the Italian Covid-19 dataset (Their dimensions are in Fig. \ref{fig:disc}):
\begin{itemize}
    \item \textbf{DX}: this community collects the official accounts and the main leaders of two Italian right-oriented political parties, `Lega'
    %(English: Northern League) 
    and `Fratelli d'Italia' 
    %(English: Brothers of Italy)
    ;
    \item \textbf{M5S}: this community contains the main politicians and the official accounts of the Italian party `Movimento 5 Stelle' (English: 5 Stars Movement), an anti-establishment political movement; 
    % It can be defined as a `big tent' party with variable positions on several topics;
    \item\textbf{IV}: this community is associated to the liberal party of `Italia Viva' (English: Italy Alive) with centre/centre-left political positions;
    \item\textbf{PD}: this cluster contains the politicians of the Italian `Partito Democratico' (English: Democratic Party), the traditional centre-left party;
    \item\textbf{FI}: this group collects the politicians and the official accounts of the Italian centre-right party of `Forza Italia' (English: Italy Forward);
    \item\textbf{MEDIA}: this type of community is present in almost all the datasets we analyzed. It contains the official accounts of newspapers, journalists, TV-channels, radio channels and in general other Italian media.
\end{itemize}

\subsection{Social bots}\label{ssec:bot}

Social bots are computer algorithms whose behaviour on social platforms is often far from being benign: malicious bots are purposely created to distribute spam, sponsor public characters and, ultimately, induce a bias within the public opinion~\cite{cresci2015fame,Ferrara2016rise,cresci2017paradigm}. Often these agents have the task of increasing the visibility of certain users~\cite{Caldarelli2020,Caldarelli2021}.

Here, we report the outcome of a study about detection of social bots in the datasets under investigation.
For bot detection, we exploit the general-purpose bot detection system based on supervised-learning presented in~\cite{de2021efficacy}. Such a system has been shown to be highly accurate, both for unveiling automated accounts that work alone and those that participate in coordinated  activities (we cannot determine phase transitions in this peculiar dynamics\cite{gabrielli2007invasion}). The bot detector is `traditional',  i.e.,  only one user per time is analyzed during the classification process~\cite{cresci2020decade}.

The classifier exploits so-called Class A features, i.e., features  that can be directly extracted from the user profile. These features were originally introduced in~\cite{cresci2015fame} and, despite their simplicity, proved to be still effective for the detection of novel bots too. Features that are known to be the most expensive to compute (mainly in terms of time needed for data gathering), namely those concerning the account's relationships (friends and followers) have been disregarded.

Hence, in order to decide about the type of the account (either a bot or not), we (i) train and evaluate different machine-learning algorithms on a dataset where bots and genuine accounts are a priori known, (ii) we select the model with the best classification performances and (iii) we apply the resulting model to the accounts of the datasets investigated in the main text of the manuscript.

To train and validate the classifier we leverage the publicly available cresci-stock-2018\footnote{\url{https://botometer.osome.iu.edu/bot-repository/datasets.html}} dataset. In particular, we use the accounts metadata of the $6842$ bots and $5880$ human that were still active at the time of data collection; data were crawled on July 2020 through the Tweepy library\footnote{\url{https://www.tweepy.org/}}.
%annotated accounts in total.  at the time of the data collection  the data-sets in Table~\ref{tab:orig-datasets} using $8520$ bots and $13820$ human annotated accounts in total.

To select the best model, we consider five algorithms, each of them belonging to a different category: MlP (Multilayer Perception)~\cite{159058}, JRip, i.e., a Java-based implementation of the RIPPER algorithm~\cite{COHEN1995115}, Naive Bayes~\cite{10.5555/2074158.2074196}, Random Forest~\cite{rf}, and the Weka~\cite{witten2016data} implementation of the Instance-based Learning Algorithms, i.e., IBk~\cite{ibk}.

%  we consider five different machine-learning algorithm families (the same proposed also in~\cite{de2021efficacy}): Random Forest~\cite{rf}, Ripper~\cite{COHEN1995115}, Multilayer Perception~\cite{159058}, Naive Bayes~\cite{10.5555/2074158.2074196}, and the Instance-based Learning Algorithms, i.e., IBk~\cite{ibk}. 

 The performances of the five different algorithms are evaluated in terms of standard metrics, such as 
 balanced accuracy, 
 precision, and f-measure.
 %and Matthew Correlation Coefficient (MCC). 
 The metrics are computed using a 10-fold cross-validation.

For all the experiments, we rely on the open source (Java-based) Weka framework that provide us the implementations of (i) the five machine-learning algorithms (for which we use the default parameter settings\footnote{\url{https://waikato.github.io/weka-wiki/documentation/}}), (ii) the evaluation metrics and (iii) the process of 10-fold cross validation.

In light of our experiments (see Table~\ref{tab:botperformance}), we select the Random Forest-based model as the classification process since it outperforms the other models.

% NOTE BOT SECTION
% \begin{itemize}
%     \item Inserire limitaizoni dell'analisi bot? (classA limitation, ...)
%     \item motivazione del perché si é utilizzato un classificatore fatto ad-hoc
% \end{itemize}

%\begin{table}
%\begin{tabular}{l||c|c|c|c|c|c}
%    \textbf{dataset} & \textbf{description} & \#bot & \#human & \#act\_bot & %\#act\_human & year \\
%    \hline
%    \hline
%	\texttt{Verified} & verified Twitter accounts & 0 & 2000 & 0 & 1974 & 2019 \\
%	\texttt{Celebrity} & celebrities & 0 & 5970 & 0 & 5649 & 2019 \\
%	\texttt{Botwiki} & self-declared bots & 704 & 0 & 664 & 0 & 2019 \\
%	\texttt{Rtbust} & genuine accounts/coordinated bots & 391 & 368 & 315 & 317 & 2019 %\\
%	\texttt{Stock} & genuine accounts/coordinated bots  & 18508 & 7479 & 6842 & 5880 & %2018 \\
%	\texttt{Vendor} & fake followers & 1088 & 0 & 699 & 0 & 2019 \\
%    \hline
%	\texttt{Final Data-set} &&&& 8520 & 13820 & \\
%    \hline
%\end{tabular}
%\caption{\small Description of the original datasets. \bf{mari: secondo me tanto vale mettere solo le colonne act bot e act human, e dire proprio bot e genuine, senza stare a specificare che sono quelli vivi al tempo della nostra analisi... Inoltre, metterei la citazione accanto al nome di ogni dataset}}\label{tab:orig-datasets}
%\end{table}

\begin{table}
\centering
\begin{tabular}{l||c|c|c|c|c}
    \textbf{Algorithm}  & \textbf{Precision} & \textbf{Recall} & \textbf{F-Measure} \\
    \hline
    \hline
    RandomForest & 0.838 & 0.828 & 0.833\\
    Ripper  & 0.819 & 0.823 & 0.821\\
    Multilayer Perceptron  & 0.709 & 0.737 & 0.723 \\
    NaiveBayes  & 0.554 & 0.986 & 0.709 2\\
    IBk  & 0.708 & 0.726 & 0.717 \\
    \hline
\end{tabular}
\caption{Performance results after 10-folds cross validation on cresci-stock-2018 data-set}\label{tab:botperformance}
\end{table}

% \begin{table}
% \begin{tabular}{l||c|c|c|c|c}
%     \textbf{Algorithm} & \textbf{Bal. Accuracy} & \textbf{Precision} & \textbf{Recall} & \textbf{F-Measure} & \textbf{MCC} \\
%     \hline
%     \hline
%     RandomForest & 0.821 & 0.838 & 0.828 & 0.833 & 0.64\\
%     Ripper & 0.805 & 0.819 & 0.823 & 0.821 & 0.61\\
%     Multilayer Perceptron & 0.692 & 0.709 & 0.737 & 0.723 & 0.385\\
%     NaiveBayes & 0.531 & 0.554 & 0.986 & 0.709 & 0.152\\
%     IBk & 0.688 & 0.708 & 0.726 & 0.717 & 0.377\\
%     \hline
% \end{tabular}
% \caption{Performance results after 10-folds cross validation on cresci-stock-2018 data-set}\label{tab:botperformance}
% \end{table}

The resulting model for bot classification is then applied to tag all the accounts involved in our study, giving an average concentration of bots that is around 23.9\% in total. In particular, if we focus on specific datasets,  we observe percentage of bots around 23.9\% for \textbf{Italian Covid-19}, 29\% for \textbf{German Covid-19}, 23.4\% for \textbf{French Covid-19}, 22.8\% for \textbf{Dutch elections}, 25.7\% for \textbf{Italian debate on migrants}, 24\% for \textbf{Italian debate on the Astrazeneca vaccine} and 18.2\% for \textbf{Italian and Turkish EURO2020} dataset. 

These are quite high values, especially if we take as a baseline measure the one provided by Varol et al.~\cite{varol2017online} in a 2017 study which estimated the percentage of active bots on the Twittersphere at between 9 and 15\%.

 However, in our research, several aspects could motivate both the high values and their variability amongst the datasets. Specifically, (i) we are looking at specific (hot) topics that might involve more significant numbers of bots than the average, (ii) we are considering datasets on significantly different topics (thus, the percentage of automated accounts might vary), and (iii) we are analyzing data collected in different time intervals, but evaluated with a single classifier (this might further affect the classification performance, due to the possible evolution of bots).

With these premises to keep in mind, we now describe how the potential bots are distributed in the discursive communities. 
They are equally distributed among the discursive communities, with a slightly higher percentage of bots in the conservative groups: for instance, in the Italian and French Covid-19 datasets, the communities with the highest percentage of bots are DX and RIGHT-WING with, respectively, the 25.5\% and the 29.7\% of suspicious accounts. In our bow-tie structures, they are basically placed in the OUT sector or in the INTENDRILS one.
\begin{figure}
    \centering
    \includegraphics[scale=0.28]{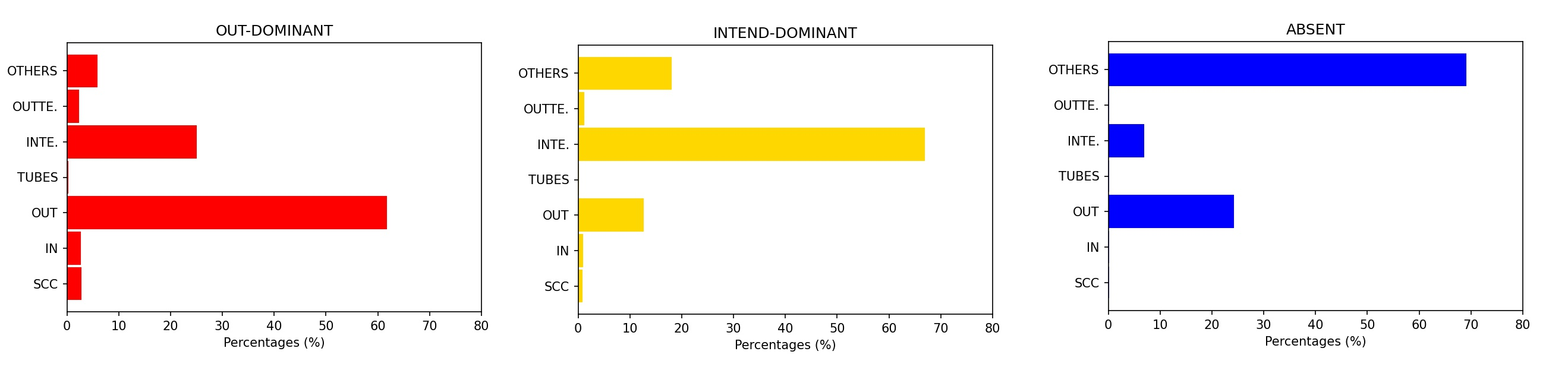}
    \caption{\textbf{Percentages of social-bots in each sector of the bow-tie structure for OUT-dominant, INTEND-dominant and not informative bow-ties}\\This figure collects the percentage of bots in every bow-tie's sector for discursive communities with OUT-dominant, INTEND-dominant and not informative bow-tie structure. It is easy to note how the highest percentages can be found in the OUT sector, in the INTENDRILS and in the OTHERS one, respectively for the case of OUT-dominant, INTEND-dominant and not informative bow-tie.}
    \label{fig:bots}
\end{figure}

%\textcolor{red}{I COMMENTI FANNO RIFERIMENTO AD UNA FIGURA CHE NON ESISTE AL MOMENTO.}
In Fig.~\ref{fig:bots} are shown the percentages of bots in a specific bow-tie sector averaged on all the discursive communities in the usual three categories. Globally, the highest percentages can be found in the OUT sector and in the INTENDRILS one: literally, social bots tend to retweet more than to be retweeted. 
In particular, in the case of OUT-dominant bow-tie, on average the  60\% of bots are placed in the OUT sector and to a lesser extent in INTENDRILS (around 25\%). %The few peaks in the INTENDRILS distribution represent the cases of weak bow-ties, in which we could find even the 60-70\% of bots in this sector.
In the case of INTEND-dominant bow-ties, we found even above the 60\% of bots in INTENDRILS sector. Instead, when  the bow-tie structure is absent, OTHERS is the block the contains the greatest number of bots.\\ 
It is worth to be mentioned is that the higher percentages of bots in the strongly connected component can be found in the right-oriented discursive communities. For instance, in the Italian Covid-19 dataset the percentage of bots in the SCC for the DX is the 7\%, while for all the others it does not overcome the 2\%. Such a situation is particularly dangerous, since the fact that social bots are able of being retweeted by human users (as it is the case for accounts in SCC) means that are able to pass off themselves as genuine accounts.

\section*{Acknowledgements}
FS ackowledge Pietro Galgani and Lizanne Dirkx for support in both the download and the analysis of the Dutch election dataset; Giulia Andrighetto, Stefano Guarino, Enrico Mastrostefano, Elena Pavan, Eugenia Polizzi and Tiziano Squartini for useful discussions. All authors acknowledge support from IMT PAI project Toffee.

\section*{Author contributions statement}
Ma.Pe. and F.S. planned the research. M.M., Ma.Pr. and F.S. performed the analyses. M.M. prepared all the figures. M.M., Ma.Pe., Ma.Pr. and F.S. wrote the main manuscript text. M.M. and F.S. wrote the Supplementary Materials. All authors reviewed the manuscript.

\section*{Additional information}
\textbf{Competing interests} The authors have no competing interests as defined by Nature Research, or other interests that might be perceived to influence the results and discussion reported in this paper.\\
\textbf{Data availability} The Twitter datasets used and analysed during the current study available from the corresponding author on reasonable request. The data about the reliability of the various news sources that support the findings of this study are available from Newsguard, but restrictions apply to the availability of these data, which were used under license for the current study, and so are not publicly available. Data are however available from the authors upon reasonable request and with permission of Newsguard. 

\bibliography{biblio.bib}

\newpage
\appendix
\section*{Supplementary Material}
\section{German Covid-19 dataset}
\begin{figure}
    \centering
    \includegraphics[scale=0.32]{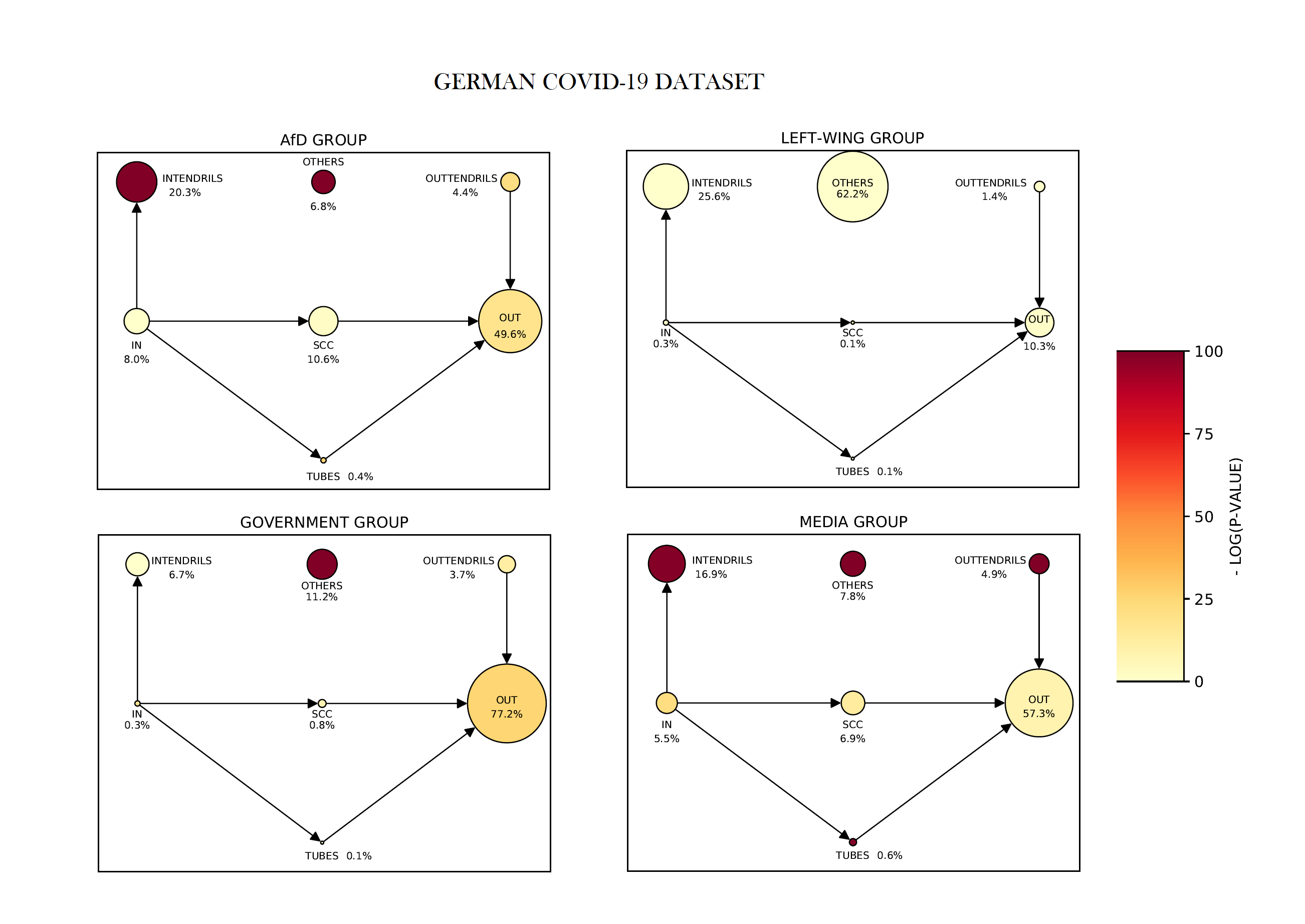}
    \caption{\textbf{The bow-tie structure of the discursive communities of the German Covid-19 dataset.} The dimension of the sectors is proportional to the number of nodes contained in them and the color \fab{indicates the distance between the observed and the predicted dimensions through $\ln(p-value)$}. The AfD, Government and MEDIA groups display an informative bow-tie structure, i.e. the OTHERS sector is the represent less than 50\% of the nodes. Considering the comparisons with the predictions of the Direct Configuration Model, the observed dimension for the OTHERS sector is significantly less numerous (considering a significance at 1\%) for all the communities, apart for the LEFT-WING one. }
    \label{covid_ger}
\end{figure}

The German Covid-19 dataset contains \fab{1,552,106} tweets shared between February 2 and April 23, 2020. \fab{we identified the following discursive communities}:
\begin{itemize}
    \item \textbf{AfD}: this group contains accounts of politicians of the German nationalist and right-wing party ``Alternative for Germany (AfD)";
    \item\textbf{LEFT-WING}: this community collects politicians of various German left-wing parties, as the ``Social Democratic Party (SPD)", ``Alliance 90/The Greens" and ``Die Linke" (literally, ``the left");
    \item\textbf{GOVERNMENT}: in this community are placed official accounts of German ministries and institutions as the Foreign, Defense or Health Ministries. It also contains politicians of the ``Christian Democratic Union of Germany (CDU)";
    \item\textbf{MEDIA}: \fab{it includes} the official accounts of the main German newspapers, blogs, TV-channels, journalists and other media in general.
\end{itemize}
\begin{figure}
    \centering
    \includegraphics[scale=0.35]{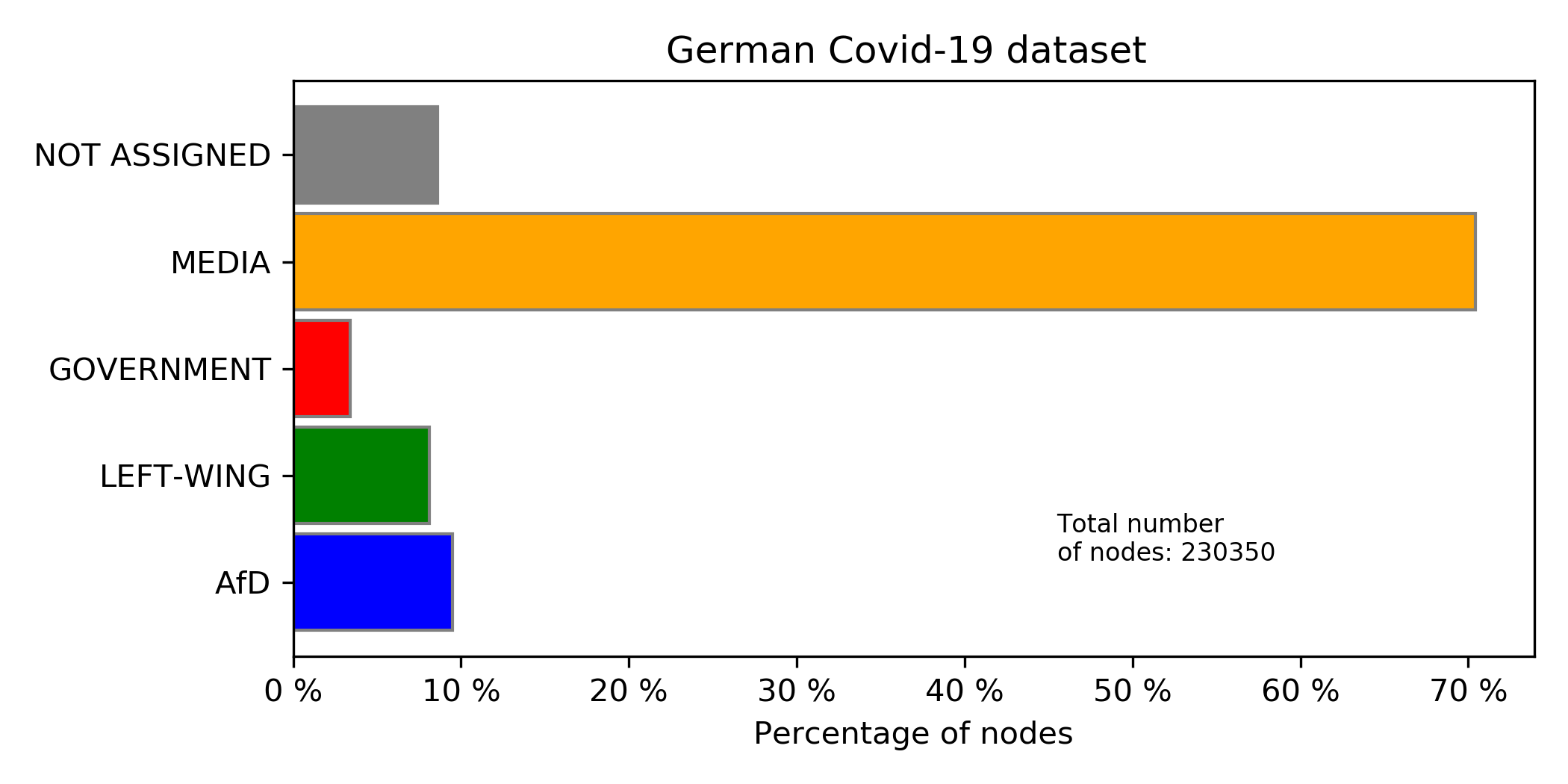}
    \caption{\textbf{The dimension of the discursive communities of the German Covid-19 dataset.}\\The bar chart displays the percentage of nodes in each discursive community. As for the other Covid-19 dataset the MEDIA group results the most numerous one, with approximately 70\% of the nodes of the entire network.}
    \label{discursive_ger}
\end{figure}
\begin{table}[]
    \centering
    \begin{tabular}{|l|r|r|}
\hline
 COMMUNITY   &   DIM. (verified accounts) &   DIM. (all accounts) \\
\hline
 AfD         &                     104 &           21903 \\
 L.-W.       &                     161 &           18674 \\
 GOV.        &                      79 &            7804 \\
 MEDIA       &                     224 &          162215 \\
\hline
\end{tabular}
    \caption{\fab{Dimension of the discursive communities of the German Covid-19 dataset,  considering verified accounts only and all users.}}
    \label{tab:dim_covid_ger}
\end{table}
\fab{Table~\ref{tab:dim_covid_ger} and the bar chart in  Fig.~\ref{discursive_ger} show the dimension of each discursive community}. As for the other Covid-19 datasets, the MEDIA group results the most numerous one, with approximately 70\% of the nodes of the entire network.\\
Fig.~\ref{covid_ger} shows the bow-tie structures for the four discursive communities. As in the main text, the dimension of the sectors is proportional to the number of nodes  and the color indicates \fab{the p-value encoding} the mismatch with the predictions of the Direct Configuration Model (described in the Methods section of the main text).
The AfD, GOVERNMENT and MEDIA groups display informative bow-tie structures; all of them are OUT-dominant, but only AfD has a strong bow-tie. In the LEFT-WING community the bow-tie is uninformative, with approximately 60\% of the vertices in the OTHERS sector. In agreement with the results of the Italian Covid-19 dataset, the OTHERS block results significantly less numerous for the three communities with an informative bow-tie, and not for the LEFT-WING.

\begin{comment}
\begin{figure}
    \centering
    \includegraphics[scale=0.37]{pageranks_covid_ger.png}
    \caption{\textbf{Top PageRanks location in the German Covid-19 dataset.}\\ In the bar charts are displayed the number top PageRanks located in each sector of the discursive communities. For each of them we selected the 100 vertices with the highest PageRanks. In those networks with a strong bow-tie (AfD, GOVERNMENT and MEDIA) the most part of the top PageRanks are placed in the SCC. The LEFT-WING is the only one in which they are more distributed across the sectors, with the OUT and the INTENDRILS sectors as the most occupied ones.}
    \label{pageranks_covid_ger}
\end{figure}

We found similar results also for the location of the top PageRanks, as it is possible to see in Fig.~\ref{pageranks_covid_ger}. In those networks with a strong bow-tie (AfD, GOVERNMENT and MEDIA) the most part of the top PageRanks are placed in the SCC. The LEFT-WING is the only one in which they are more distributed across the sectors, with the OUT and the INTENDRILS sectors as the most occupied ones. Again, all the top PageRanks for each community result significantly higher than what we expect from the DCM. 
\end{comment}

\begin{figure}
    \centering
    \includegraphics[scale=0.75]{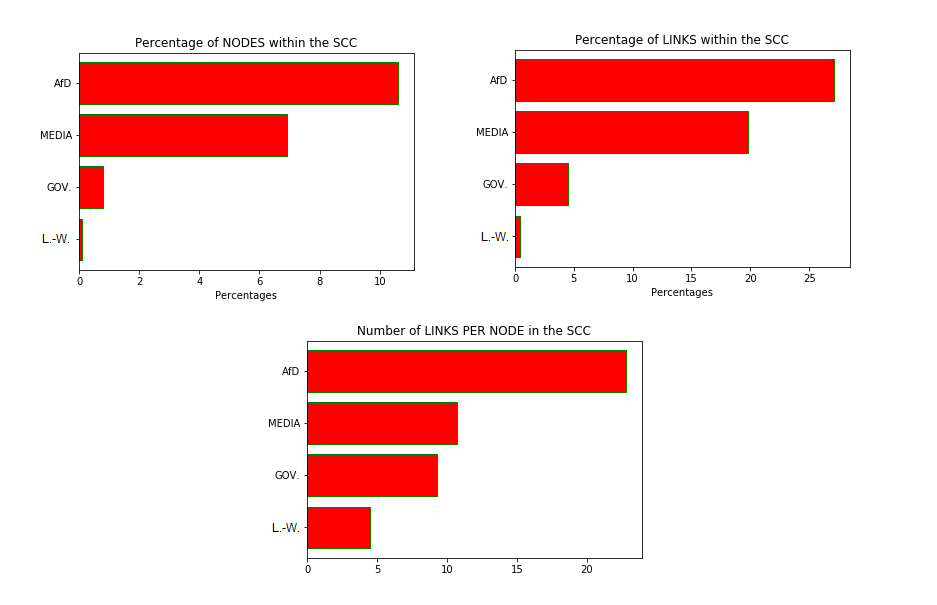}
    \caption{\textbf{Percentage of nodes and edges in the SCC for the communities in the German Covid-19 dataset.} As for the other datasets, in the German Covid-19 one the conservatives and right-oriented discursive community (AfD) has more numerous and denser SCCs, as it is displayed in the two top panels. In the bottom panel, it can be seen that also considering the fraction of links per node in the SCC, the AfD group results again the first one.}
    \label{conservatives_ger}
\end{figure}

Also in this dataset, the AfD \fab{discursive community}, which contains right-oriented and conservatives accounts, shows a more numerous and denser SCC. It is the only community with above  10\% of the nodes and  25\% of the links within the SCC, in which each vertex has over 20 links on average attached to it (see Fig.~\ref{conservatives_ger}). The accounts in the AfD discursive community are those who retweets the most urls of web-pages indicated by Newsguard as untrustworthy. Indeed, we found approximately 3,500 retweets of this type in AfD, about 200 in MEDIA, 20 in LEFT-WING and even none in GOVERNMENT. For AfD, 30\% of them originates from the SCC and ends in the OUT sector,  25\% between IN and OUT and 20\% remains in the SCC. Therefore, in over 50\% of the cases an user shares untrustworthy contents from the SCC.

\section{French Covid-19 dataset}
\begin{figure}
    \centering
    \includegraphics[scale=0.35]{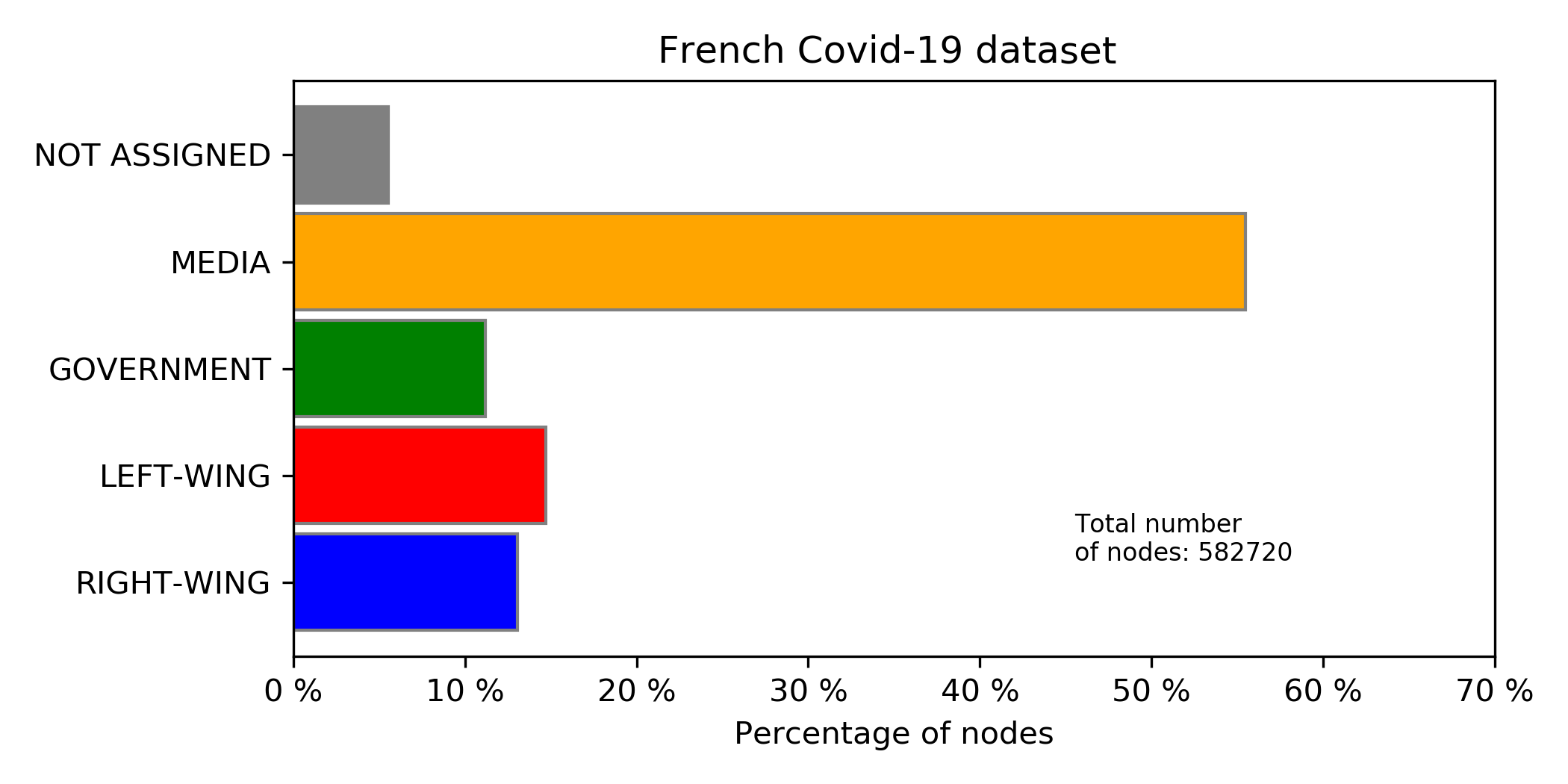}
    \caption{\textbf{The dimension of the discursive communities of French Covid-19 dataset.}\\\fab{The bar chart displays} the percentage of nodes in each discursive community. The MEDIA group results the most numerous one, with approximately 60\% of the nodes, as it happens in the other Covid-19 datasets.}
    \label{discursive_fr}
\end{figure}

The French Covid-19 dataset consists in \fab{3,052,708} posts published between March 23 and April 7 about the epidemic. We identified 4 different discursive communities:
\begin{itemize}
    \item \textbf{RIGHT-WING}: it collects conservatives and right-oriented accounts from French parties like ``Rassemblement National", ``Les Républicains" and ``Les Identitaires";
    \item\textbf{LEFT-WING}: in this community there are the politicians and the supporters of center-left French parties like ``La France Insoumise" or the socialists party (``Parti Socialiste");
    \item\textbf{GOVERNMENT}: it collects accounts of institutions and ministries like the official account of the French government or that of ``Ministère des solidarités et de la santé" (Ministry of Solidarity and Health). It also contains politicians from the party ``La République En Marche", whose leader is president Macron;
    \item\textbf{MEDIA}: this is the usual community containing official accounts of various media and journalists.
\end{itemize}
\begin{table}
    \centering
    \begin{tabular}{|l|r|r|}
\hline
 COMMUNITY   &   DIM. (verified accounts) &   DIM. (all accounts) \\
\hline
 R.-W.       &                     194 &           76243 \\
 L.-W.       &                     228 &           85750 \\
 GOV.        &                     397 &           65190 \\
 MEDIA       &                     303 &          323280 \\
\hline
\end{tabular}
    \caption{\fab{Dimension of the discursive communities of the French Covid-19 dataset,  considering verified accounts only and all users.}}
    \label{tab:dim_covid_fr}
\end{table}
As for the other Covid-19 datasets, the MEDIA community results the most numerous one, with approximately 60\% of the nodes of the network (Fig.~\ref{discursive_fr}). \fab{Tab.~\ref{tab:dim_covid_fr} shows the communities' dimensions.}

For this dataset we could not make the comparisons with the predictions of the \fab{entropy-based null-}model because of the huge dimension of its discursive communities. For these groups, the computation time for generating \fab{a sample of the graph} ensemble and analysing their bow-tie structure became too long. Therefore, in the following \fab{plots} there will be no information about the \fab{p-values}.%, as the color of the nodes in Fig.~\ref{covid_fr}.% or of the bars in Fig.~\ref{pageranks_fr}.
\begin{figure}
    \centering
    \includegraphics[scale=0.28]{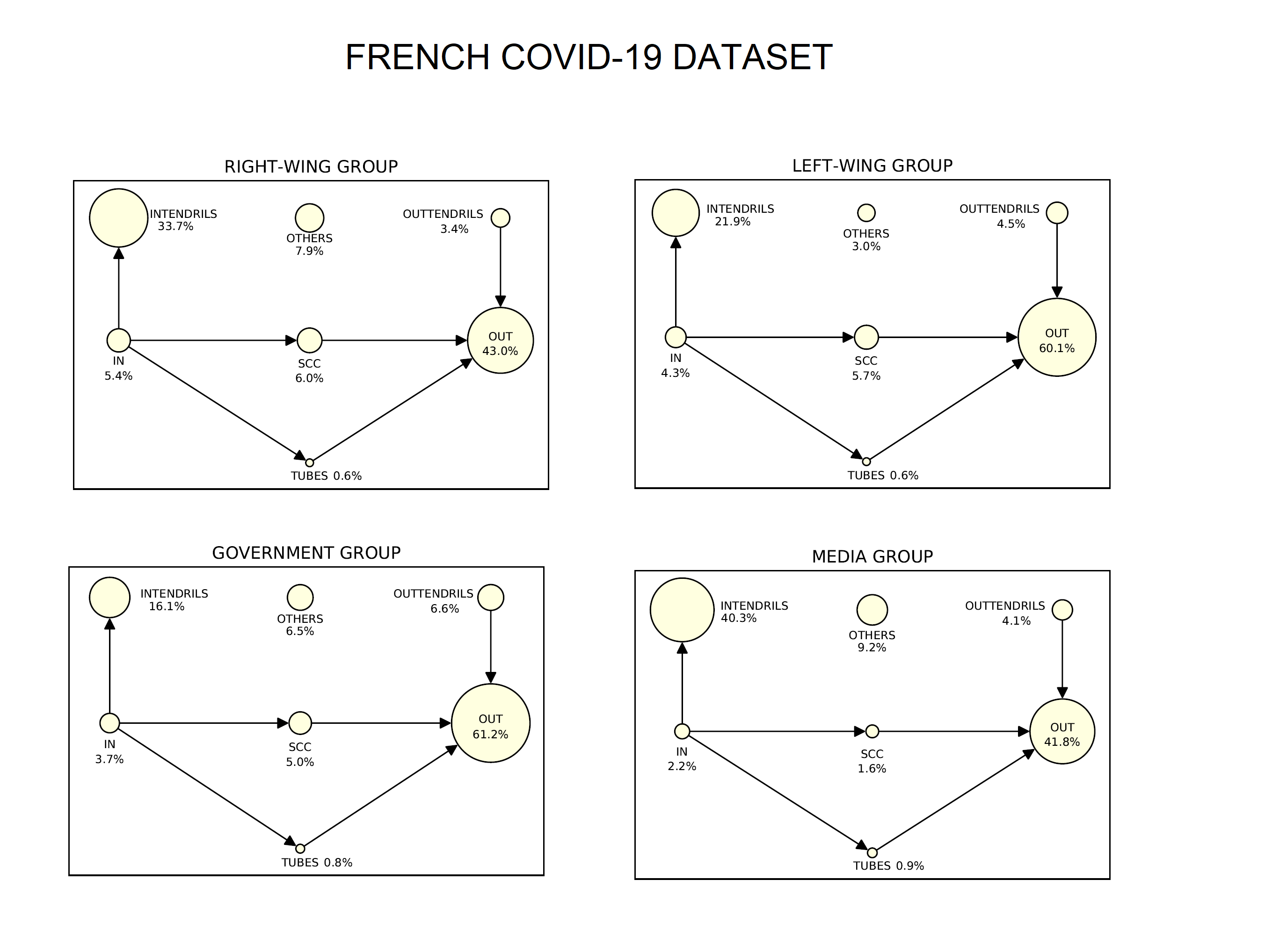}
    \caption{\textbf{The bow-tie structure of the discursive communities of the French Covid-19 dataset.} \fab{All} discursive communities have informative bow-tie structures and all of them are OUT-dominant.}
    \label{covid_fr}
\end{figure}

\begin{figure}
    \centering
    \includegraphics[scale=0.4]{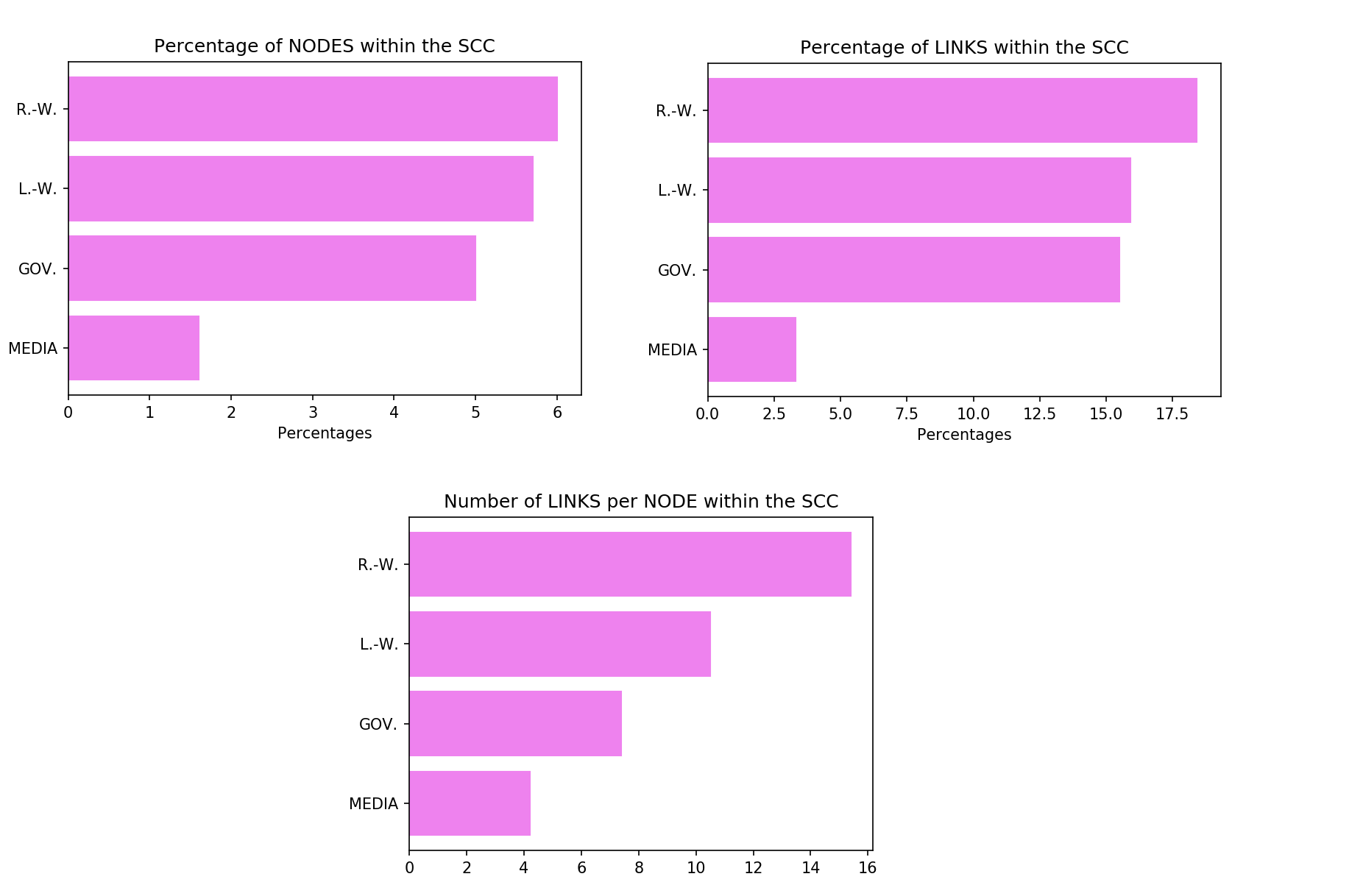}
    \caption{\textbf{Percentage of nodes and edges in the SCC for the communities in the French Covid-19 dataset.} The right-wing community has more links and nodes in the SCC, even if this time in much less extent respect to the other datasets.}
    \label{conservatives_fr}
\end{figure}

\fab{Fig.~\ref{covid_fr} shows} that each discursive community displays an informative bow-tie structure. Remarkably, each of them are OUT-dominant ones, with no less of 40\% of the nodes in every OUT sector. %This affects also the results about the location of the top PageRanks (Fig.~\ref{pageranks_fr}): they are all located in the SCC for every community. 
The mismatch in the number of nodes and links in the SCC between the right-wing community and the others still holds, but at much less extent, see Fig.~\ref{conservatives_fr}.
\fab{When matching Newsguard's data with Twitter's ones, we get results similar to the ones observed in the other datasets}: we found 979 retweets with urls to untrustworthy web-pages in the right-wing community, 103 in the left-wing and none in the others. For the former community 25\% are located between SCC and OUT, 22\% between IN and OUT, 14\% in the SCC, 12\% between IN and SCC and much less between the other sectors. In the case of the left wing, 45\% of these retweets are located between SCC and OUT and 31\% in the SCC.

\section{Dutch elections dataset}
The Dutch elections dataset consists in \fab{1,002,499} tweets posted between February 2 and March 31 2021. In this case almost each discursive community has the name of a specific Dutch politcal party: ``GroenLinks" (center-left, green), ``Christian Democratic Appeal (CDA)" (center, Christian-democratic), ``Democrats 66 (D66)" (center/center-left, liberal), ``People's Party for Freedom and Democracy (VVD)" (conservative-liberal) and ``Labour Party (PvdA)" (center-left, social-democratic). Then we have the CONSERVATIVES community which collects accounts from right-oriented parties like ``Party for Freedom" or ``Forum for Democracy" and the MEDIA \& S.P., which is the usual MEDIA community with a couple of accounts belonging to the Dutch ``Socialist Party".
\begin{figure}
    \centering
    \includegraphics[scale=0.35]{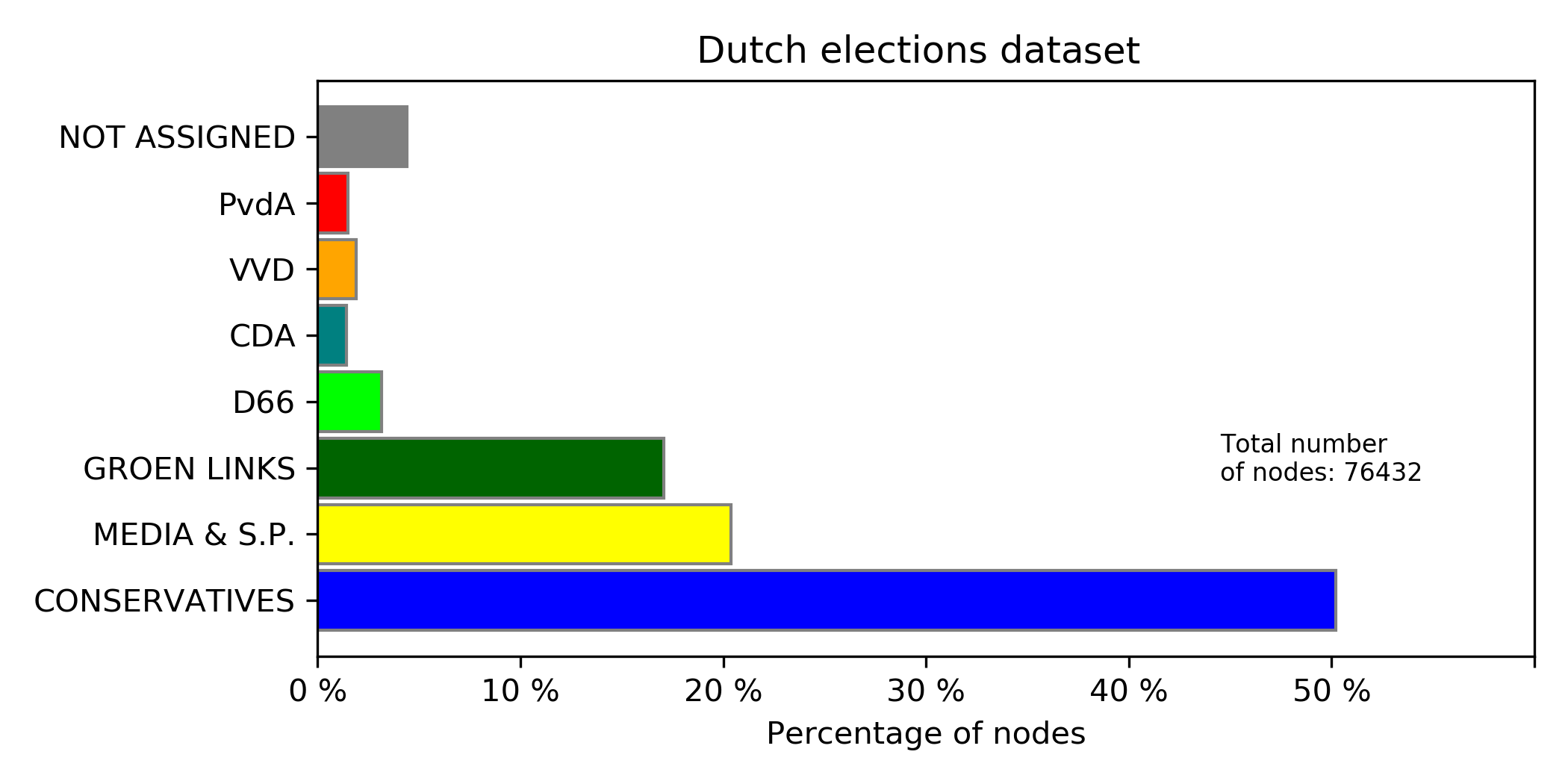}
    \caption{\textbf{The dimension of the discursive communities of the Dutch elections dataset.}\\The bar chart displays the percentage of nodes in each discursive community. The CONSERVATIVES group results by far the most numerous one, with approximately 50\% of the nodes of the entire network.}
    \label{discursive_dutch}
\end{figure}
Fig.~\ref{discursive_dutch} and \fab{Table~\ref{tab:dim_dutch_el} show the dimension of these seven discursive communities}.
\begin{table}[]
    \centering
    \begin{tabular}{|l|r|r|}
\hline
 COMMUNITY    &   DIM. (verified accounts) &   DIM. (all accounts) \\
\hline
 CONS.        &                      41 &           38366 \\
 MEDIA \& S.P. &                      39 &           15570 \\
 GROEN LINKS  &                      36 &           13057 \\
 D66          &                      31 &            2409 \\
 CDA          &                      18 &            1081 \\
 VVD          &                      15 &            1453 \\
 PvdA         &                      11 &            1133 \\
\hline
\end{tabular}
    \caption{\fab{Dimension of the discursive communities of the Dutch Elections dataset, before and after the label propagation procedure (i.e., considering verified accounts only and all users)}.}
    \label{tab:dim_dutch_el}
\end{table}
\begin{figure}
    \centering
    \includegraphics[scale=0.27]{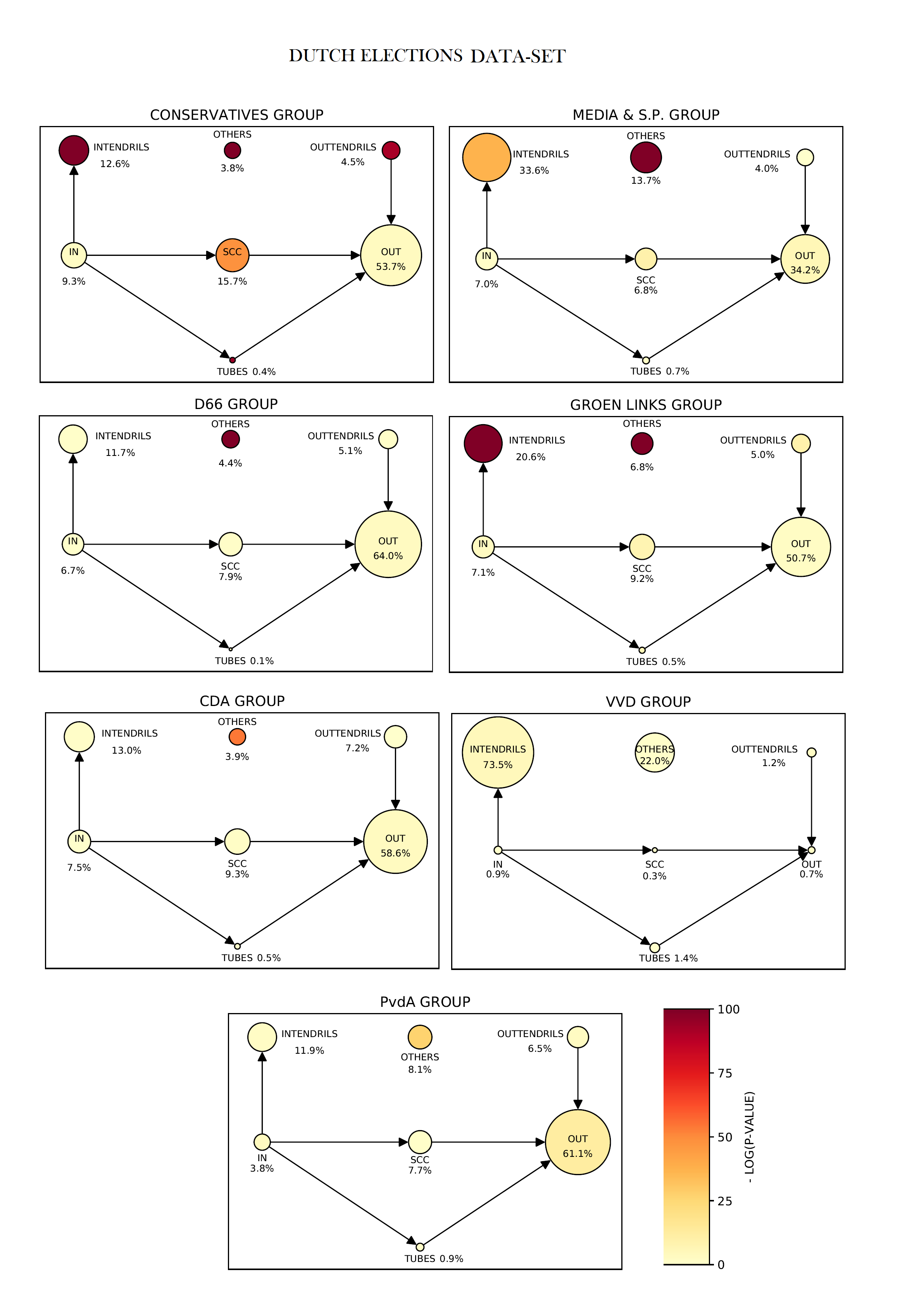}
    \caption{\textbf{The bow-tie structure of the discursive communities of the Dutch elections dataset.}\\ All the discursive communities in this dataset display a strong bow-tie structure, but the VVD one.}
    \label{dutch_el}
\end{figure}
As it is possible to observe in Fig.~\ref{dutch_el}, all the discursive communities show an informative bow-tie structure, with the only exception of VVD.% displays a weak bow-tie, with over the 70\% of the nodes in the INTENDRILS. \\

\begin{comment}
Also the location of the top PageRanks in this case follows our previous results: in the discursive communities with a strong bow-tie they are mainly located in the SCC (only in the MEDIA \& S.P. group there is a balanced situation between SCC and OUT, see Fig.~\ref{pageranks_dutch}) while for the VVD group are principally placed in the OUTTENDRILS (remember that for the assigning the PageRanks we inverted the direction of the edges, therefore it corresponds to the previous INTENDRILS).
\begin{figure}
    \centering
    \includegraphics[scale=0.3]{pageranks_dutch_el.png}
    \caption{\textbf{Top PageRanks location in the Dutch elections dataset.}\\ In the bar charts are displayed the number top PageRanks located in each sector of the discursive communities. For each of them we selected the 100 vertices with the highest PageRanks. In almost all the networks with a strong bow-tie the most part of the top PageRanks are placed in SCC. The VVD is the only one in which they are more distributed across the sectors, with the OUTTENDRILS sector as the most occupied ones.}
    \label{pageranks_dutch}
\end{figure}
\end{comment}

\begin{figure}
    \centering
    \includegraphics[scale=0.75]{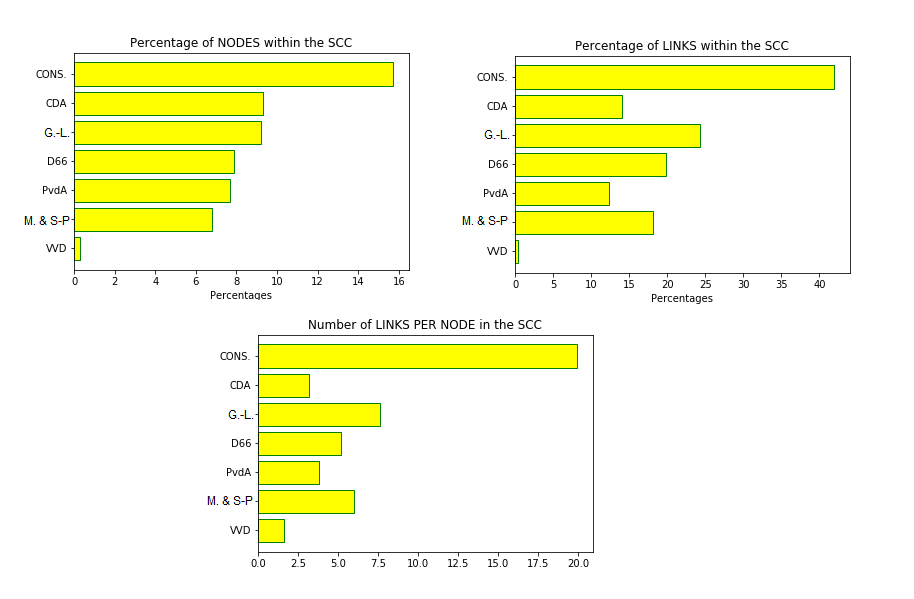}
    \caption{\textbf{Percentage of nodes and edges in the SCC for the communities in the Dutch elections dataset.}\\The conservative and right-oriented discursive community (CONS.) has more numerous and denser SCCs, as it is displayed in the two top panels. In the bottom panel, it can be seen that also considering the fraction of links per node in the SCC, the CONS. group outperforms other communties.}
    \label{conservatives_dutch}
\end{figure}

The CONSERVATIVES group results again that community with the highest percentages of nodes and links within SCC (see Fig.~\ref{conservatives_dutch}).
\fab{Indeed, SCC contains more than 40\% of the links of the entire network}. \\
In this dataset are present \fab{just a} few retweets containing urls to untrustworthy web-pages (Newsguard). However, they are all located in the CONSERVATIVES community: 153 in total, whose 55\% in the SCC and 45\% between SCC and OUT.

\section{Italian debate on migrants dataset}
\begin{figure}
    \centering
    \includegraphics[scale=0.35]{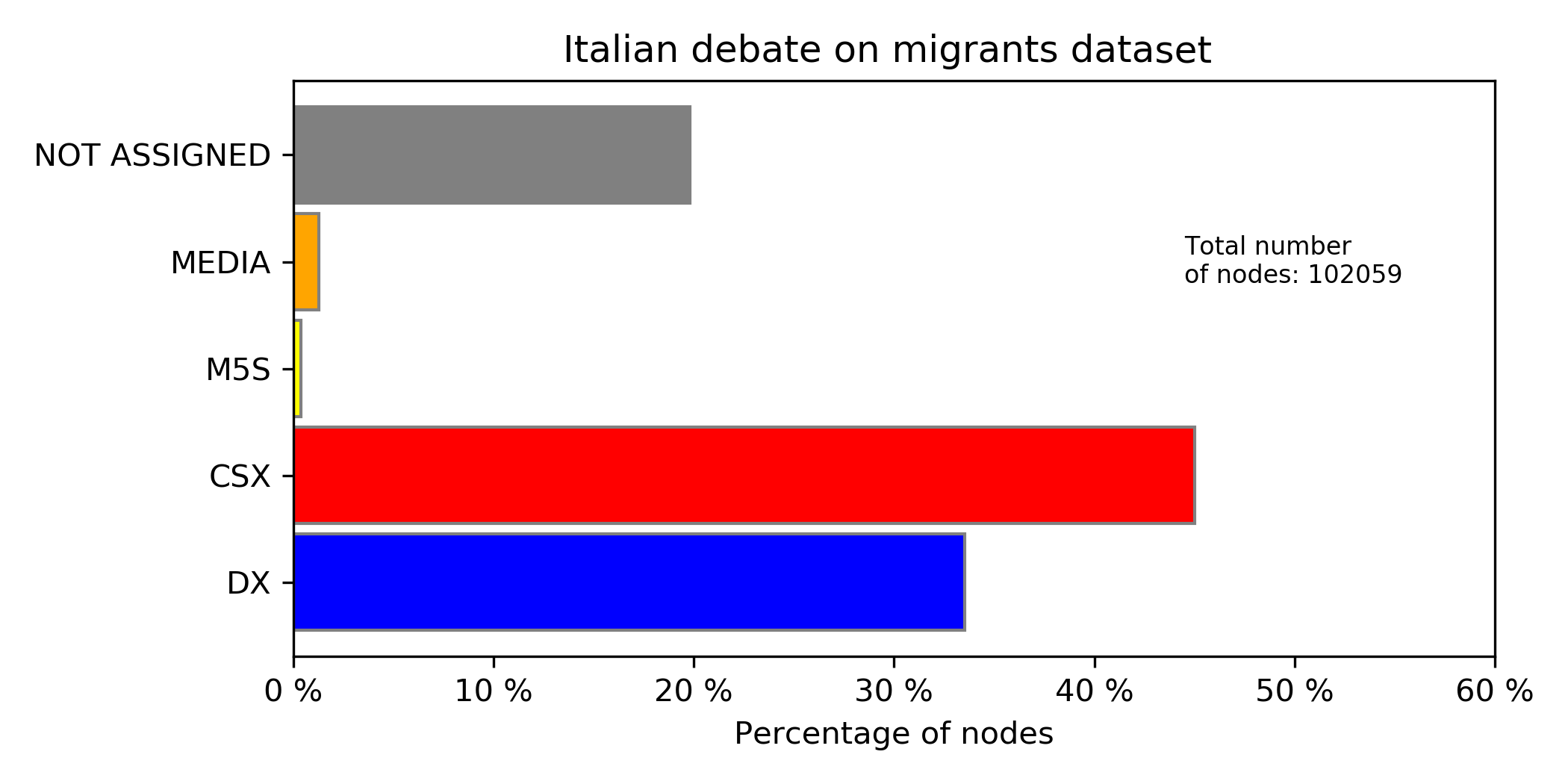}
    \caption{\textbf{The dimension of the discursive communities of Italian debate on migrants dataset.}\\The bar chart displays the percentage of nodes in each discursive community. The DX and CSX groups result the most numerous ones, with between 30\% and 50\% of the nodes. \fab{Among the ones we analysed, t}his is the only dataset in which the percentage of nodes not assigned to a discursive community by the label propagation procedure overcome 15\%.}
    \label{discursive_migr}
\end{figure}

\begin{figure}
    \centering
    \includegraphics[scale=0.3]{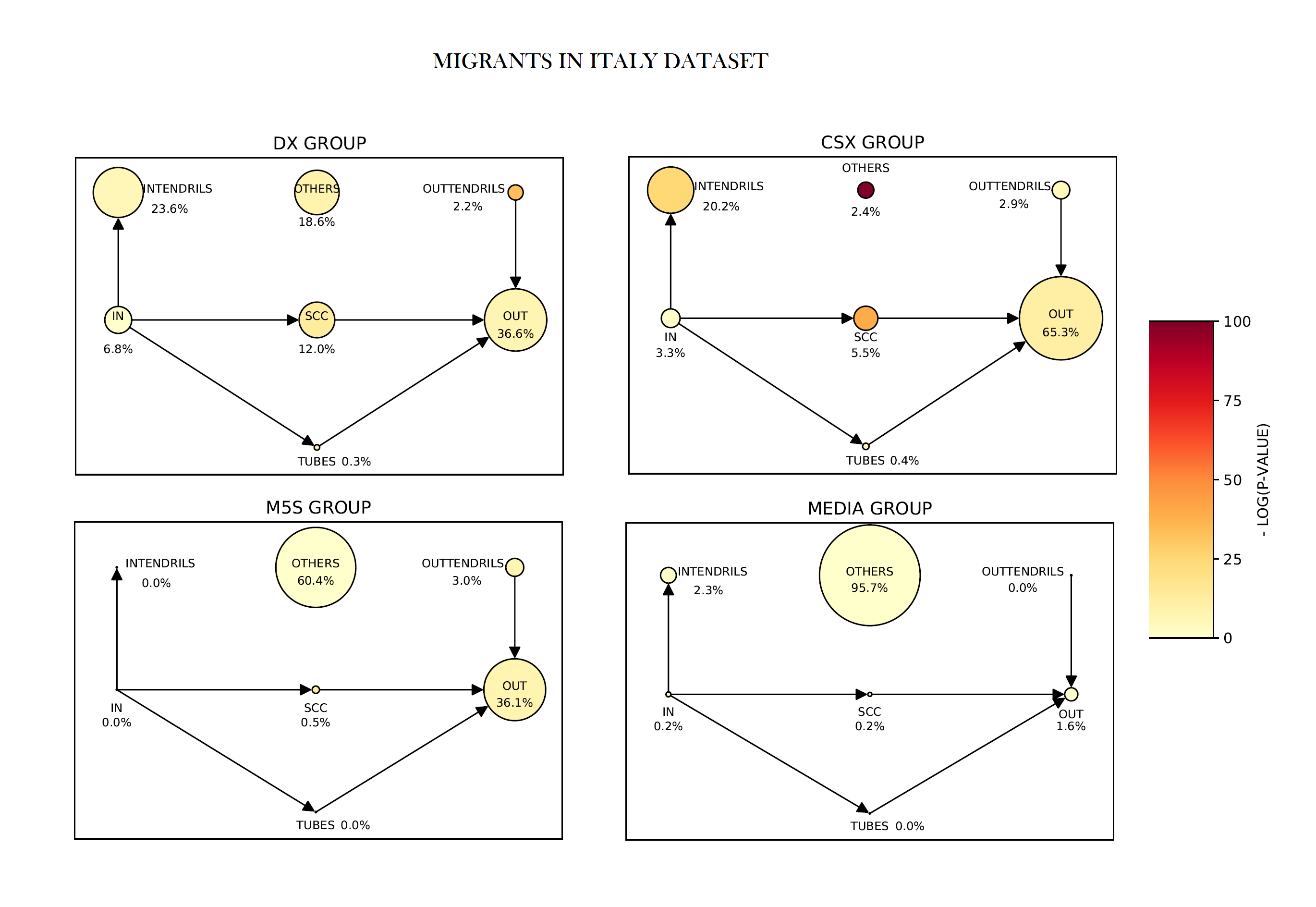}
    \caption{\textbf{The bow-tie structure of the discursive communities of the Italian debate on migrants dataset.} \\ The most numerous community of DX and CSX display informative bow-ties (only the CSX one is strong), while in the smaller ones of M5S and MEDIA the nodes are mostly located in the OTHERS sector. Looking to the colors in the graphs, \fab{M5S and MEDIA are} more in agreement with the predictions of the DCM.}
    \label{migrants}
\end{figure}

This Italian dataset contains Twitter posts about the migration flows from Northern Africa. The dataset consists of \fab{1,081,780} posts published between January 23, 2019 and February 22, 2019.  \\
\fab{Using Louvain communitiy detection algorithm (see the main text), the network was partitioned in the following communities:} DX (right-oriented Italian parties as ``Lega Nord"), CSX (left-oriented Italian parties as the Democratic Party and other minor center-left parties), M5S (``Five Star Movement" party) and the usual MEDIA community. The first two communities result the most numerous ones (see Fig.~\ref{discursive_migr} and \fab{Table~\ref{tab:dim_migrants}}).
\begin{table}[]
    \centering
   \begin{tabular}{|l|r|r|}
\hline
 COMMUNITY   &   DIM. (verified accounts) &   DIM. (all accounts) \\
\hline
 DX          &                      17 &           34235 \\
 CSX         &                      51 &           45940 \\
 M5S         &                      14 &             396 \\
 MEDIA       &                      25 &            1293 \\
\hline
\end{tabular}
    \caption{\fab{Dimension of the discursive communities in the dataset about the Italian debate on migrants, considering verified accounts only and all accounts}.}
    \label{tab:dim_migrants}
\end{table}
\\
In Fig.~\ref{migrants} there are the bow-tie structures for the four discursive communities in this dataset.
The most numerous community of DX and CSX display informative bow-ties while in the two smaller ones the nodes are mostly located in the OTHERS sector, especially for the MEDIA community (above 95\%). Looking to the colors in the graphs, in general, the latter ones result more in agreement with the Direct Configuration Model.

\begin{comment}
Differences between the first two communities and the other two can be seen also in the location of the top PageRanks. Indeed, in DX and CSX they are mostly located in the SCC while in MEDIA and M5S in the OTHERS sector, as we could expect (see Fig.~\ref{pageranks_mig}).
\begin{figure}
    \centering
    \includegraphics[scale=0.4]{pageranks_migrants.png}
    \caption{\textbf{Top PageRanks location in the Italian debate on migrants dataset.}\\ In the bar charts are displayed the number top PageRanks located in each sector of the discursive communities. For each of them we selected the 100 vertices with the highest PageRanks. In the first two communities (DX and CSX) with strong bow-ties the most part of the top PageRanks are placed in the SCC. For M5S and MEDIA they are placed in the OTHERS and there is also a relevant percentage of blue-bars, i.e. of vertices with a PageRanks significantly lower than what we expect from the DCM.}
    \label{pageranks_mig}
\end{figure}
In the latter ones there is also a relevant percentage of blue-bars, i.e. of vertices with a PageRanks significantly lower than the predictions. \\
\end{comment}

\begin{figure}
    \centering
    \includegraphics[scale=0.75]{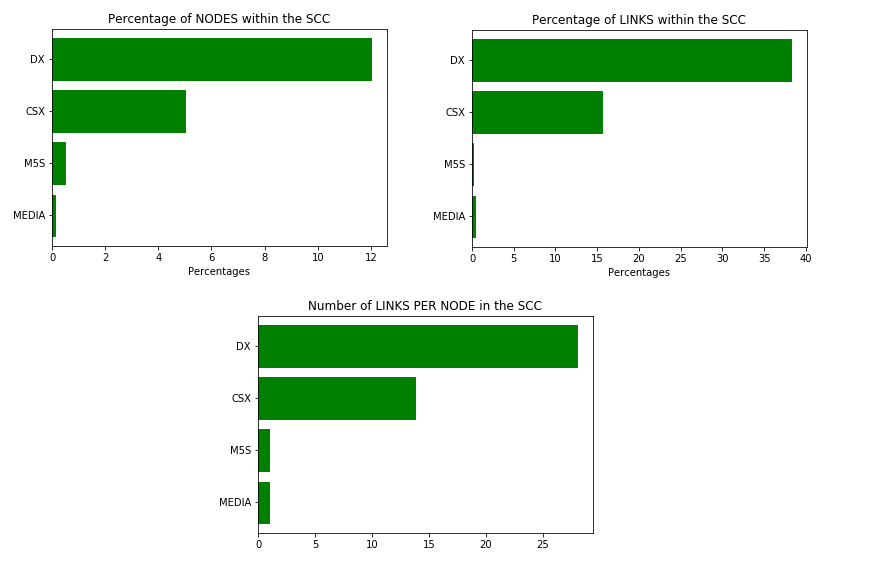}
    \caption{\textbf{Percentage of nodes and edges in the SCC for the communities in the Italian debate on migrants dataset.}\\Again, the conservative and right-oriented discursive community (DX) has more numerous and denser SCCs, as it is displayed in the two top panels. In the bottom panel, it can be seen that also considering the fraction of links per node in the SCC, the DX group results again the first one.}
    \label{conservatives_mig}
\end{figure}

The DX community, which again contains politicians of right-oriented Italian parties, has the most numerous and denser SCC (Fig.~\ref{conservatives_mig}), such that on average a node therein has over 25 links.\\
Newsguard data \fab{tell us} that 15,160 retweets in the DX network contain the urls of untrustworthy web-pages, while only 14 for the CSX, 3 for MEDIA and none for M5S. In the case of the DX community, 59\% of them can be found inside the SCC and 36\% between the SCC and OUT.

\section{Italian debate on Astrazeneca vaccine}
\begin{figure}
    \centering
    \includegraphics[scale=0.35]{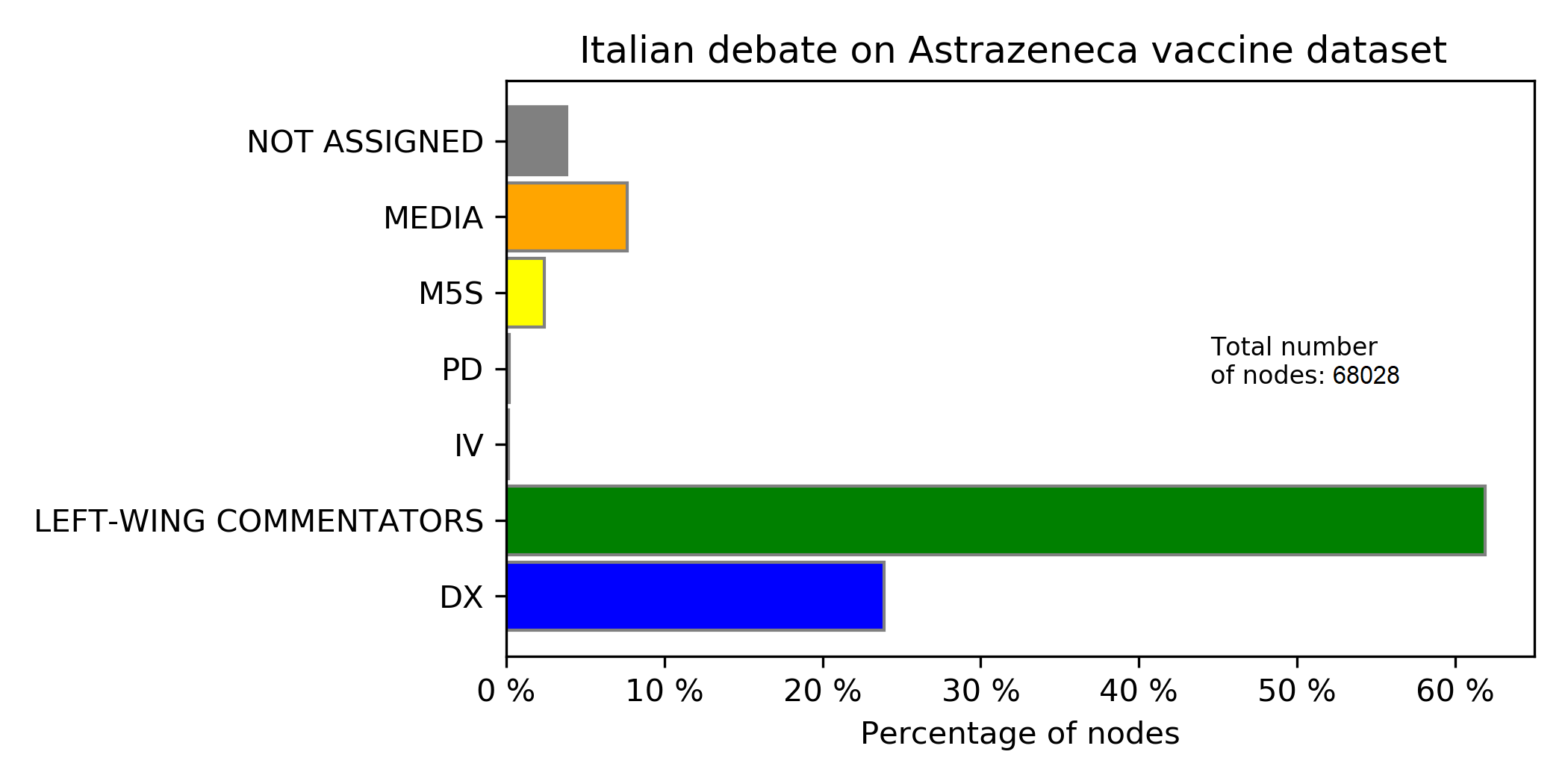}
    \caption{\textbf{The dimension of the discursive communities of Italian debate on Astrazeneca vaccine dataset.}\\In this bar chart the percentage of nodes in each discursive community is displayed. All communities but MEDIA and PD display informative bow-ties, and, among them, the DX and IV ones are strong. The LEFT-WING COMMENTATORS group results the most numerous one, with over 60\% of the nodes.} %This is probably because it contains the accounts of the most famous Italian epidemiologist.}
    \label{discursive_astra}
\end{figure}

\begin{figure}
    \centering
    \includegraphics[scale=0.27]{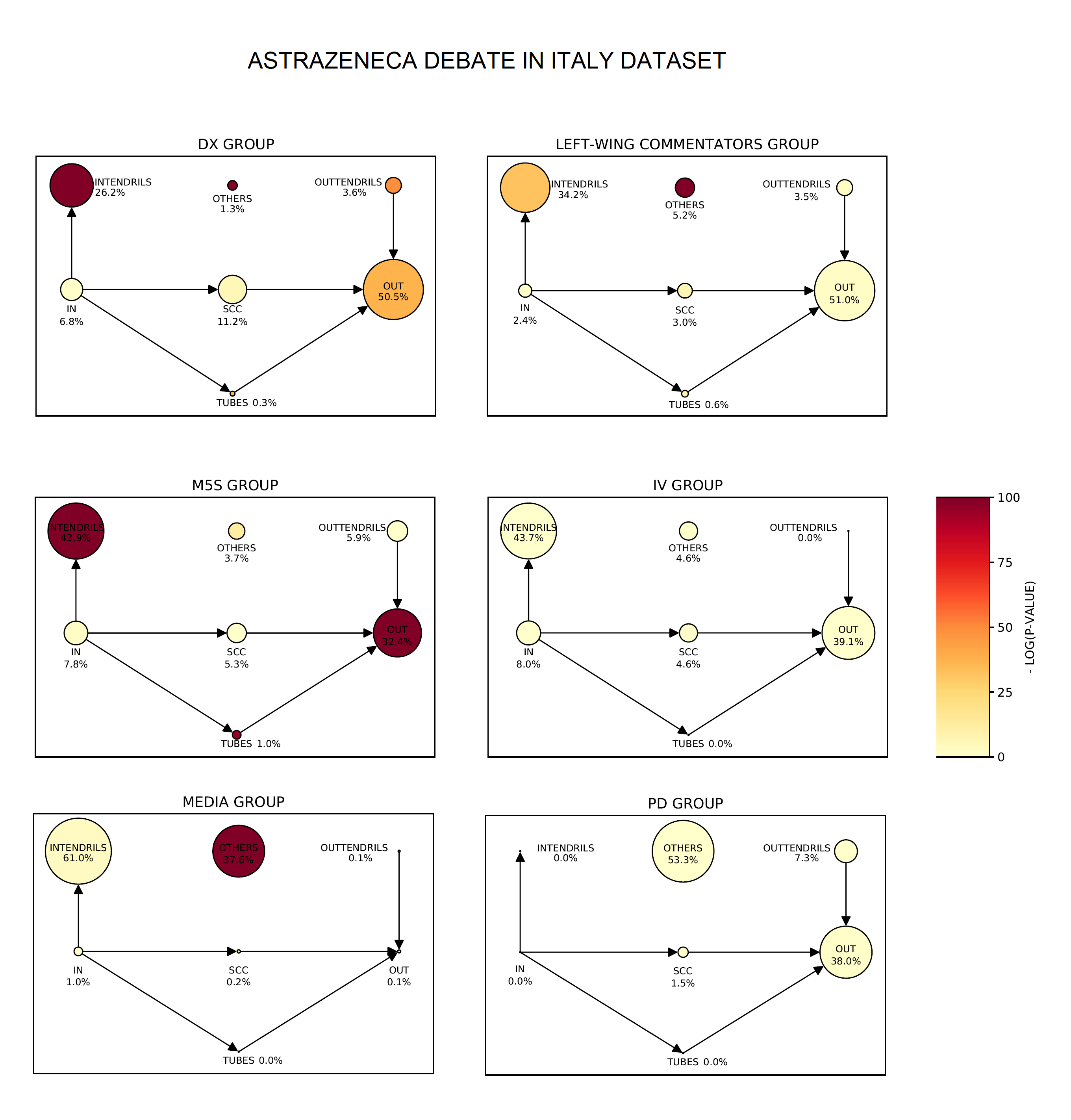}
    \caption{\textbf{The bow-tie structure of the discursive communities of the Italian debate on Astrazeneca vaccine dataset.}\\ The DX and LEFT-WING COMMENTATORS communities display informative bow-tie structure; nevertheless they are both OUT-dominant ones. In the weak bow-ties of M5S and IV there is a balanced situation between INTENDRILS and OUT as the dominant sector. The PD community does not display an informative bow-tie structure, with over 50\% of the nodes in the OTHERS.}
    \label{astrazeneca}
\end{figure}

This dataset contains \fab{583,236} Twitter posts published in Italy and regarding the discussion about the safety of Astrazeneca vaccine against Covid-19. The dataset contains posts shared between March 15, 2021 and May 15, 2021. \fab{we identified the following discursive communities}:
\begin{itemize}
    \item \textbf{DX}: this is the usual right-oriented and conservatives community found even in the other Italian datasets, i.e. it contains accounts from the ``Lega" and ``Fratelli d'Italia" parties;
    \item\textbf{PD}: the Italian Democratic Party (center-left);
    \item\textbf{IV}: it collects the politicians of the ``Italia Viva" party (center-left);
    \item\textbf{LEFT-WING COMMENTATORS}: this particular community is formed by several well-known personalities, often left-oriented, which are not politicians but journalists, blogger, actors or entertainers. This community contains also the most famous Italian epidemiologist Roberto Burioni;
    \item\textbf{M5S}: the Italian populist party ``Movimento 5 Stelle";
    \item\textbf{MEDIA}: the usual community containing official accounts of newspaper, blog, TV-channels, radio and others.
\end{itemize}
The distribution of the nodes in these six communities is showed in Fig.~\ref{discursive_astra} \fab{and the related dimensions are shown in Table~\ref{tab:dim_astraz}. }
\begin{table}[]
    \centering
    \begin{tabular}{|l|r|r|}
\hline
 COMMUNITY   &   DIM. (verified accounts) &   DIM. (all accounts) \\
\hline
 DX          &                      22 &           16249 \\
 L.-W. COMM. &                      17 &           42112 \\
 M5S         &                      19 &            1635 \\
 MEDIA       &                      34 &            5197 \\
 IV          &                      11 &              88 \\
 PD          &                      12 &             140 \\
\hline
\end{tabular}
    \caption{\fab{Dimension of the discursive communities in the dataset on the Italian debate on Astrazeneca vaccine, before and after the label propagation procedure (i.e., considering verified accounts only or all accounts).}}
    \label{tab:dim_astraz}
\end{table}
The two biggest communities, DX and LEFT-WING COMMENTATORS, show a respectively strong and weak bow-tie structures (Fig.~\ref{astrazeneca}), denoting, once more, that the strength of the structure does not depend on its dimension. Nevertheless, thery are both OUT-dominant. In the M5S and IV ones there is a nearly balanced situation between INTENDRILS and OUT as the dominant sector. While MEDIA bow-tie is poorly informative, the PD community is not informative, with over 50\% of the nodes in the OTHERS.

\begin{comment}
For the DX community, as usual for communities with strong bow-ties, the top PageRanks are mostly placed in the SCC (Fig.~\ref{pageranks_astra}). In LEFT-WING COMMENTATORS, IV and M5S groups, the SCC and the OUT sectors contain very similar number of vertices with high PageRanks. The most unusual situations can be found in the PD and MEDIA communities, in which the top PageRanks are placed respectively in the INTENDRILS an in the OTHERS sectors.
\begin{figure}
    \centering
    \includegraphics[scale=0.4]{pageranks_astrazeneca.png}
    \caption{\textbf{Top PageRanks location in the Italian debate on Astrazeneca vaccine dataset.}\\ In the bar charts are displayed the number top PageRanks located in each sector of the discursive communities. For each of them we selected the 100 vertices with the highest PageRanks. In the DX group, with a strong bow-tie, the most part of the top PageRanks are placed in the SCC. In LEFT-WING COMMENTATORS, IV and M5S groups, the SCC and the OUT sectors contain almost the same number of vertices with high PageRanks. The most unusual situations can be found in the PD and MEDIA communities, in which the top PageRanks are placed respectively in the INTENDRILS an in the OTHERS sectors.}
    \label{pageranks_astra}
\end{figure}\\
\end{comment}

\begin{figure}
    \centering
    \includegraphics[scale=0.4]{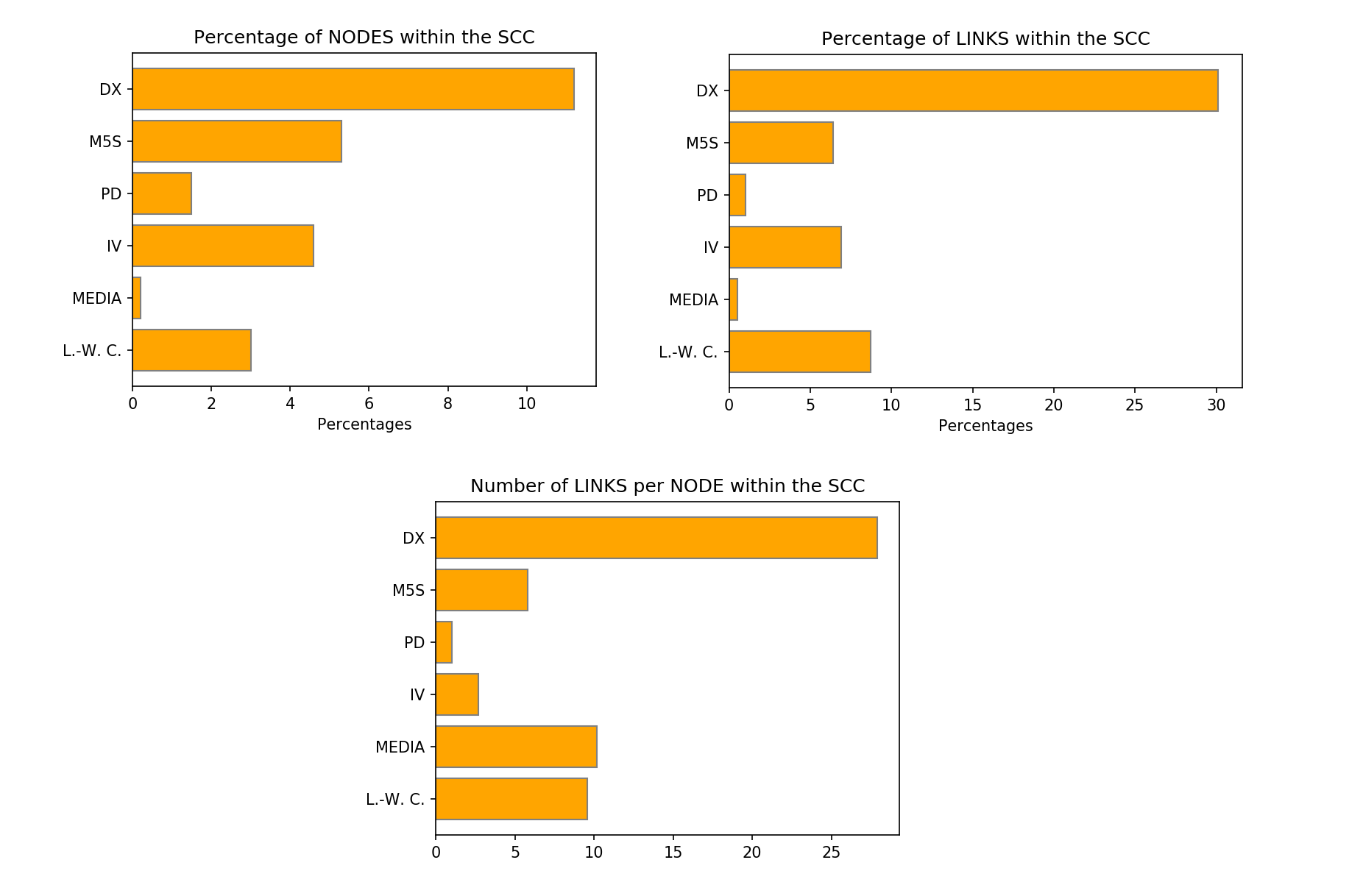}
    \caption{\textbf{Percentage of nodes and edges in the SCC for the communities in the Italian debate on Astrazeneca vaccine dataset.}\\Again, the conservative and right-oriented discursive community (DX) has more numerous and denser SCCs, as it is displayed in the two top panels. In the bottom panel, it can be seen that also considering the fraction of links per node in the SCC, the DX group results again the first one.}
    \label{conservatives_astra}
\end{figure}

The DX community results again the community with the most numerous and denser strongly connected component; \fab{in DX community, the average degree of nodes in SCC is greater} than 27 links, while in the other communities there are always less than 10 links per node (Fig.~\ref{conservatives_astra}). \fab{According to Newsguard}, 728 retweets \fab{in DX community} contained urls to untrustworthy pages. They are distributed as follow: 43\% between SCC and OUT, 26\% in the SCC, 13\% between IN and SCC, 12\% between IN and OUT and much less between the other sectors. We found only two retweets of this type in the M5S community and none in the others.

\end{document}